\documentclass[prb]{revtex4-2}

\usepackage{mathtools}
\usepackage{amsmath,bm}
\usepackage{amssymb}
\usepackage{subfigure}
\usepackage{setspace}
\usepackage{dcolumn}
\usepackage{footnote}
\usepackage{color}

\usepackage[hidelinks]{hyperref}

\begin{document}

%Title of paper
\title{Adiabatic edge-to-edge transformations in time-modulated elastic lattices and non-Hermitian shortcuts}

\author{Emanuele Riva$^{a}$}
\email[]{emanuele.riva@polimi.it}
\author{Gianmaria Castaldini$^{a}$}
\author{Francesco Braghin$^{a}$}
\affiliation{ $^a$ Department of Mechanical Engineering, Politecnico di Milano, Italy, 20156}

\definecolor{rev}{rgb}{0.0, 0.0, 1.0}
\newcommand{\ER}{\textcolor{black}}

%\homepage[]{Your web page}
%\thanks{}
%\altaffiliation{}

\date{\today}

\begin{abstract}
The temporal modulation of a relevant parameter can be employed to induce modal transformations in Hermitian elastic lattices. 
When this is combined with a proper excitation mechanism, it allows to drive the energy transfer across the lattice with tunable propagation rates. 
Such a modal transformation, however, is limited by the adiabaticity of the process, which dictates an upper bound for the modulation speed. In this manuscript, we employ a non-Hermitian shortcut by way of a tailored gain and loss to violate the adiabatic limit and, therefore, to achieve superfast modal transformations. A quantitative condition for adiabaticity is firstly derived and numerically verified for a pair of weakly coupled time-dependent mechanical oscillators, which can be interpreted in the light of modal interaction between crossing states.  It is shown that for sufficiently slow time-modulation, the elastic energy can be transferred from one oscillator to the other. A non-Hermitian shortcut is later induced to break the modal coupling and, therefore, to speed-up the modal transformation. The strategy is then generalized to elastic lattices supporting topological edge states. We show that the requirements for a complete edge-to-edge energy transfer are lifted from the adiabatic limit toward higher modulation velocities, opening up new opportunities in the context of wave manipulation and control.
\end{abstract}

\maketitle
\section{Introduction}
The emergence of novel wave behaviors in physics has been lately accompanied by growing interest in the field of mechanics. This is motivated by a number of potential technologically relevant  applications concerning wave motion and vibrations, such as nondestructive evaluation and imaging \cite{zhu2011holey,datta2020model,tian2020selective}, acoustic/elastic insulation \cite{romero2016perfect,sugino2017general,PhysRevB.84.165136,d2016modeling} and communication \cite{cha2018experimental,cha2018electrical}, among others \cite{hussein2014dynamics}.
As such, the quest for new mechanisms to manipulate elastic waves has involved several research groups and has led to the discovery of elastic/acoustic analogies of important behaviors observed in physics. Topical examples include analogies to the Quantum Hall effect (QH) \cite{PhysRevLett.115.104302,nash2015topological,chen2019mechanical}, Quantum Spin Hall effect (QSH) \cite{miniaci2018experimental,chen2018elastic,susstrunk2015observation}, and Quantum Valley Hall effect (QVH) \cite{pal2017edge,riva2018tunable,liu2018tunable}, which have been employed for the fabrication of scattering-immune waveguides with unprecedented energy transfer capabilities. Also, wave focusing \cite{tol2017phononic}, mode conversion \cite{colombi2017elastic,alan2019programmable}, as well as cloaking \cite{norris2008acoustic,chen2017broadband,zhang2011broadband,quadrelli2021elastic} have recently been observed in mechanics and acoustics.\\ 
An emerging trend leverages temporal (active) modulations of elastic or physical parameters \cite{wang2020tunable} to accomplish different tasks, such as nonreciprocity \cite{nassar2020nonreciprocity,riva2020non,riva2019generalized,marconi2020experimental,attarzadeh2018non,attarzadeh2020experimental,vila2017bloch,trainiti2016non}, parametric amplification \cite{trainiti2019time}, frequency conversion \cite{PhysRevB.98.054109} and edge-to-edge pumping \cite{PhysRevLett.126.095501,PhysRevB.102.014305,grinberg2020robust,PhysRevB.102.174312,liu2020topological,PhysRevLett.125.224301,ni2019observation}. This great variety of behaviors can be systematically observed in the same structure when the relevant parameters are biased with a different modulation speed. A fast modulation leads to scattering of energy to bulk modes and frequency conversion, which is generally undesired for communication and energy transfer purposes and limits the applicability of pumping protocols to a family of slow transfer mechanisms \cite{chaunsali2016stress,PhysRevB.102.174312}.
Such a family involves sufficiently slow, or adiabatic, temporal modulations, which activate a modal transformation that is capable of driving the transfer of energy across a mechanical structure \cite{PhysRevLett.126.095501,PhysRevB.102.014305,PhysRevLett.126.054301,PhysRevB.101.094307,rosa2019edge}. The concept is \textit{de facto} general and is often applied to a pair of coupled (crossing) modes, or states, that undergo a shape transformation when the relevant parameter is varied in an adiabatic manner, i.e. when the modulation speed is lower than a threshold, usually quantified in physics through established tools \cite{ibanez2014adiabaticity,tong2010quantitative,amin2009consistency,yukalov2009adiabatic,yi2007adiabatic,guery2019shortcuts}. In the field of mechanics, however, the theoretical tools to address the conundrum between adiabatic and non-adiabatic transformations are not well established yet. \\
Another line of work explores non-Hermitian structures, that belong to a family of non-conservative systems with non-unitary time evolution of the states, and in many cases are characterized by balanced loss and gain (damping and anti-damping). 
The rich dynamical behavior and the underlying physics makes the study of non-Hermitian systems very attractive and applicable to several realms of physics \cite{ge2018breaking,el2018non,ozdemir2019parity}. 
Relevant examples of non-Hermitian physics include the emergence of skin modes \cite{PhysRevLett.125.118001,PhysRevResearch.2.023173,scheibner2020odd} and nonreciprocal wave propagation in the form of directional amplification \cite{rosa2020dynamics}. 
Also, $\mathcal{PT}$ symmetric non-Hermitian systems characterized by balanced gain/loss have been shown to exhibit a real spectrum, whereby the strength of the balanced action delineates a transition between broken (complex spectrum) and unbroken (real spectrum) $\mathcal{PT}$ symmetric regimes. Such a transition has been employed in recent studies for invisibility and enhanced sensing purposes \cite{wu2019asymmetric,fleury2015invisible,PhysRevLett.116.207601,rosa2021exceptional}.
In the present work, we leverage a non-Hermitian physics to functionally modify the behavior of a mechanical lattice, activating a shortcut for nonadiabatic edge-to-edge transitions \cite{torosov2013non}.\\
The manuscript starts with the analysis of modal transformations in a pair of weakly coupled oscillators, which are biased in stiffness and parametric with a phase parameter $\phi$. A limiting condition for adiabaticity is firstly derived and numerically verified via integration of the equation of motion for different rates $\partial\phi/\partial t$. Then, a non-Hermitian shortcut is induced to increase the speed of energy transfer between the oscillators.
Following the work done in quantum systems \cite{torosov2013non}, the non-Hermitian terms are induced by way of a balanced gain and loss in the form of damping and anti-damping, in which the gain/loss pairs are tailored to decouple the crossing states and therefore to speed-up the transformation. This balanced action inhibits the energy stored in a state to leak to the neighboring modes, which is the key to activate superfast transitions and to drive a complete energy transfer from one oscillator to the other independently of the speed of modulation of the relevant parameter, thereby creating a shortcut to adiabaticity. In the second part of the paper, we pursue the same scope in the context of temporally modulated 1D elastic lattices, whereby each lattice site is equipped with tailored a gain/loss action capable of decoupling an arbitrary pair of crossing states. It is demonstrated that a chain supporting topological boundary modes can be locally biased to achieve superfast edge-to-edge transitions, expanding the range of available pumping protocols in mechanics. \\
The manuscript is organized as follows. In Chapter {\rm II} the theoretical framework is presented and applied to a pair of weakly coupled oscillators, starting from the limiting speed for adiabaticity to the non-Hermitian shortcut. The theoretical framework is extended to elastic lattices in Chapter {\rm III}, in which non-hermitian shortcuts are applied to different lattice configurations. Concluding remarks are reported in chapter {\rm IV}.

\section{Role of non-hermitian shortcuts in non-adiabatic transformations}
\begin{figure}[t]
	\centering
	\hspace{-0.75cm}\subfigure[]{\includegraphics[width=0.22\textwidth]{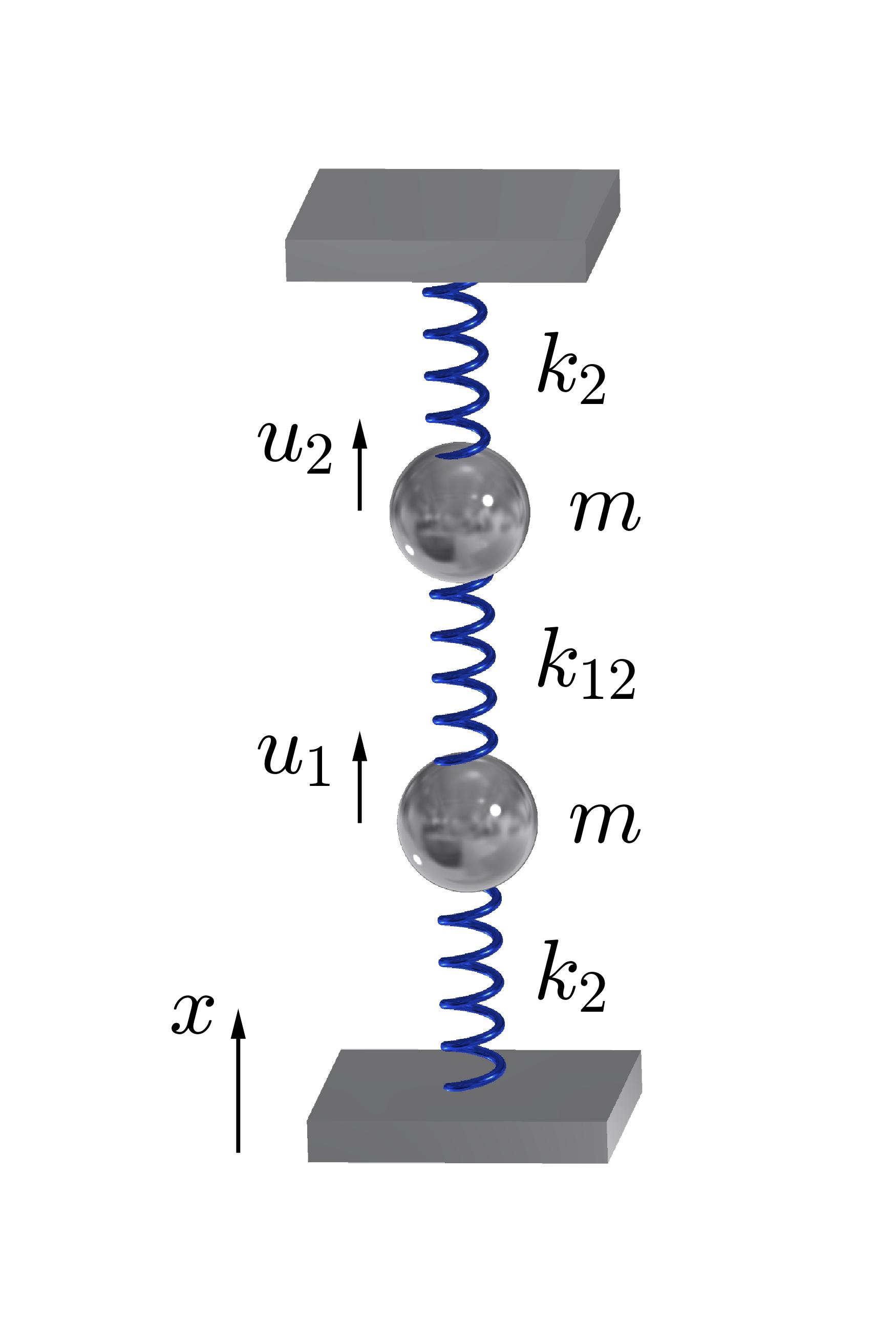}}\hspace{-0.5cm}
	\subfigure[]{\includegraphics[width=0.53\textwidth]{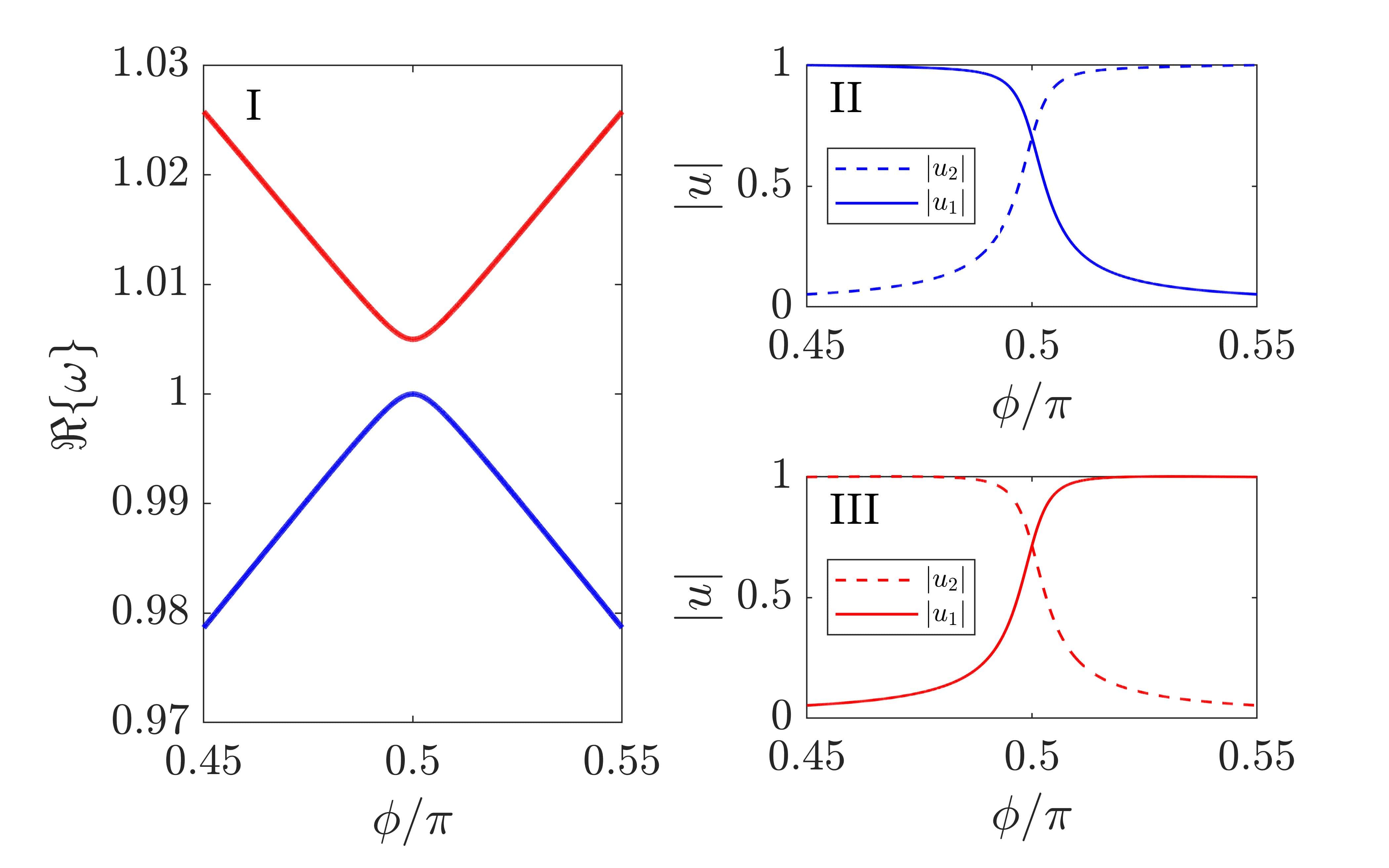}}
	\subfigure[]{\includegraphics[width=0.265\textwidth]{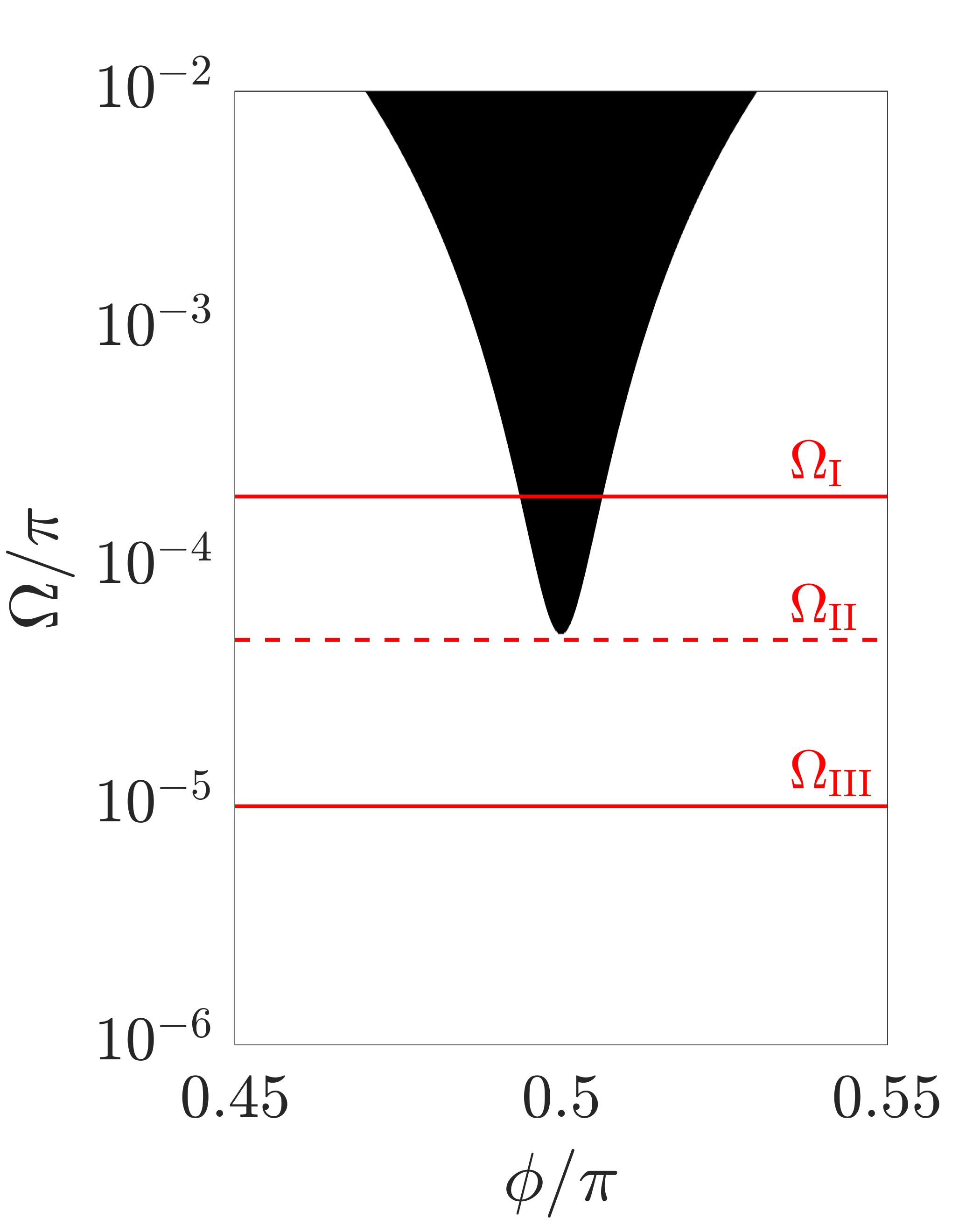}}\\
	\subfigure[]{\includegraphics[width=0.32\textwidth]{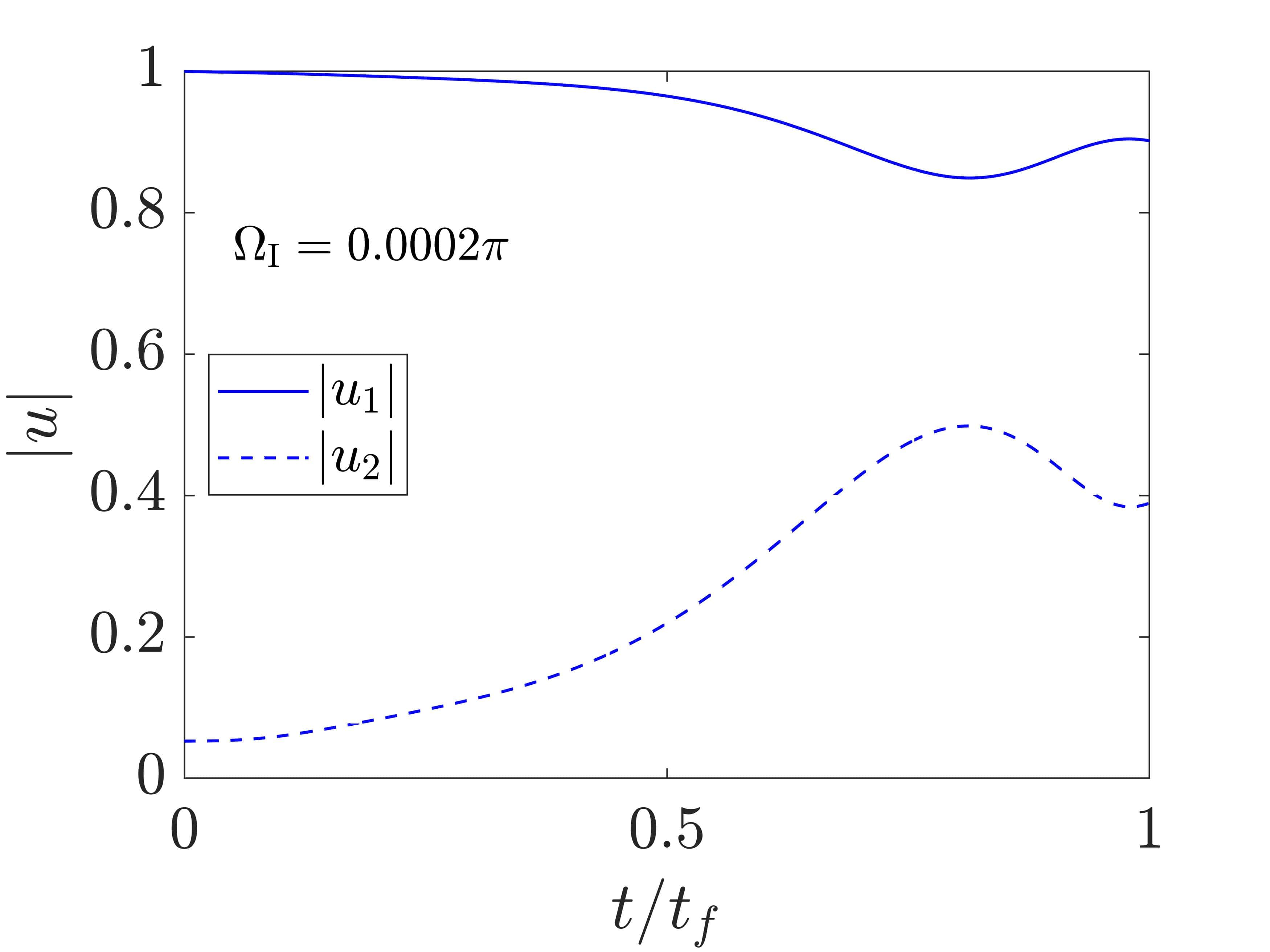}}
	\subfigure[]{\includegraphics[width=0.32\textwidth]{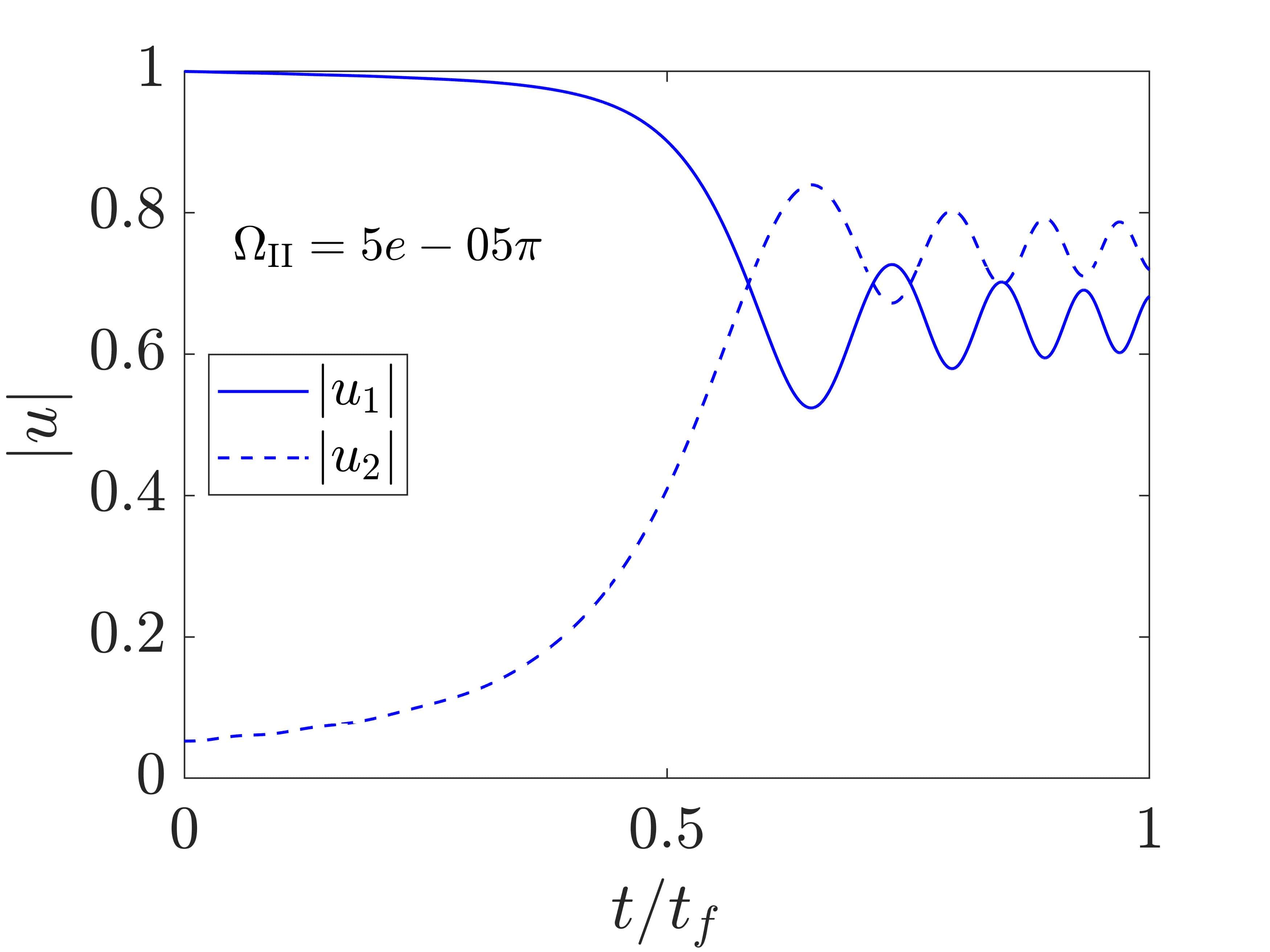}}
	\subfigure[]{\includegraphics[width=0.32\textwidth]{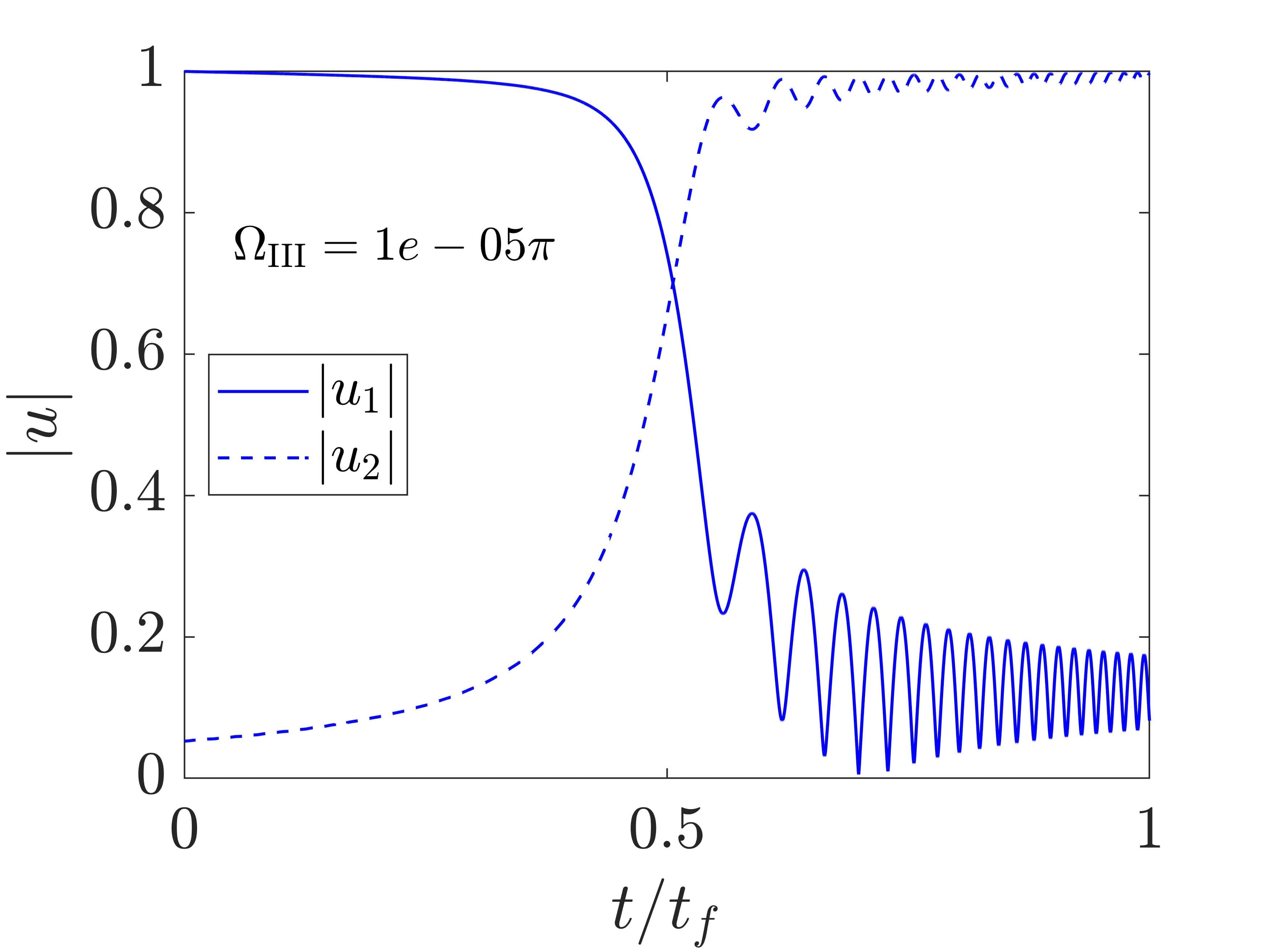}}\\
	\subfigure[]{\includegraphics[width=0.32\textwidth]{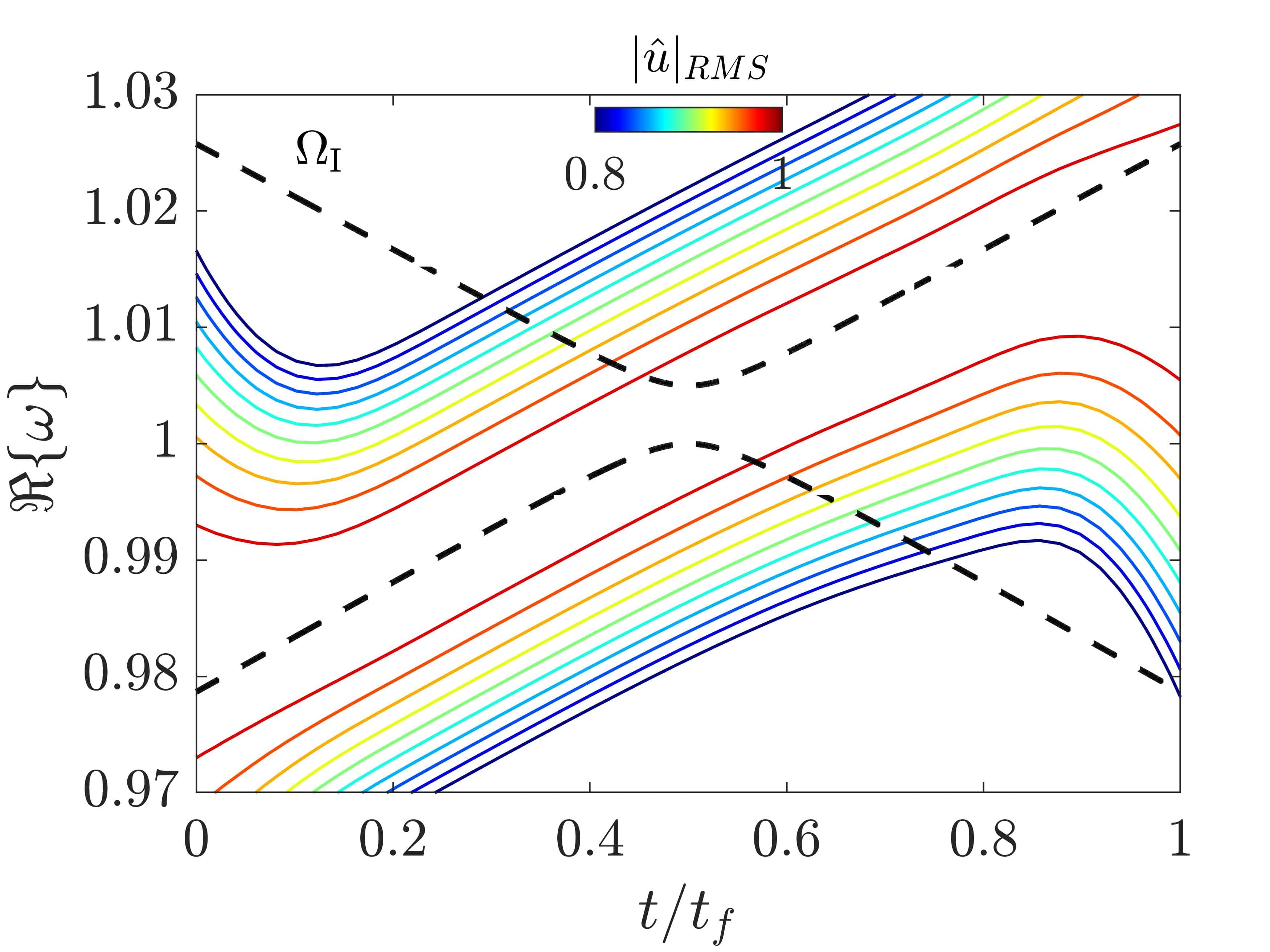}}
	\subfigure[]{\includegraphics[width=0.32\textwidth]{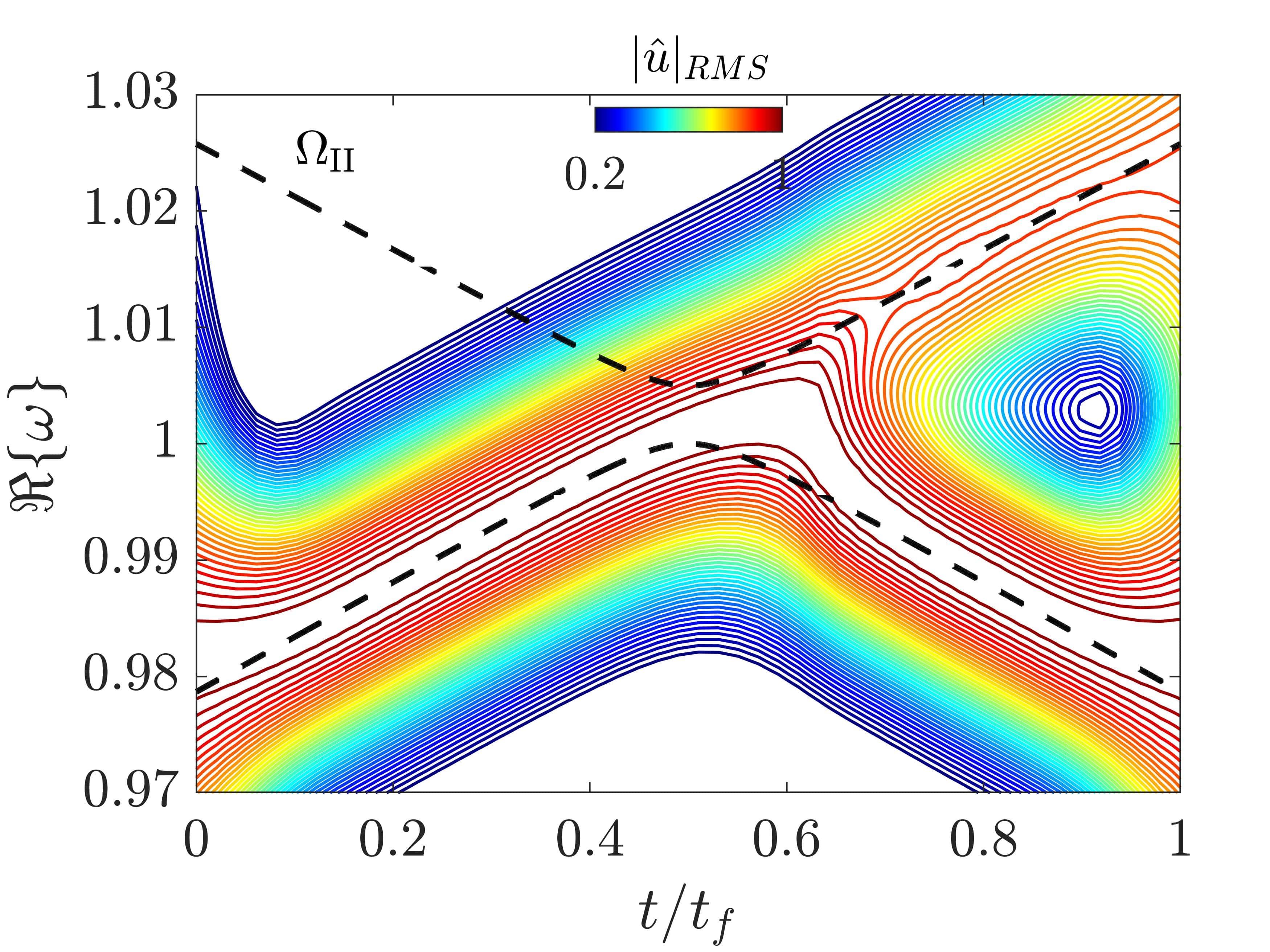}}
	\subfigure[]{\includegraphics[width=0.32\textwidth]{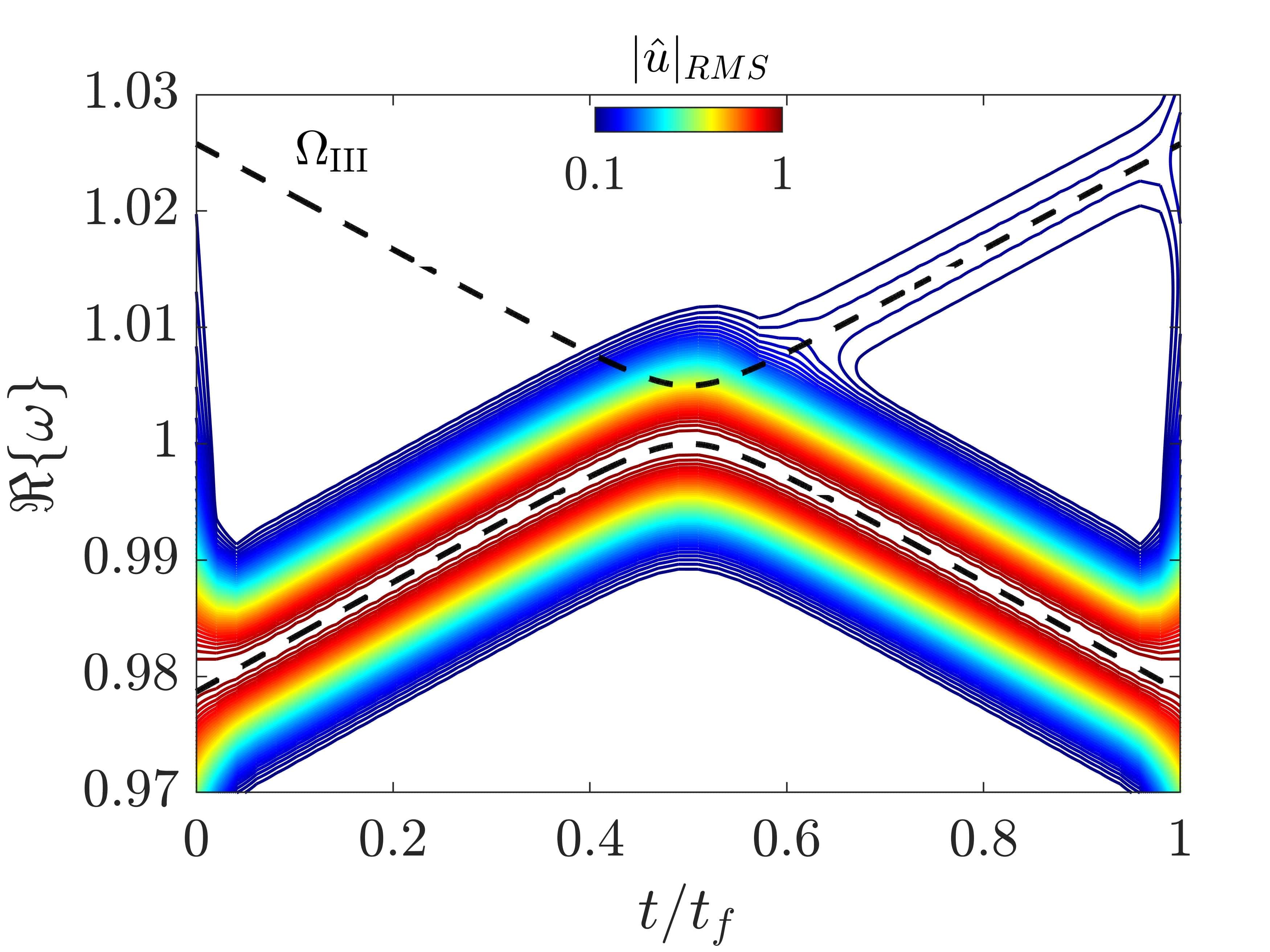}}
	\caption{(a) Graphic representation of a system of weakly coupled oscillators. (b)-I Locus of the eigenfrequencies for a system of coupled oscillators upon varying the modulation phase $\phi$. (b)-II-III The corresponding modal amplitudes are shown alongside the spectrum. (c) Graphic representation of the limiting modulation speed for an adiabatic transformation for different $\phi$ values. The black region identify non-adiabatic transformations, and vice-versa for the white regions. The vertical red lines correspond to the modulation speeds employed in the numerical simulations. (d-f) Transient response to initial conditions for fast, medium and slow modulations, respectively. (g-i) Corresponding spectrograms. The dashed red line describe locus of the available states upon varying $\phi$, while the colored diagram is the Fourier-transformed displacement field. }
	\label{Fig1}    
\end{figure}

We start the discussion considering the system of oscillators illustrated in Fig. \ref{Fig1}(a). Two point masses $m_1=m_2=m=1$ are connected to ground though linear parametric springs $k_1=k_0\left(1-\alpha\cos\left(\phi\right)\right)$ and $k_2=k_0\left(1+\alpha\cos\left(\phi\right)\right)$, where $\alpha=0.3$ is the modulation amplitude (constant), $k_0=1$ and $\phi=\phi\left(t\right)$ is the modulation phase, also called phason, which is a linear function of time $\phi\left(t\right)=\phi_i+\Omega t$. $\Omega$ being the angular velocity. A linear spring $k_{12}=\epsilon k_0$ is placed between the point masses and represents a weak coupling between the displacements $u_1$ and $u_2$, for a sufficiently small value for $\epsilon=0.005$. In absence of damping, the equation of motion writes: 
\begin{equation}
	\ddot{\bm{u}}+D_2\left(\phi\right)\bm{u}=0\hspace{1cm}D_2=\frac{1}{m}\begin{bmatrix}
		k_1\left(\phi\right)+k_{12}&-k_{12}\\[5pt]
		-k_{12}&k_2\left(\phi\right)+k_{12}
	\end{bmatrix}
	\label{eq:dynamic}
\end{equation}
which can be easily re-casted as a first order differential equation in the form $\dot{\bm{\psi}}=H\left(\phi\right)\bm{\psi}$. The quasi-static spectrum of such a system, i.e. the locus of the natural frequencies $\omega_s$ upon varying the phase $\phi$, is obtained solving the instantaneous eigevalue problem ${\rm i}\omega_s\bm{\psi}_s^R=H\left(\phi\right)\bm{\psi}_s^R$ within the relevant portion of parameter space $\phi\in\left[\phi_i,\phi_f\right]$.
The eigenvalues $\omega_s$ are displayed in Figure \ref{Fig1}(b)-I upon varying the modulation phase $\phi$. 
Interestingly, the spectrum in the neighborhood of $\phi=\pi/2$ is characterized by an avoided crossing between otherwise degenerate states, which is a behavior observable for standard Landau-Zener models in physics and suitably employed to engineer modal transformations. As such, the displacements $u_1$ and $u_2$, corresponding to the displacement components of the right eigenvector $\bm{\psi}_s^R$ undergo a change in amplitude such that the energy, initially stored in the first (second) mass, migrates to the second (first) for smooth modulations of $\phi$, as shown in Fig. \ref{Fig1}(b) II-III,.
In the remainder of the paper we focus on the blue branch, but opposite behavior can be observed following the red branch.
Under the assumption that the energy is initially stored in correspondence of the mode below (blue line), a complete temporal transformation occurs only for slow variations of the phason $\phi$, whereby the limiting speed $\partial\phi/\partial t$, which dictates the transition between adiabatic and non-adiabatic transformations, is given by the following condition:
\begin{equation}
	\left|\displaystyle\frac{\left\langle\bm{\psi}_r^L\left|\displaystyle\frac{\partial H}{\partial\phi}\right|\bm{\psi}_s^R\right\rangle}{\left(\omega_s-\omega_r\right)^2}\frac{\partial\phi}{\partial t}\right|<<1
	\label{eq:02}
\end{equation}
where $\bm{\psi}_r^L$ is the $r^{th}$ left eigenvector and $r,s$ denote the $r^{th}$ and $s^{th}$ crossing states involved in the modal transformation. The mathematical steps and relevant assumptions employed to get to Equation \ref{eq:02} are reported in Appendix A. The limiting condition is graphically illustrated in Fig. \ref{Fig1}(c). In the figure, the white portion corresponds to the region of parameter space that satisfies Equation \ref{eq:02}. In contrast, the black region denotes too fast modulations which are not compliant with an adiabatic process. 
\\
To validate the above arguments, the free response to initial conditions $\bm{\psi}_0$ with different angular velocities $\Omega=2\cdot10^{-4}\pi,\;5\cdot10^{-5}\pi,\;1\cdot10^{-5}\pi$ rad/s is hereafter discussed, corresponding to the red vertical lines I-III in Figure \ref{Fig1}(b). The initial conditions are chosen in such a way that all the energy is injected in the branch below, i.e. $\bm{\psi}_0$ corresponds to the right eigenvector evaluated for the starting phase value $\bm{\psi}_0=\bm{\psi}_1^R\left(\phi_i\right)$. The resulting time histories are displayed in Figures \ref{Fig1}(d-f). As expected, Figure \ref{Fig1}(d) illustrates a non-adiabatic (fast) transformation, in which the energy remains confined in correspondence of the first mass. As such, the corresponding spectrogram represented in Figure \ref{Fig1}(g) illustrates a frequency transition characterized by a jump from the bottom to the top state, demonstrating that the modulation is too fast for the state to perform a modal transformation along the lower branch. 
The spectrogram is computed by windowing the time history with a moving Gaussian function $g(t)={\rm e}^{-(t-t_0)^2/2c^2}$ centered in $t_0\in\left[t_i,t_f\right]$, where $c$ is a parameter that controls the shape of the Gaussian function. Then, the amplitudes in reciprocal space $\hat{\bm{u}}\left(\omega,\kappa,t_0\right)$ are evaluated through a 2D Fourier transform of the displacement field. For ease of visualization, the second dimension is eliminated by taking the RMS value along $\kappa$ and normalized to unity for the probed time instants $t_0$, yielding the spectrogram $\hat{\bm{u}}\left(\omega,t_0\right)$ displayed in the figure. An incomplete modal transformation is achieved for an intermediate modulation speed, see Figure \ref{Fig1}(e), which corresponds to the minimum limiting speed identified with the dashed vertical line in Fig. \ref{Fig1}(c). Hence, the corresponding spectrogram displayed in Figure \ref{Fig1}(h) illustrates that part of the energy follows the lower branch, while a relevant amount of energy is scattered to the neighboring state. Finally, the last example displayed in Figure \ref{Fig1}(f) and Figure \ref{Fig1}(i) describes an almost adiabatic transformation along the lower branch, whereby only a small amount of energy is scattered to the upper branch. It is worth mentioning that the behavior observed in these examples is consistently described by the limiting condition expressed in Equation \ref{eq:02}. Such a condition is violated in the following section through a suitable design of a non-Hermitian action and, as a result, a complete adiabatic transition will be achieved through the same modulation velocities employed for the Hermitian counterparts.\\
\\
It is now assumed that two damping and stiffness parameters $\gamma,\beta\neq0$ are added to locally alter the motion of the coupled oscillators. These additional components can be interpreted as co-located velocity and position feedback control actions employed to modify the system parameters. The new equation of motion writes:
\begin{equation}
	\ddot{\bm{u}}+D_1^\gamma\dot{\bm{u}}+D_2^\beta\left(\phi\right)\bm{u}=0\hspace{0.8cm}with:\hspace{0.2cm}D_1^\gamma=\Gamma\hspace{0.8cm}D_2^\beta=D_2\left(\phi\right)+B\hspace{0.8cm}\Gamma=\frac{1}{m}\begin{bmatrix}
		\gamma&0\\
		0&-\gamma
	\end{bmatrix}\hspace{0.8cm}B=\frac{1}{m}\begin{bmatrix}
		\beta&0\\
		0&-\beta
	\end{bmatrix}
	\label{eq:03}
\end{equation}
Note that the new control parameters $\gamma$ and $\beta$, free to vary in time, are provided in the form of damping/stiffness modulation acting on the degrees of freedom $u_1$ and $u_2$, with same amplitude but opposite sign.
Such a condition limits our analysis to one of the available shortcuts, and will be later relaxed for the study of topological chains. Now, following the work done in quantum systems \cite{torosov2013non} a change of coordinates is performed $\bm{u}=R\left(\theta\right)\bm{q}$, such that:
\begin{equation}
	R^{-1}D_2R=\begin{bmatrix}
		\lambda_-&0\\
		0&\lambda_+
	\end{bmatrix}\;\;with:\;\;R=\begin{bmatrix}
		\cos{\theta}&\sin{\theta}\\
		-\sin{\theta}&\cos{\theta}
	\end{bmatrix}
	\label{eq:04}
\end{equation}
where $\lambda_{\pm}$ are the instantaneous solutions of the eigenvalue problem $\lambda\bm{u}=D_2\left(t\right)\bm{u}$, and $\bm{q}$ are the generalized modal coordinates. For such a simple system, $R$ corresponds to the $2\times2$ eigenvector matrix. Due to the time-dependent springs, the angle $\theta$ that makes Equation \ref{eq:04} diagonal is also time dependent:
\begin{equation}
	\theta\left(t\right)=\displaystyle\frac{1}{2}\arctan\left(\displaystyle\frac{2k_{12}}{k_1\left(t\right)-k_2\left(t\right)}\right)
\end{equation}
With implied time dependence, $\bm{u}=R\bm{q}$ is differentiated twice:
\begin{equation}
\ddot{\bm{u}}=\frac{\partial^2R}{\partial\theta^2}\dot{\theta}^2\bm{q}+\frac{\partial R}{\partial\theta}\ddot{\theta}\bm{q}+2\frac{\partial R}{\partial\theta}\dot{\theta}\dot{\bm{q}}+R\ddot{\bm{q}}
\end{equation}
Plugging this expression in the governing Equation \ref{eq:03} and left-multiplying by $R^{-1}$ gives:
\begin{equation}
\ddot{\bm{q}}=-R^{-1}\left(2\displaystyle\frac{\partial R}{\partial\theta}\dot{\theta}+D_1^\gamma R\right)\dot{\bm{q}}-R^{-1}\left(\displaystyle\frac{\partial^2R}{\partial\theta^2}\dot{\theta}^2+\frac{\partial R}{\partial\theta}\ddot{\theta}+D_1^\gamma\displaystyle\frac{\partial R}{\partial\theta}\dot{\theta}+D_2^\beta R\right)\bm{q}
\end{equation}
yielding a second-order differential equation in the new coordinate set, which can be recasted in a first-order differential form assuming that $\bm{z}=\left[\dot{\bm{q}};\bm{q}\right]$:
\begin{equation}
\dot{\bm{z}}=H\bm{z}\hspace{1cm}H=\begin{bmatrix}
	-R^{-1}\left(2\displaystyle\frac{\partial R}{\partial\theta}\dot{\theta}+D_1^\gamma R\right)&-R^{-1}\left(\displaystyle\frac{\partial^2R}{\partial\theta^2}\dot{\theta}^2+\frac{\partial R}{\partial\theta}\ddot{\theta}+D_1^\gamma\displaystyle\frac{\partial R}{\partial\theta}\dot{\theta}+D_2^\beta R\right)\\[8pt]
	I&0
\end{bmatrix}
\label{eq:07}
\end{equation}
where:
\begin{equation}
\begin{split}
	R^{-1}\left(2\displaystyle\frac{\partial R}{\partial\theta}\dot{\theta}+D_1^\gamma R\right)&=\begin{bmatrix}
		\displaystyle\frac{\gamma\cos{2\theta}}{m}&2\dot{\theta}+\displaystyle\frac{\gamma\sin\left(2\theta\right)}{m}\\[8pt]
		-2\dot{\theta}+\displaystyle\frac{\gamma\sin\left(2\theta\right)}{m}&-\displaystyle\frac{\gamma\cos{2\theta}}{m}
	\end{bmatrix}\\[8pt]
	R^{-1}\left(\displaystyle\frac{\partial^2R}{\partial\theta^2}\dot{\theta}^2+\frac{\partial R}{\partial\theta}\ddot{\theta}+D_1^\gamma\displaystyle\frac{\partial R}{\partial\theta}\dot{\theta}+D_2^\beta R\right)&=\begin{bmatrix}
		\lambda_--\dot{\theta}^2-\dot{\theta}\displaystyle\frac{\gamma\sin{2\theta}}{m}+\displaystyle\frac{\beta\cos{2\theta}}{m}&\ddot{\theta}+\dot{\theta}\displaystyle\frac{\gamma\cos{2\theta}}{m}+\displaystyle\frac{\beta\sin{2\theta}}{m}\\[8pt]
		-\ddot{\theta}-\dot{\theta}\displaystyle\frac{\gamma\cos{2\theta}}{m}+\displaystyle\frac{\beta\sin{2\theta}}{m}&\lambda_+-\dot{\theta}^2+\dot{\theta}\displaystyle\frac{\gamma\sin{2\theta}}{m}-\displaystyle\frac{\beta\cos{2\theta}}{m}
	\end{bmatrix}
\end{split}
\label{eq:08}
\end{equation}
Equations \ref{eq:07}-\ref{eq:08} describe the dynamic evolution of the states, which are projected in the quasi-static reference frame associated to the generalized coordinates $\bm{q}$. 
\begin{figure}[t]
	\centering
	\subfigure[]{\includegraphics[width=0.45\textwidth]{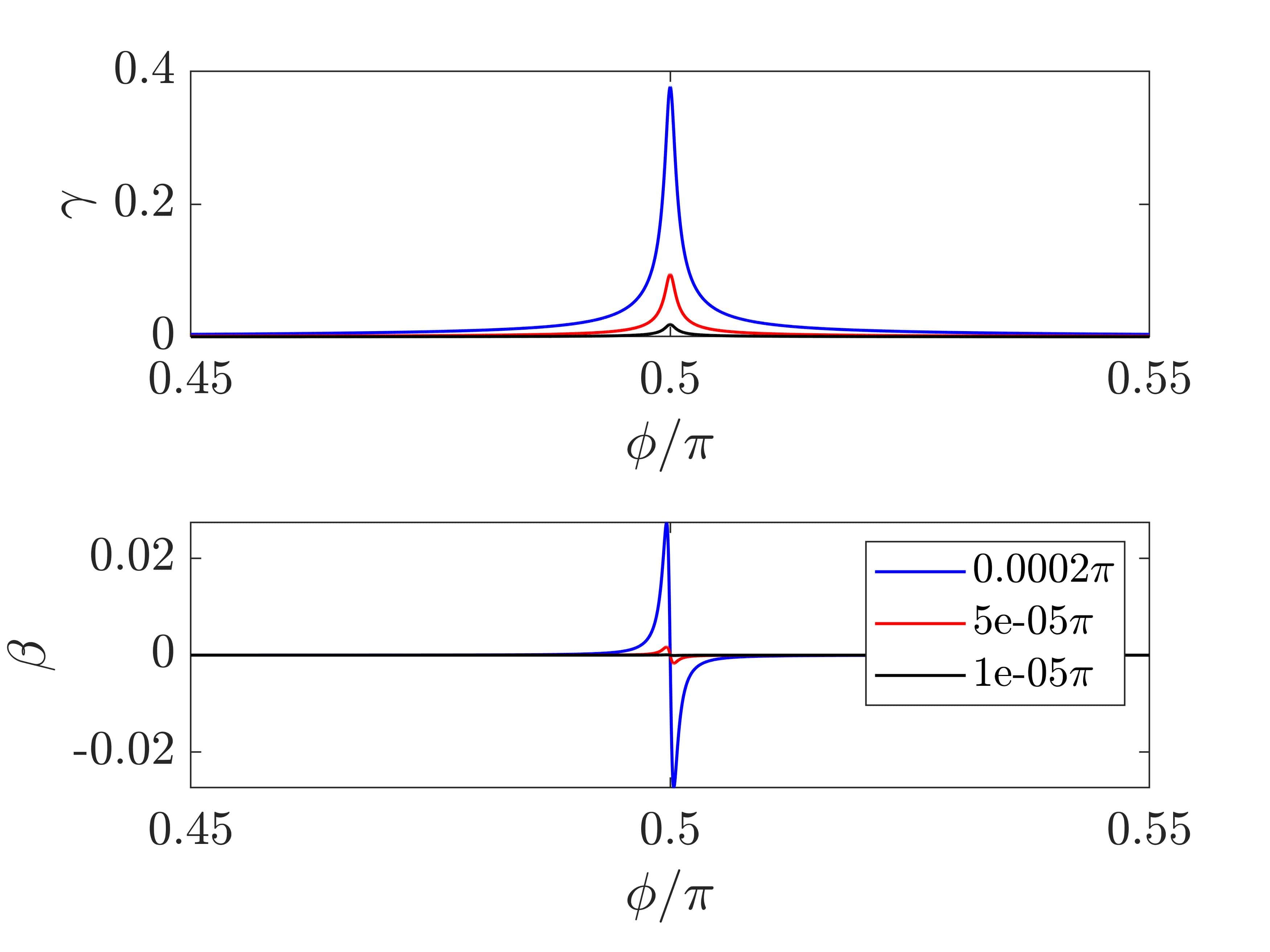}}\hspace{0.2cm}
	\subfigure[]{\includegraphics[width=0.45\textwidth]{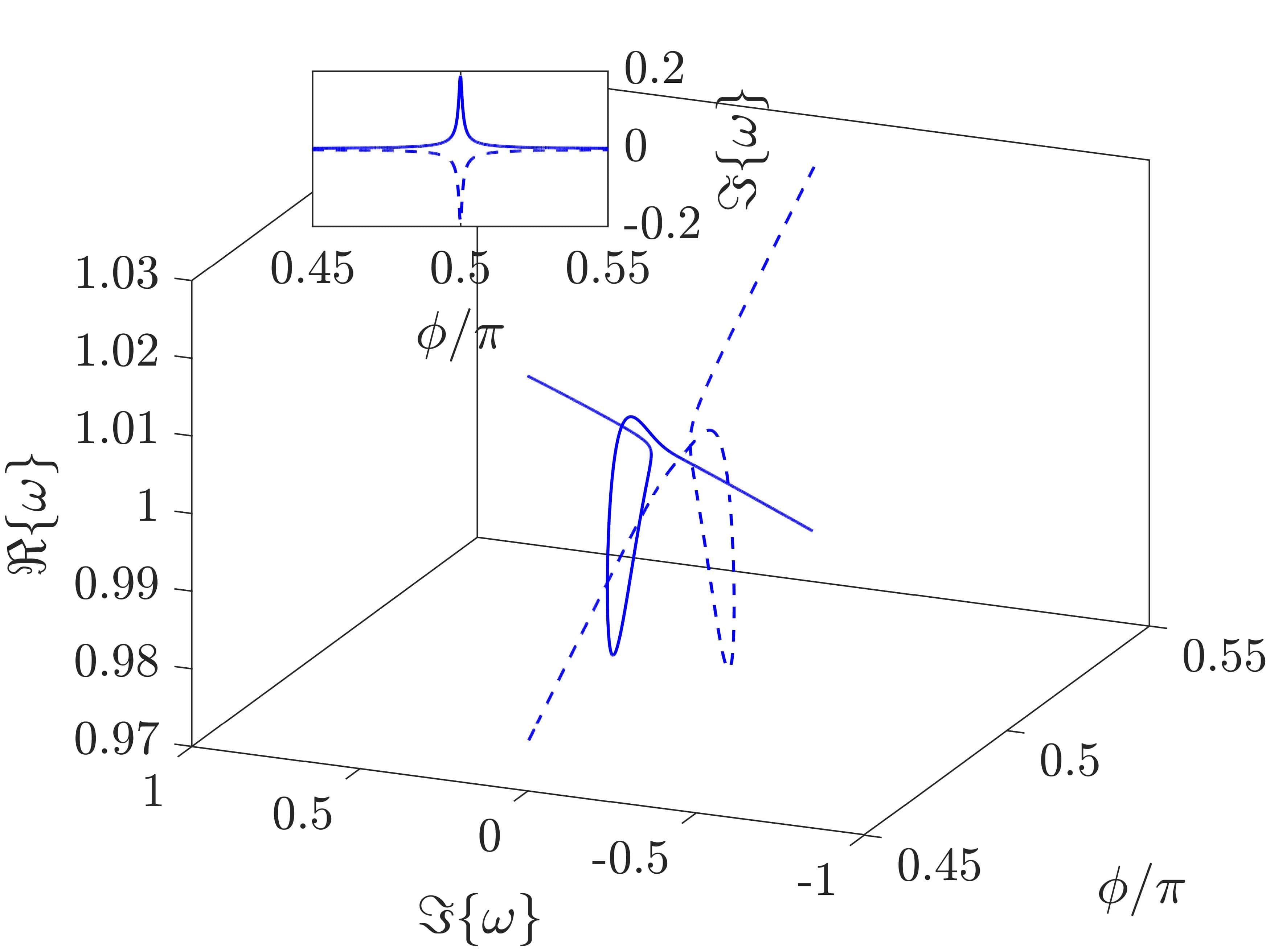}}
	\subfigure[]{\includegraphics[width=0.45\textwidth]{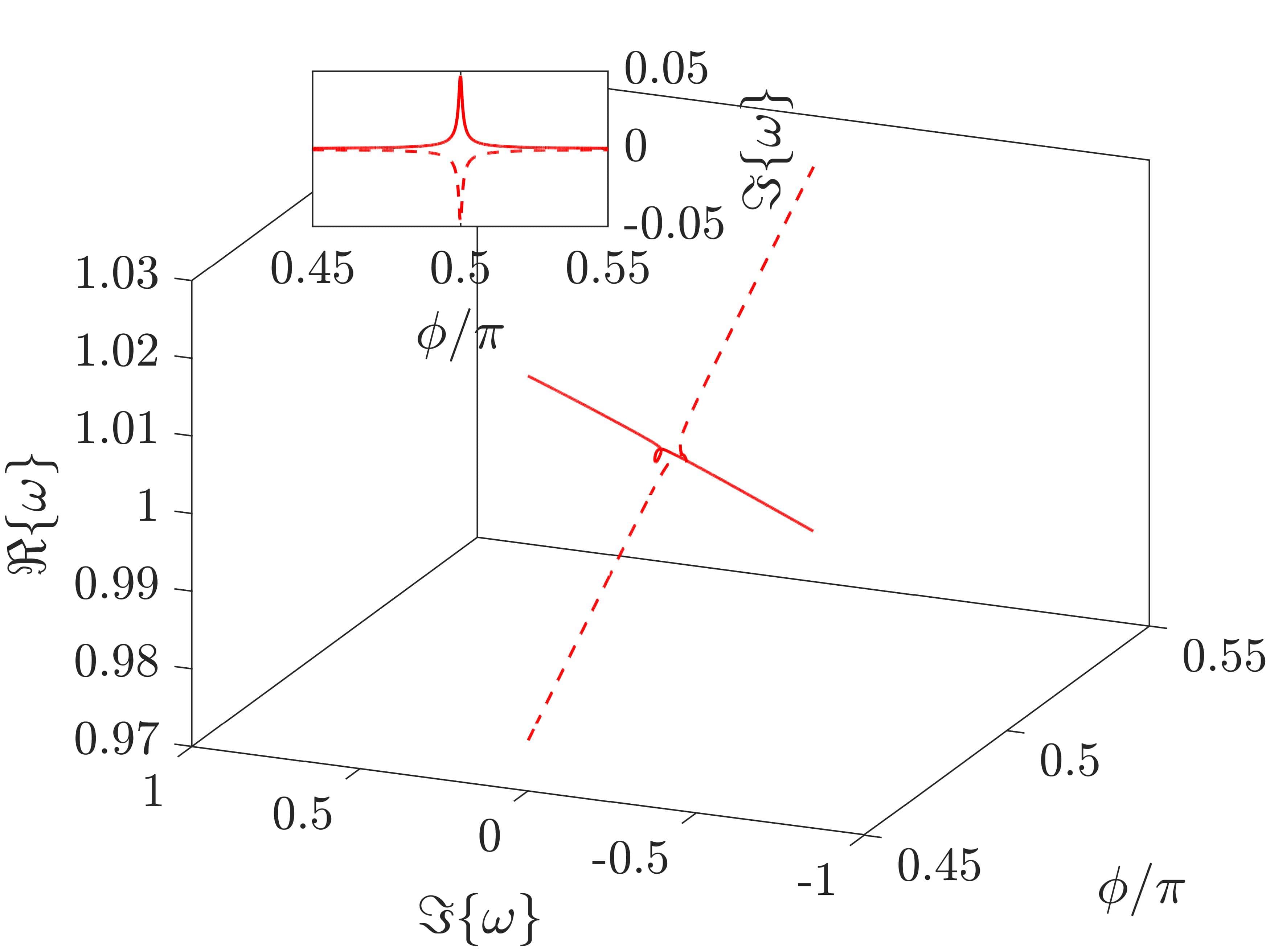}}
	\subfigure[]{\includegraphics[width=0.45\textwidth]{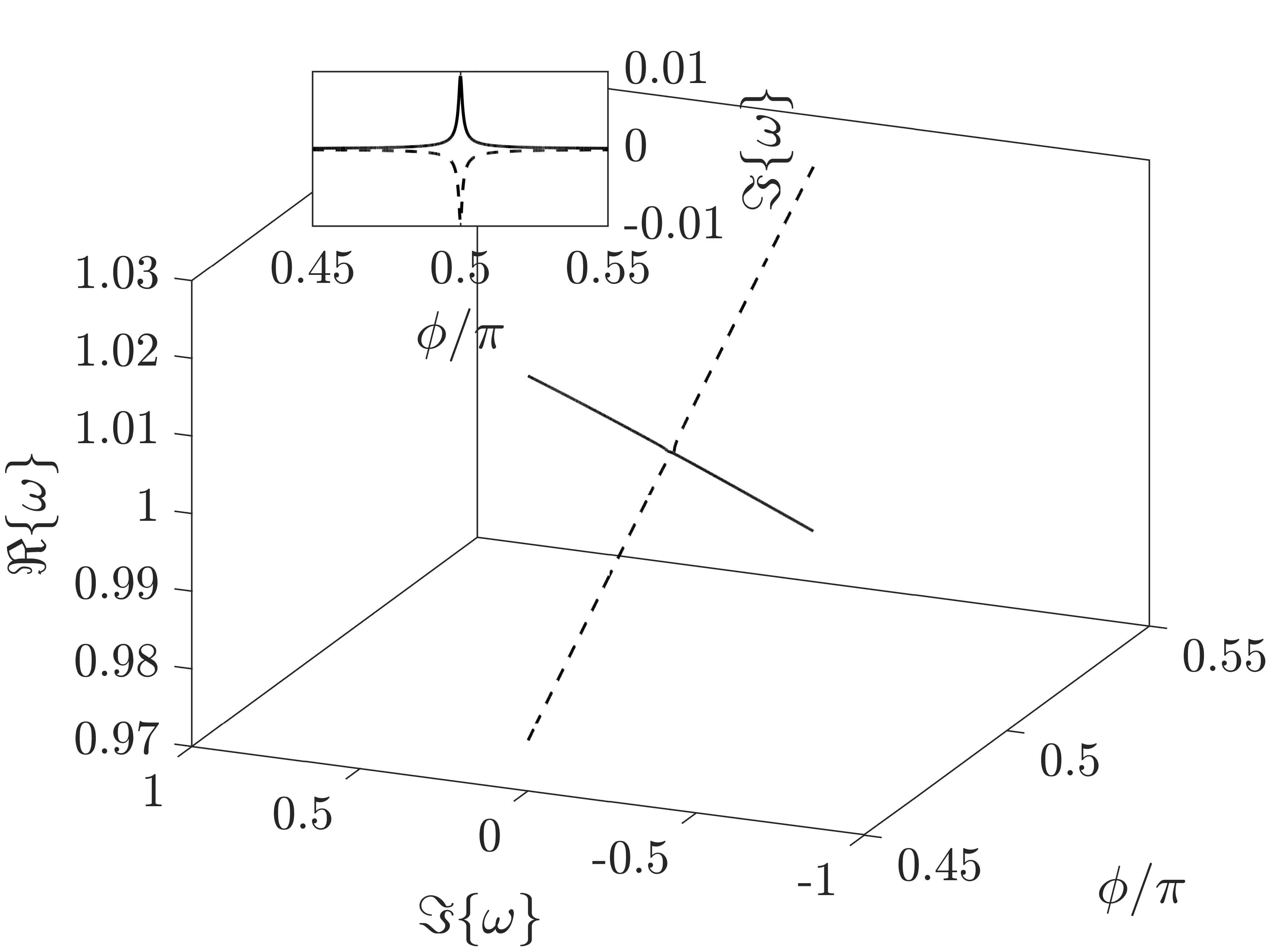}}	
	\caption{(a) Evolution of the control parameters $\gamma$ and $\beta$ as a function of the phason $\phi$. (b-d) Corresponding locus of the complex eigenfrequencies upon varying $\phi$ for fast, medium and slow modulations speed $\Omega$, respectively. The inset represents a top view of the diagram.  }
	\label{Fig2}     
\end{figure}
By comparing the off-diagonal terms in Equation \ref{eq:04} and Equation \ref{eq:08}, it is straightforward to conclude that the modulation velocity $\dot{\theta}$ and acceleration $\ddot{\theta}$ are responsible for the coupling between states, as the generalized coordinates $q_1$ and $q_2$ would have been uncoupled in absence of $\gamma,\beta$ and if $\dot{\theta},\ddot{\theta}=0$. Such a coupling can be weak or strong depending on the modulation speed, which reflects into a limiting conditions for adiabaticity.
Interestingly, this limit can be broken upon a proper choice of $\gamma$ and $\beta$, i.e. when $\gamma,\beta$ are tailored to nullify the off-diagonal terms. This is the key factor to activate modal transformations without leakage of energy to the neighboring modes. We get to an expression for $\gamma,\beta$:
\begin{equation}
	\begin{split}
		\gamma&=-2\dot{\theta}\displaystyle\frac{m}{\sin{2\theta}}\\[5pt]
		\beta&=-\ddot{\theta}\displaystyle\frac{m}{\sin{2\theta}}-\dot{\theta}\frac{\gamma}{\tan{2\theta}}
	\end{split}
	\label{eq:sol}
\end{equation}
Provided initial conditions  $\bm{\psi}_0=\bm{\psi}_-^R\left(\phi_i\right)$, where the subscript $(\cdot)_-$ stands for lower branch of the spectrum, the transient response is therefore no longer limited by the modulation speed $\Omega$ as the modal coupling between neighboring states is broken. 
As such, the dynamics of the time dependent system of oscillators in the generalized coordinate frame is characterized by the following dynamic matrices:
\begin{equation}
	\begin{split}
		2R^{-1}\displaystyle\frac{\partial R}{\partial\theta}\dot{\theta}&=\begin{bmatrix}
			-\displaystyle\frac{2\dot{\theta}}{\tan{2\theta}}&0\\[8pt]
			-4\dot{\theta}&\displaystyle\frac{2\dot{\theta}}{\tan{2\theta}}
		\end{bmatrix}\\[8pt]
		R^{-1}\left(\displaystyle\frac{\partial^2R}{\partial\theta^2}\dot{\theta}^2+\frac{\partial R}{\partial\theta}\ddot{\theta}+D_1^\gamma\displaystyle\frac{\partial R}{\partial\theta}\dot{\theta}+D_2^\beta R\right)&=\begin{bmatrix}
			\lambda_-+\dot{\theta}^2-\displaystyle\frac{\ddot{\theta}}{\tan{2\theta}}+\displaystyle\frac{2\dot{\theta}^2}{\displaystyle\tan^2{2\theta}}&0\\[8pt]
			-2\ddot{\theta}+\displaystyle\frac{4\dot{\theta}^2}{\tan{2\theta}}&\lambda_+-3\dot{\theta}^2+\displaystyle\frac{\ddot{\theta}}{\tan{2\theta}}-\frac{2\dot{\theta}^2}{\tan^2{2\theta}}
		\end{bmatrix}
	\end{split}
	\label{eq:nonhermitian}
\end{equation}
which are now inherently uncoupled by the choice of $\gamma$ and $\beta$.
\begin{figure}[t]
	\centering
	\subfigure[]{\includegraphics[width=0.32\textwidth]{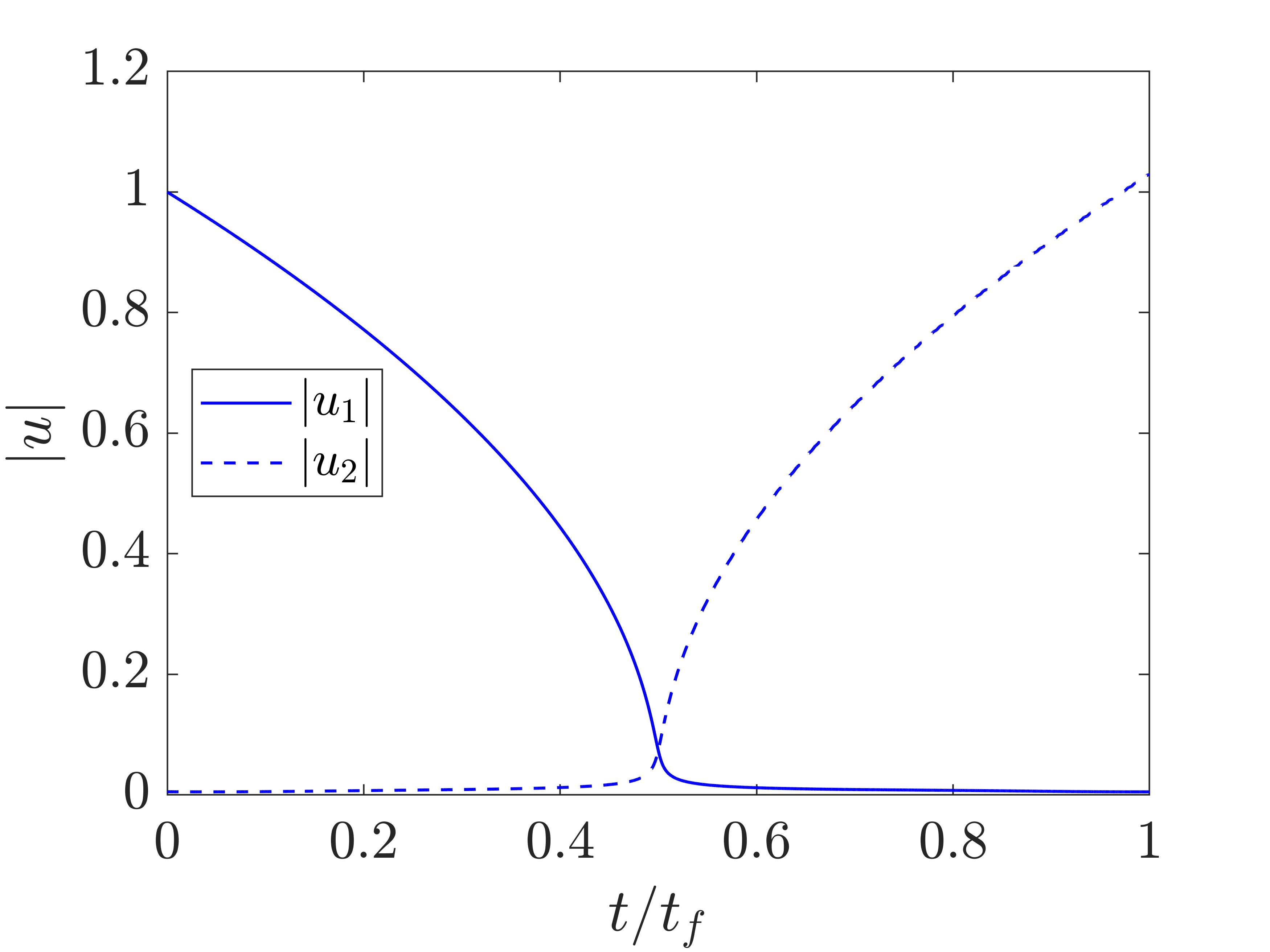}}
	\subfigure[]{\includegraphics[width=0.32\textwidth]{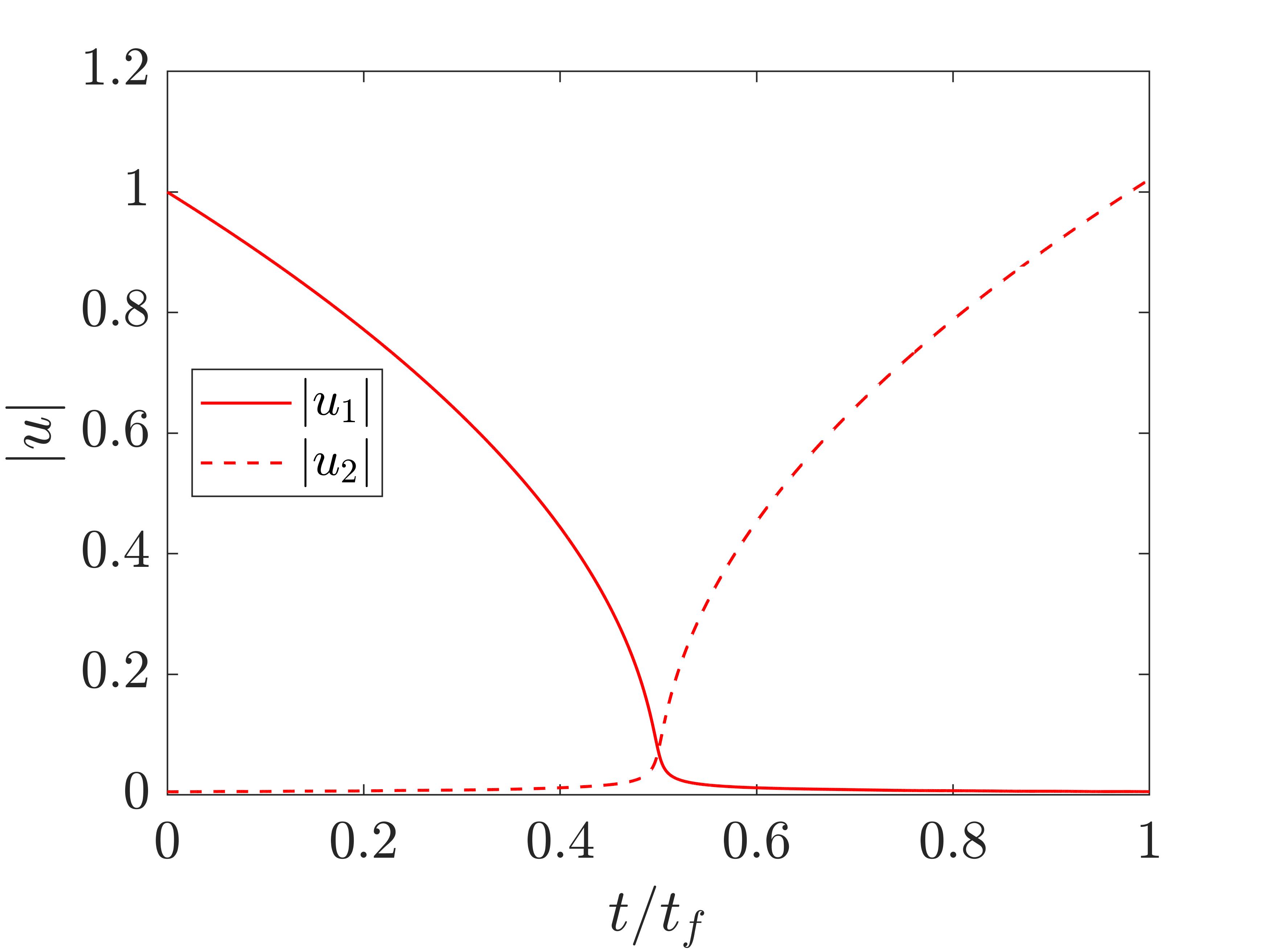}}
	\subfigure[]{\includegraphics[width=0.32\textwidth]{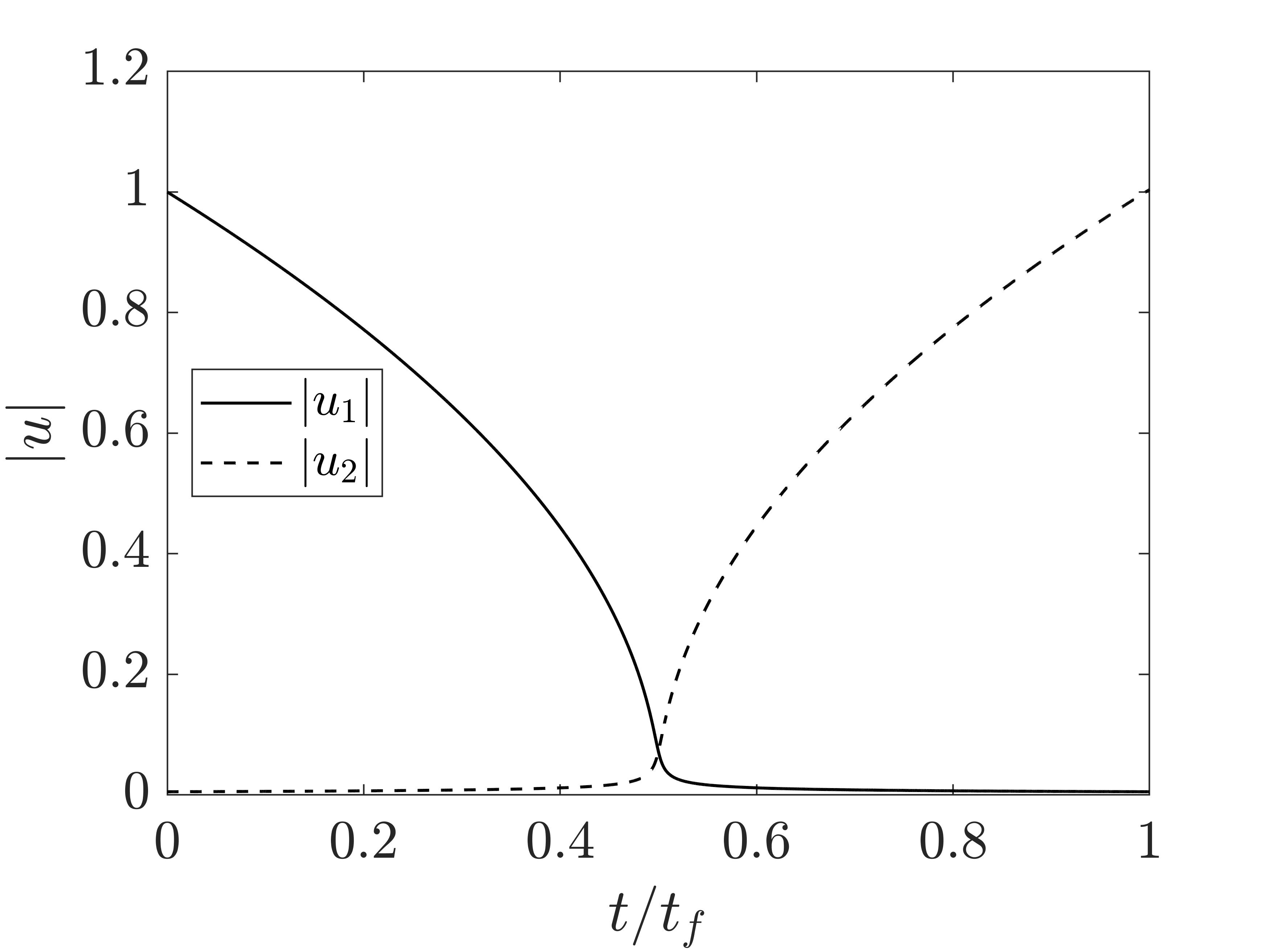}}\\
	\subfigure[]{\includegraphics[width=0.32\textwidth]{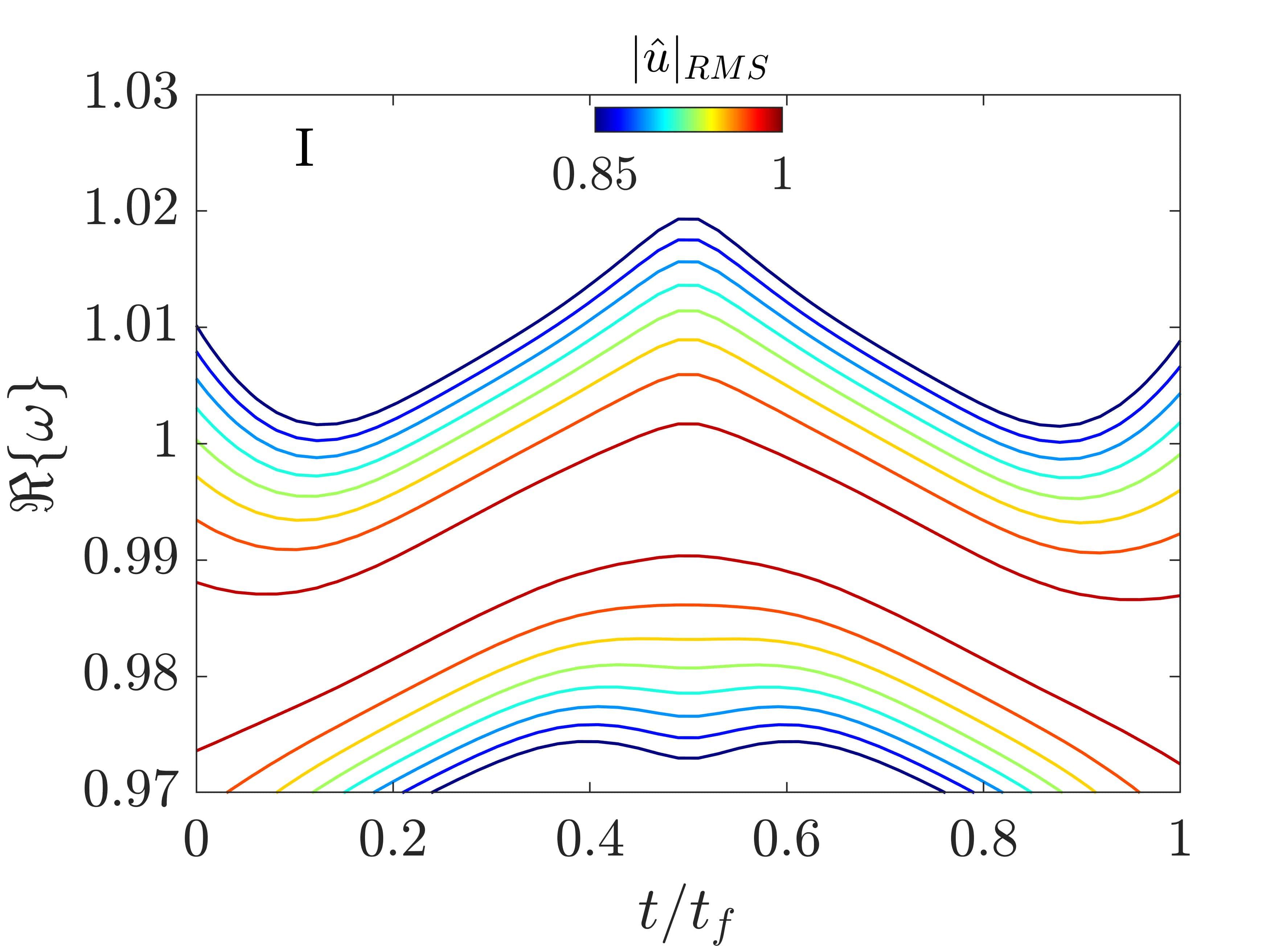}}	
	\subfigure[]{\includegraphics[width=0.32\textwidth]{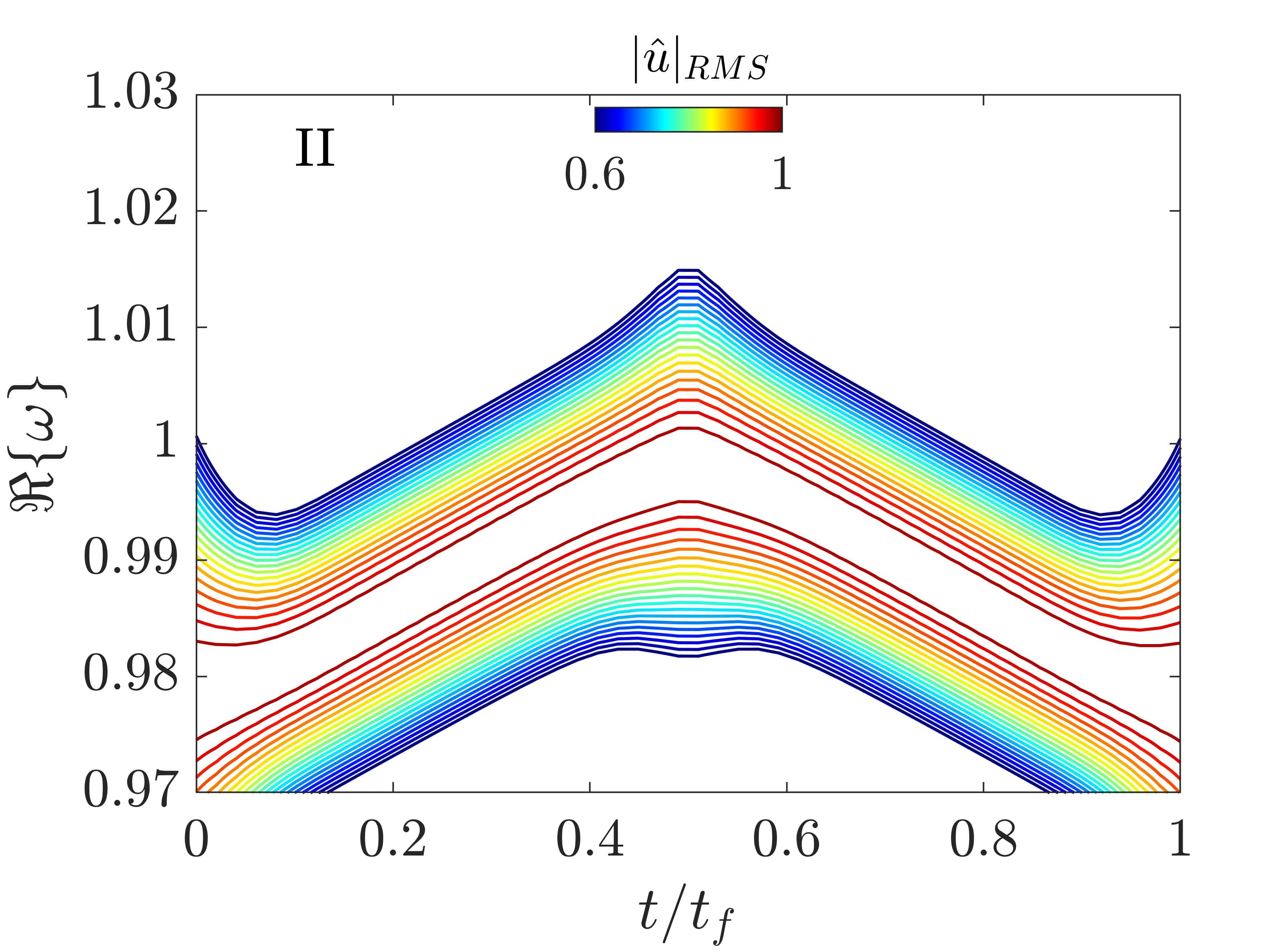}}		\subfigure[]{\includegraphics[width=0.32\textwidth]{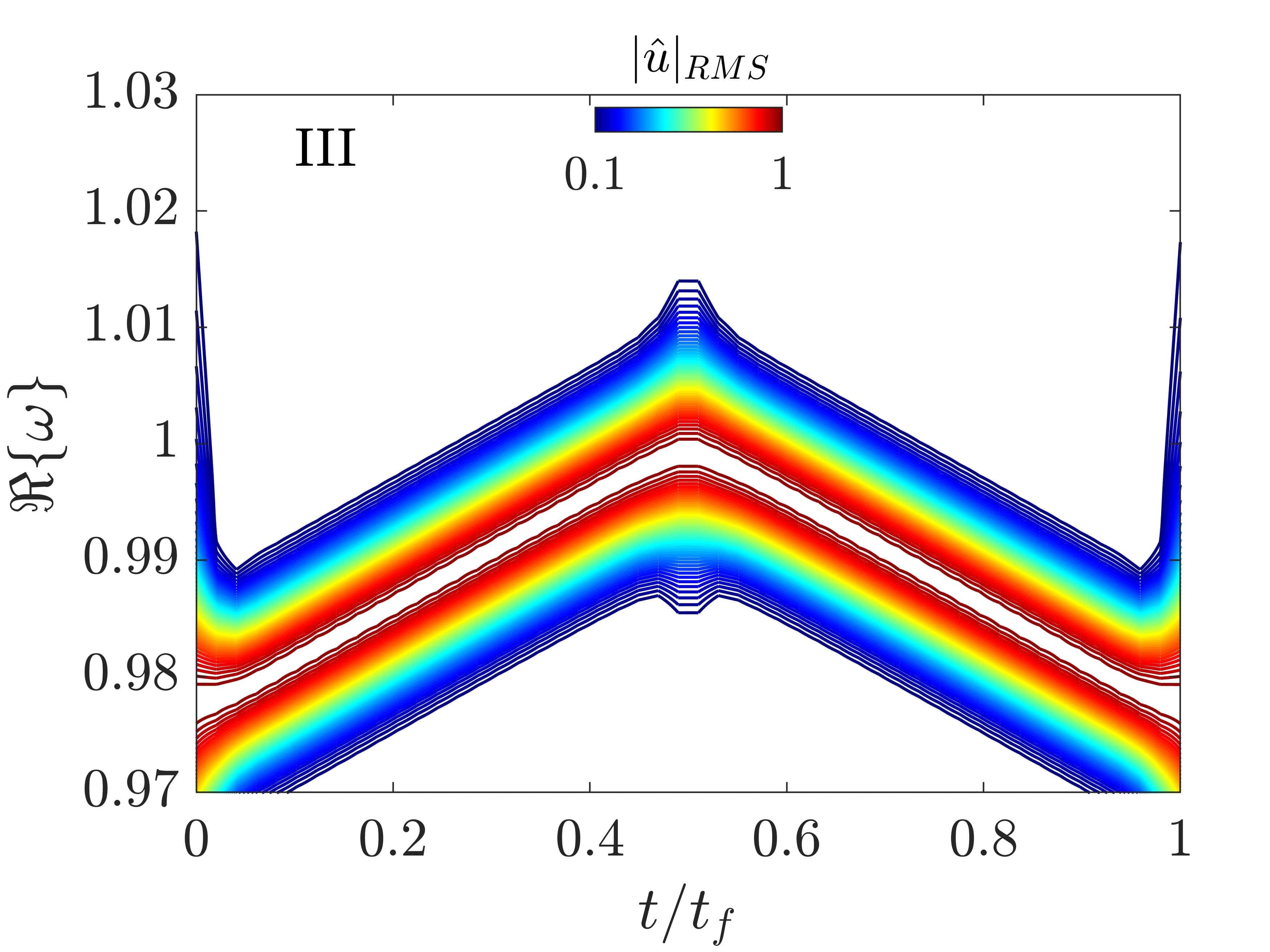}}
	\caption{(a-c) Time response to initial conditions for fast, intermediate and slow modulations in case $\gamma$ and $\beta$ are tailored to activate the non-hermitian shortcut. (d-f) Corresponding spectrograms showing that the modal transition is independent of the speed of modulation.}
	\label{Fig3}    
\end{figure}
Fig. \ref{Fig2}(a) illustrates the temporal evolution of $\gamma$ and $\beta$ for the three modulation velocities $\Omega$ previously employed for the analysis of the Hermitian counterpart. 
In the case at hand, stronger gain/loss pairs are necessary if greater modulation velocities are probed and, in addition, we observe that the amplitudes of $\gamma,\beta$ are maximum in the neighborhood of the crossing point, i.e. where the coupling is stronger. Due to the balanced gain and loss, the instantaneous eigenvalue problem ${\rm i}\omega_s\bm{\psi}_s^R=H\bm{\psi}_s^R$ is characterized by complex valued $\omega_s$, as the system is now nonconservative. The locus of the instantaneous eigenvalues, corresponding to the different modulation velocities, is shown in Figure \ref{Fig2}(b-d). The figures describe a solution with opposite signed imaginary parts of growing amplitude. Practically, a rapid instability is induced in the form of gain/loss in correspondence of the crossing point which, according to the above arguments, enables a superfast transformation of the associated state. 
We remark that this procedure yields a different condition as compared to a direct nullification of the coupling $k_{12}$, which would have induced a pair of uncoupled but degenerate states. For such a condition, an infinitely slow modulation would have been required, according to the expression in Eq. \ref{eq:02}. It is of paramount importance that the spectrum represented in Figures \ref{Fig2}(b-d) describe a non degenerate transition and the eigenvalue pair never cross.\\
To validate the theoretical claims, the transient response to initial conditions is illustrated in Fig. \ref{Fig3}(a-c), along with the corresponding spectrograms in Fig. \ref{Fig3}(d-e). It is observed that a complete transformation is achieved for all the probed modulation velocities and the energy, initially stored in the motion of $u_1$, is transferred to $u_2$. Such a transformation occurs independently of $\Omega$ along the bottom branch of the spectrum, consistently with the desired frequency transformation and without modal scattering to the neighboring mode, demonstrating that the transformation occurs as expected. 
As such, the limit for adiabaticity has been violated, and a complete modal transformation has been accomplished through the tailored shortcut mechanism. As a final remark, it can also be observed that the transition is analog but different with respect to what is observed in quantum non-Hermitian shortcuts, which have been shown to exhibit linearly varying modal amplitude in time \cite{torosov2013non}. 

\section{Superfast edge-to-edge pumping in a topological lattice}
\begin{figure}[t]
	\centering
	\hspace{-2cm}
	\subfigure[]{\includegraphics[width=0.58\textwidth]{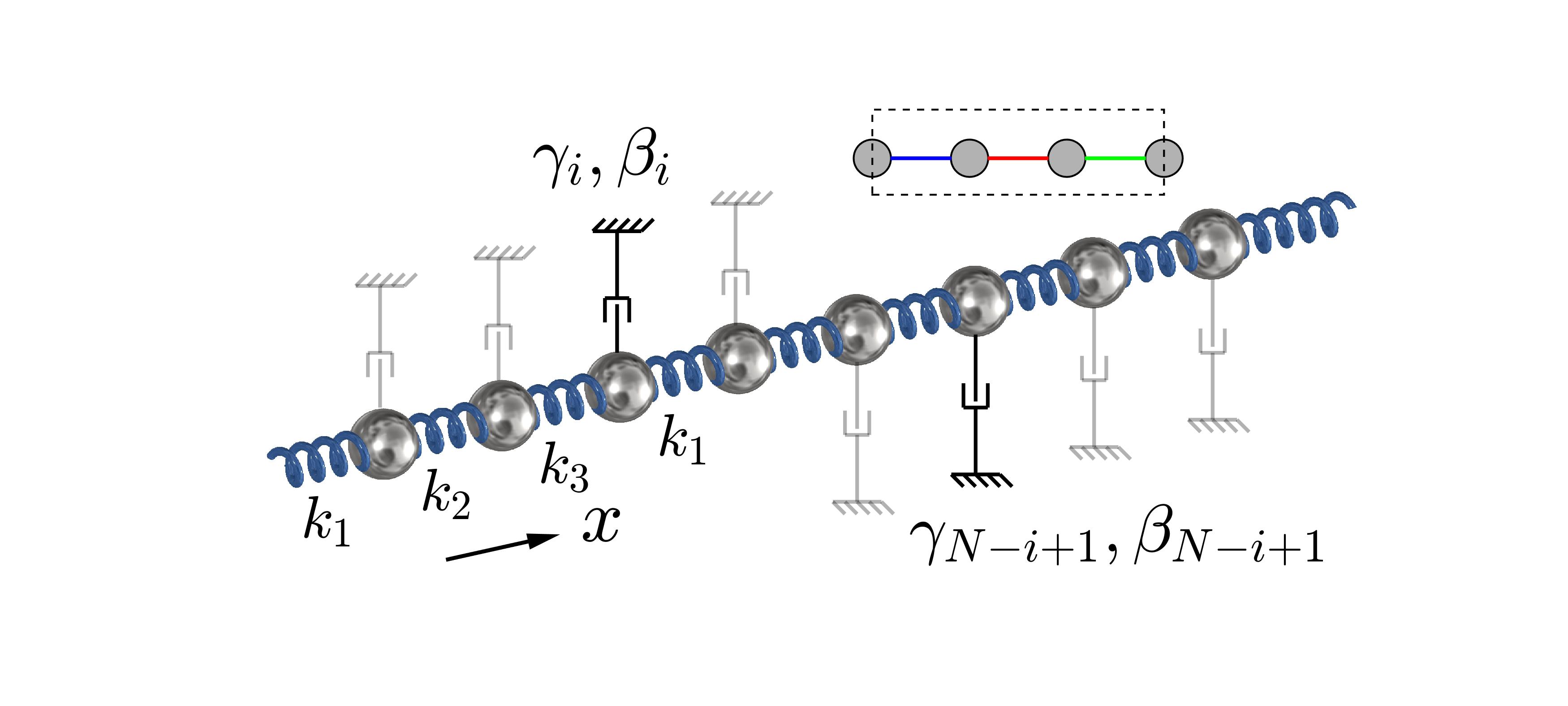}}\hspace{-1cm}
	\subfigure[]{\includegraphics[width=0.26\textwidth]{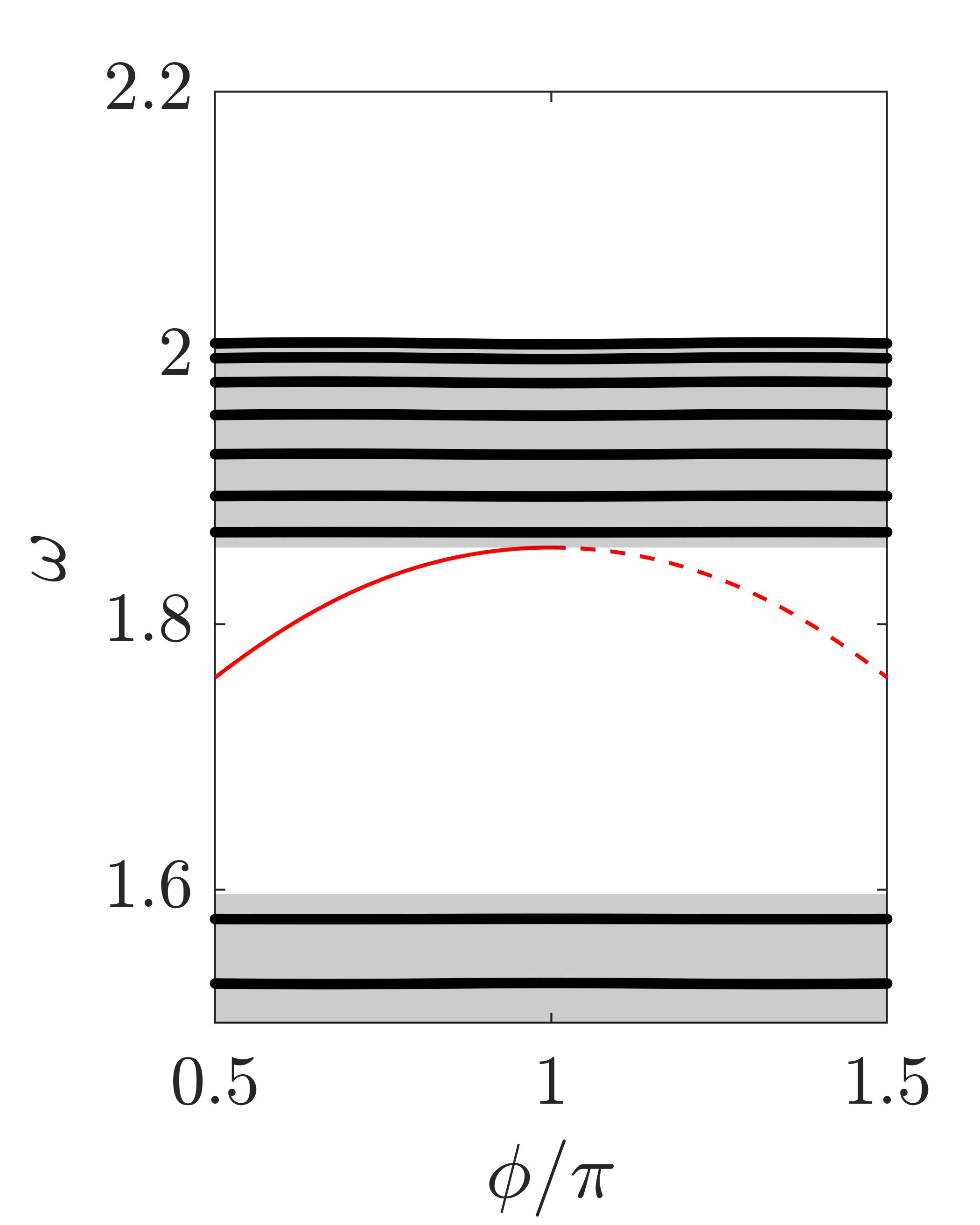}}
	\subfigure[]{\includegraphics[width=0.26\textwidth]{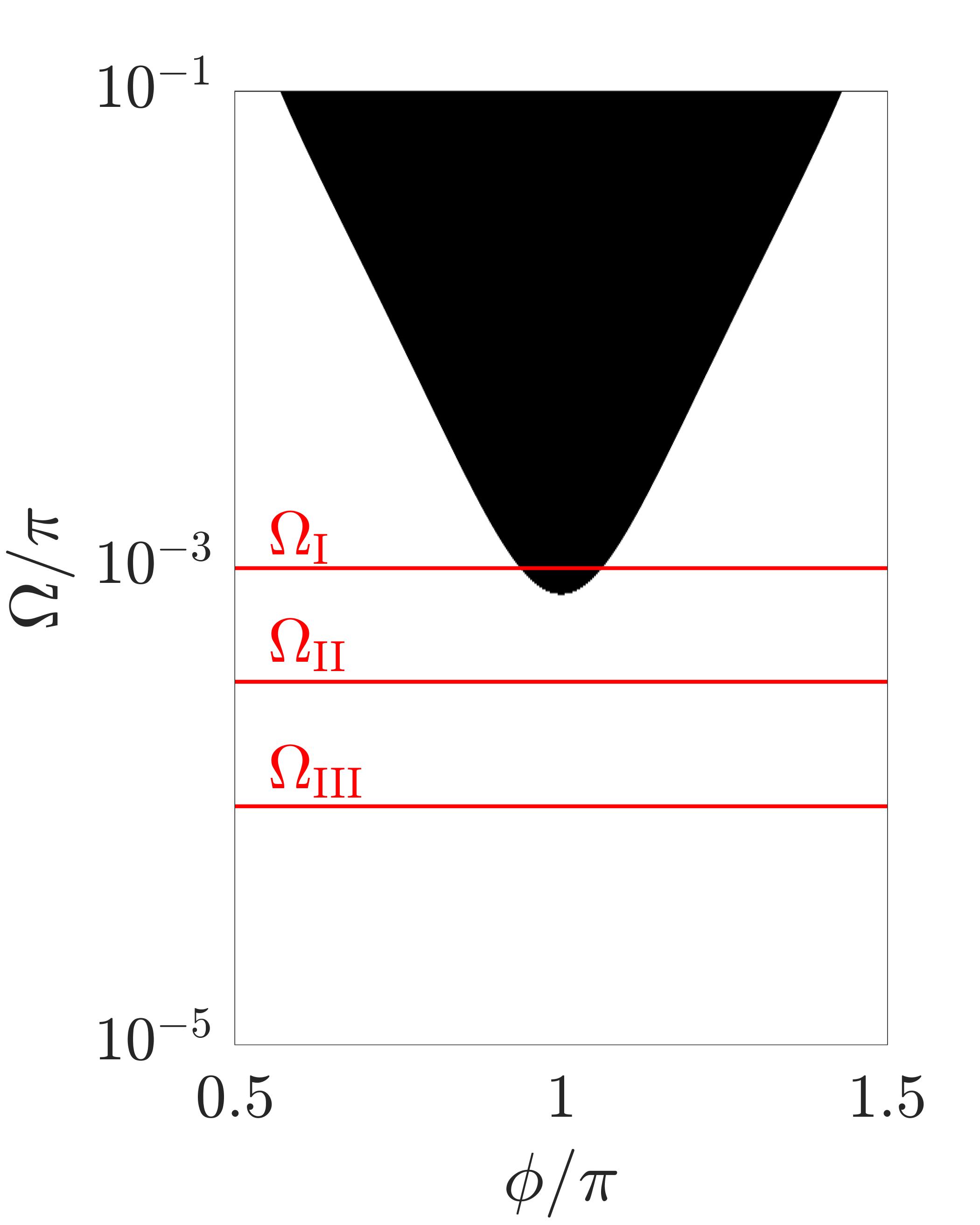}}	\caption{(a) Schematic of the mechanical lattice employed to induce topological edge states. The system is made of a chain of masses connected through a periodic arrangement of springs. A group of three mass-spring defines the unit cell displayed in the inset. The additional terms $\gamma,\beta$ employed to generate the shortcut are represented by dashpots in black. (b) Evolution of the bulk (black) and topological (red) modes upon varying the phase parameter $\phi$. Gray regions represent the upper and lower limits of the dispersion relation, which are populated by bulk states. The red solid line denotes a right localized mode spanning the gap, while the red dashed line describes a left localization of the edge state. (c) The adiabaticity limit upon varying the phase parameter $\phi$ and the modulation velocity $\Omega$. The vertical lines correspond to the $\Omega$ values employed in the Hermitian and non-Hermitian configurations. }
	\label{Fig4}    
\end{figure}

We now pursue a modal transformation in a topological lattice made of $N=24$ mass elements connected through an array of springs, according with the schematic displayed in Figure \ref{Fig4}(a). Consecutive stiffness values $k_i$ are generated through the following law:
\begin{equation}
	k_i=k_0\left[1+\alpha\cos\left(2\pi\theta i+\phi\right)\right]
\end{equation}
which defines a family of modulations with fixed modulation amplitude $\alpha=0.3$ and central stiffness $k_0=1$. $\theta$ is a projection parameter that defines the periodicity of the underlying medium and $\phi$ is the phason, which is a free parameter in analogy to that employed in section II. Such a stiffness modulation induces edge states localized at one of the boundaries, depending on the parameter $\theta$ and $\phi$, as extensively shown in prior studies \cite{rosa2019edge}.
For simplicity, we consider $\theta=1/3$, but different values (either rational or irrational) can be employed for the same purpose. A value of $\theta=1/3$ implies a unit cell of three sub-elements, which encourages the dispersion analysis through the application of Bloch conditions in the form $u\left(\omega,x+3a\right)=u\left(\omega,x\right){\rm e}^{{\rm i}\kappa3a}$, where $a$ is the distance between consecutive masses and $\kappa$ is the wavenumber. The dispersion relation is written in terms of reduced mass $M_{R}$ and stiffness $K_{R}$:
\begin{equation}
	K_R\left(\phi,\mu\right)\bm{u}=\omega^2M_R\left(\mu\right)\bm{u}
\end{equation}
where $\mu=\kappa 3a$ is the normalized wavenumber. The analysis of the band topology in wavenumber-parameter space reveals the non trivial nature of the modulation. Such analysis is defined over a two-dimensional torus $\left[\mu,\phi\right]\in D=\left[0,2\pi\right]\times\left[0,2\pi\right]$, where the Chern number $C$ is the relevant topological invariant, contextually evaluated as:
\begin{equation}
	C_n=\frac{1}{2\pi {\rm i}}\int_D\nabla\times\left(\bm{u}^*\cdot\nabla\bm{u}\right)dD
\end{equation} 
A label for the $r^{th}$ gap $C_g^{(r)}=\sum_{n=1}^{r}C_n$ is defined as the sum of the Chern numbers of the dispersion branches below the gap, which is inherently linked to the existence of a topological mode spanning the $r^{th}$ gap. In the case at hand, $C_1=1$ and $C_2=-2$ for the first two dispersion bands, which imply $C_g^{(1)}=1$ and $C_g^{(2)}=-1$. We focus only on the second topological gap with $C_g^{(2)}=-1$, which is represented in Figure \ref{Fig4}(b) in terms of finite spectrum (natural frequencies of the lattice) under free-free boundary conditions. 
\begin{figure}[t]
	\centering
	\subfigure[]{\includegraphics[width=0.3\textwidth]{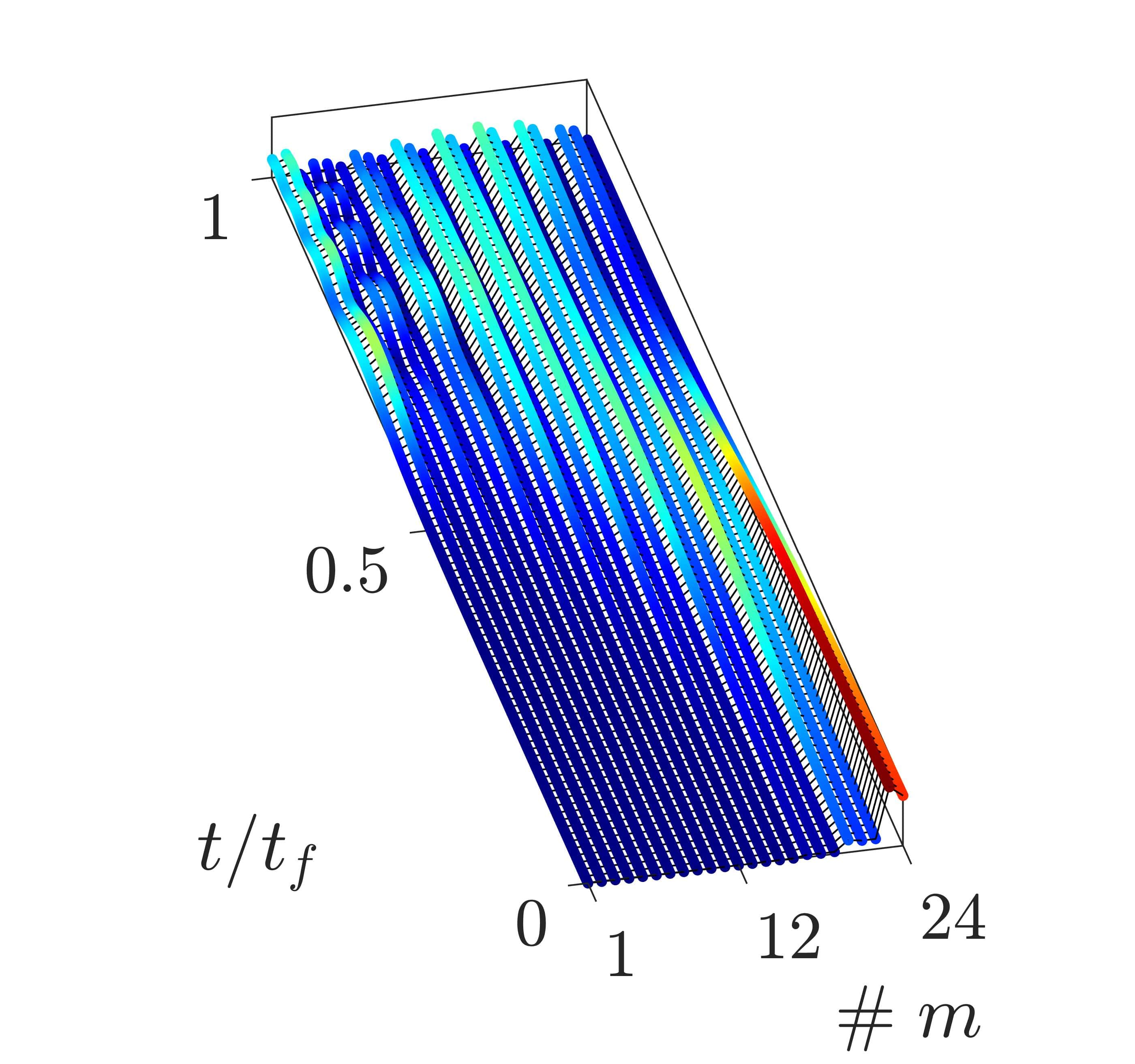}}\hspace{-0.25cm}
	\subfigure[]{\includegraphics[width=0.3\textwidth]{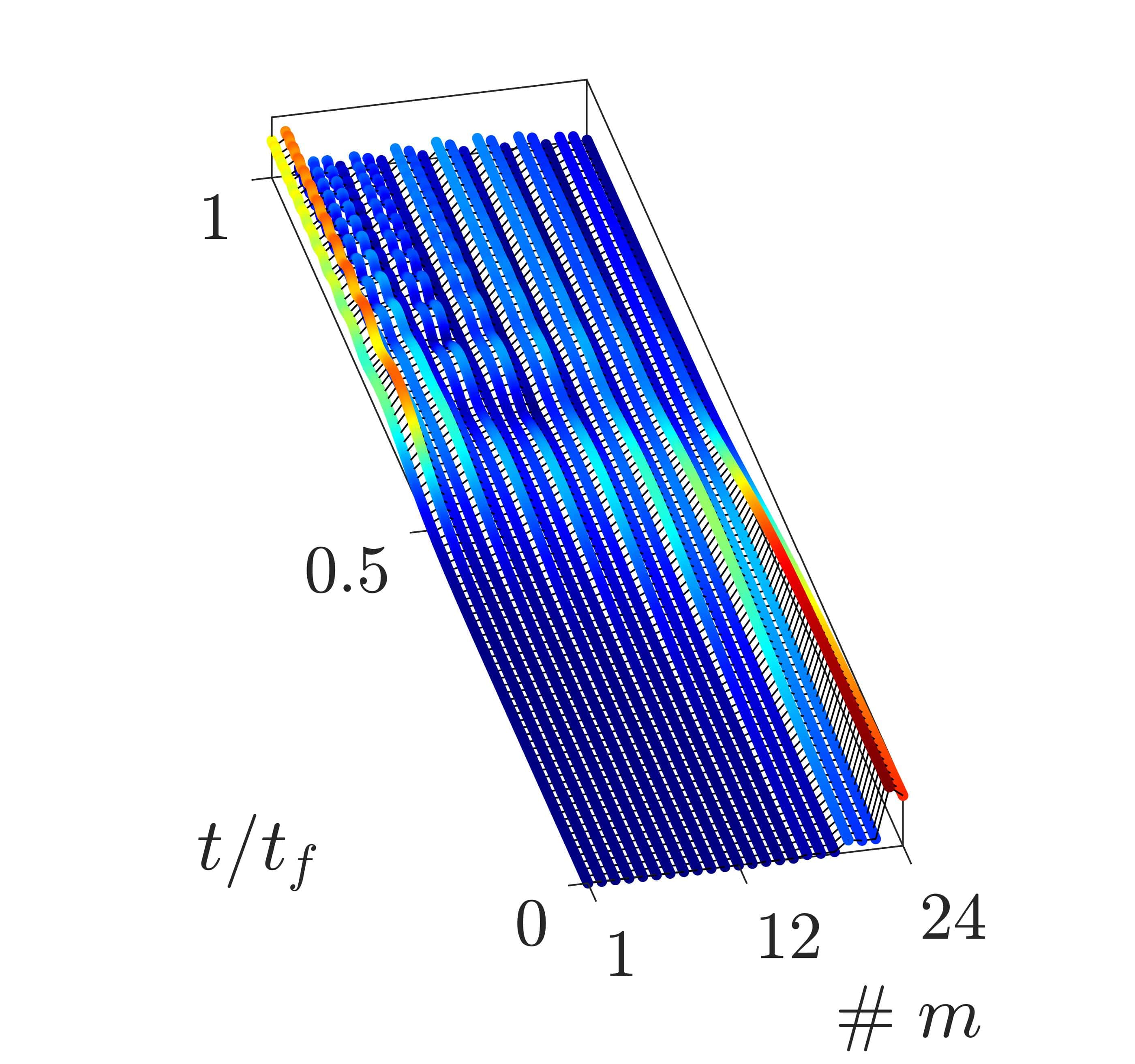}}\hspace{-0.25cm}
	\subfigure[]{\includegraphics[width=0.3\textwidth]{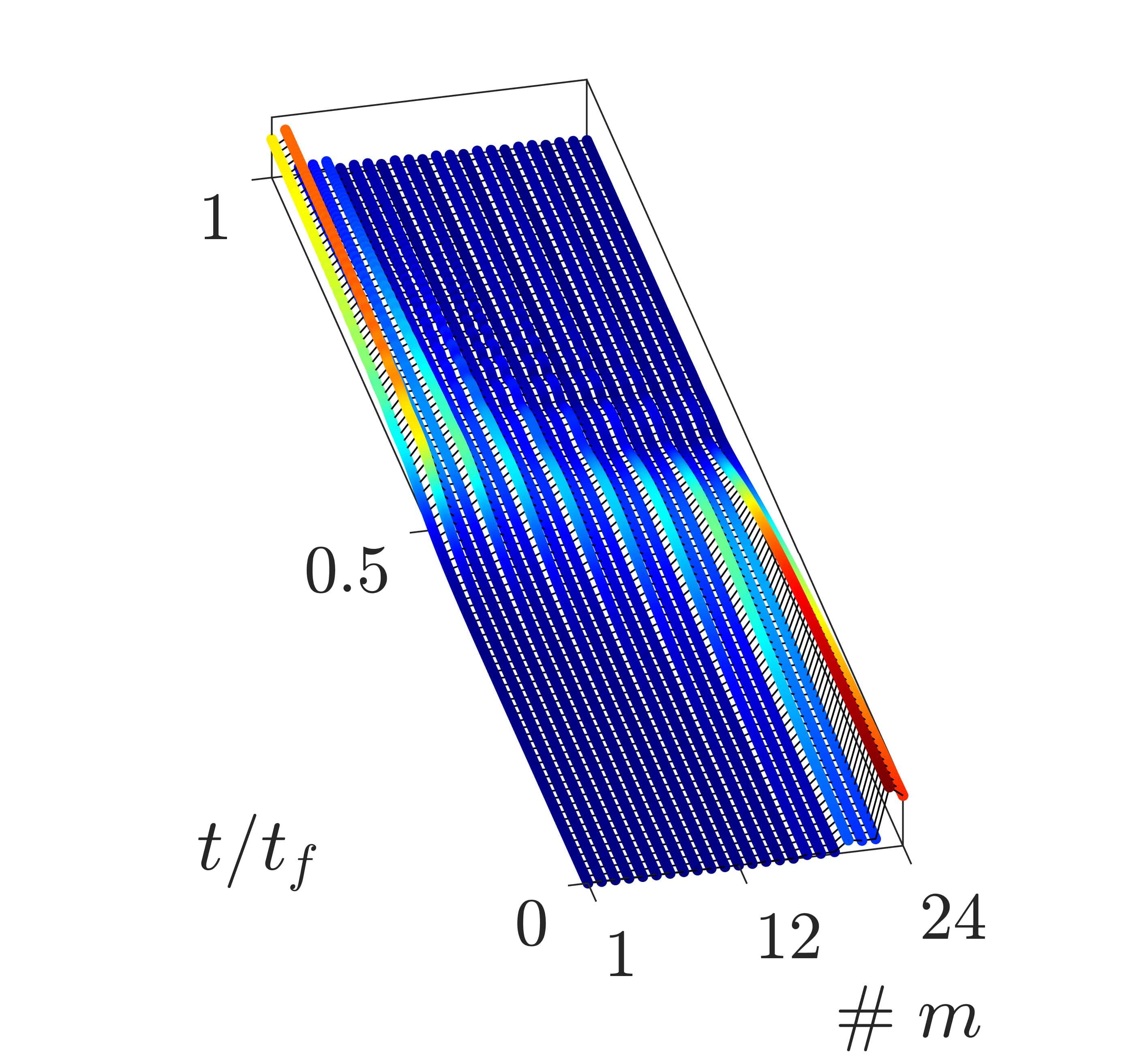}}\\
	\subfigure[]{\includegraphics[width=0.32\textwidth]{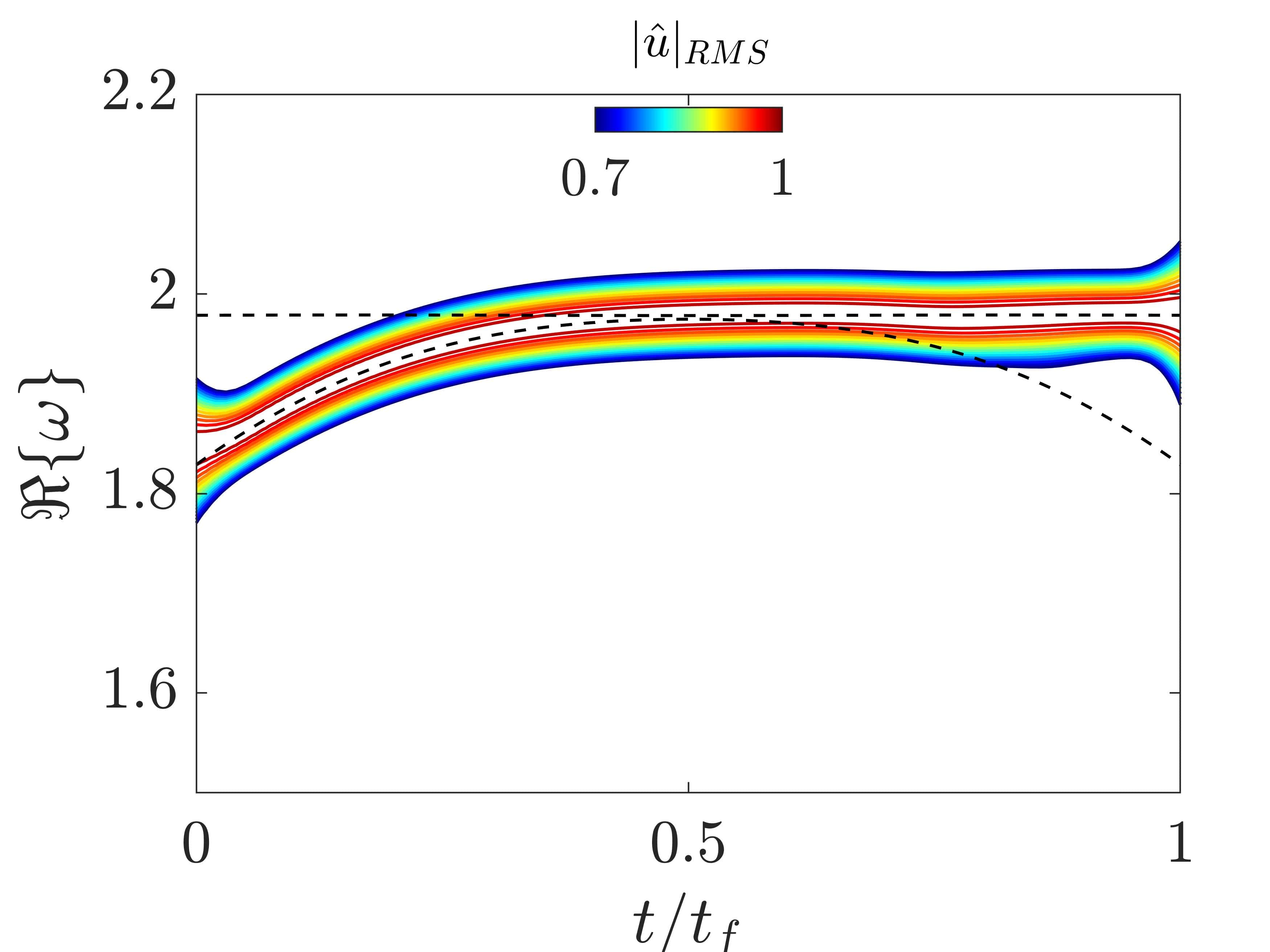}}\hspace{-0.25cm}	
	\subfigure[]{\includegraphics[width=0.32\textwidth]{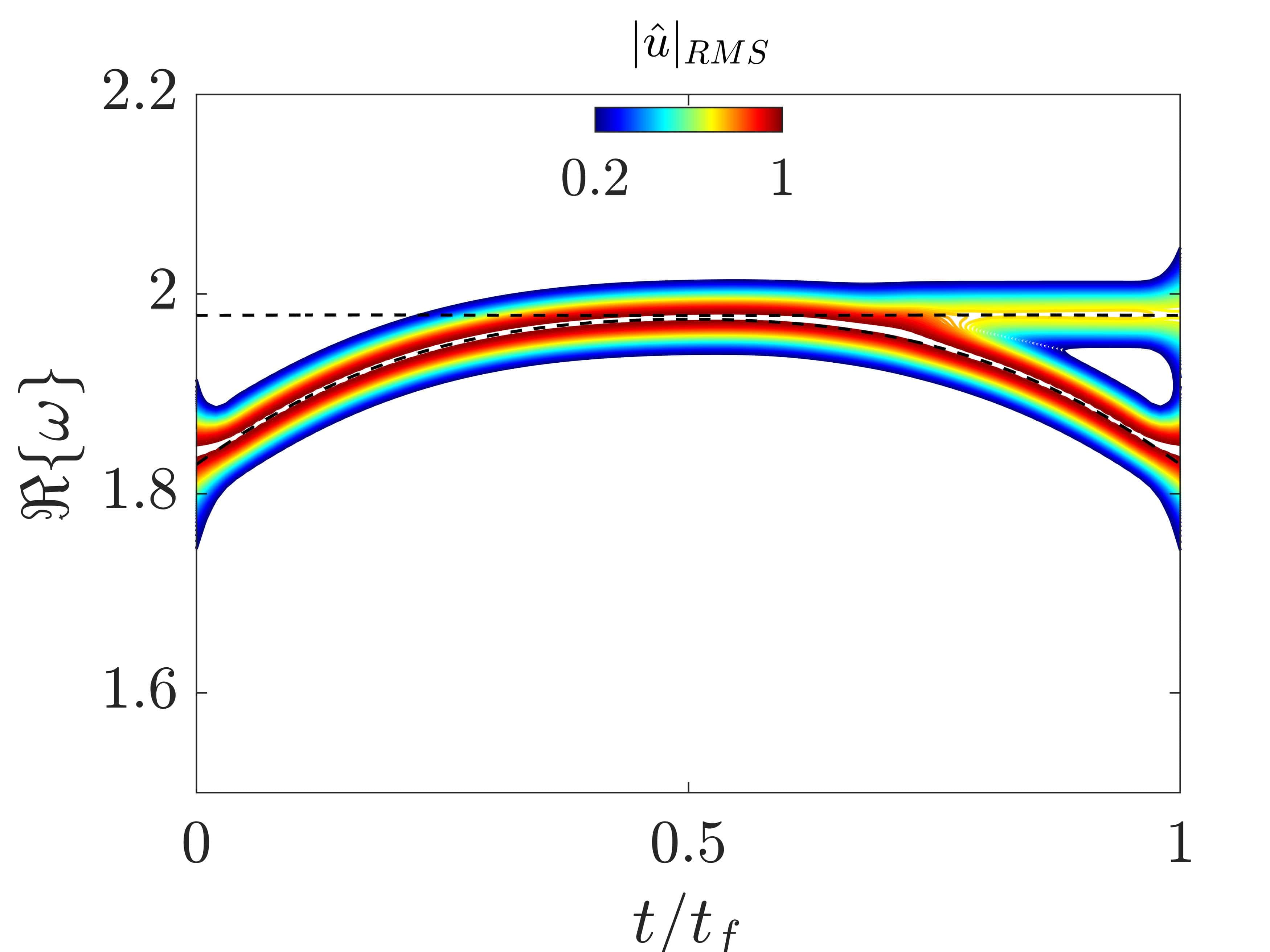}}\hspace{-0.25cm}	\subfigure[]{\includegraphics[width=0.32\textwidth]{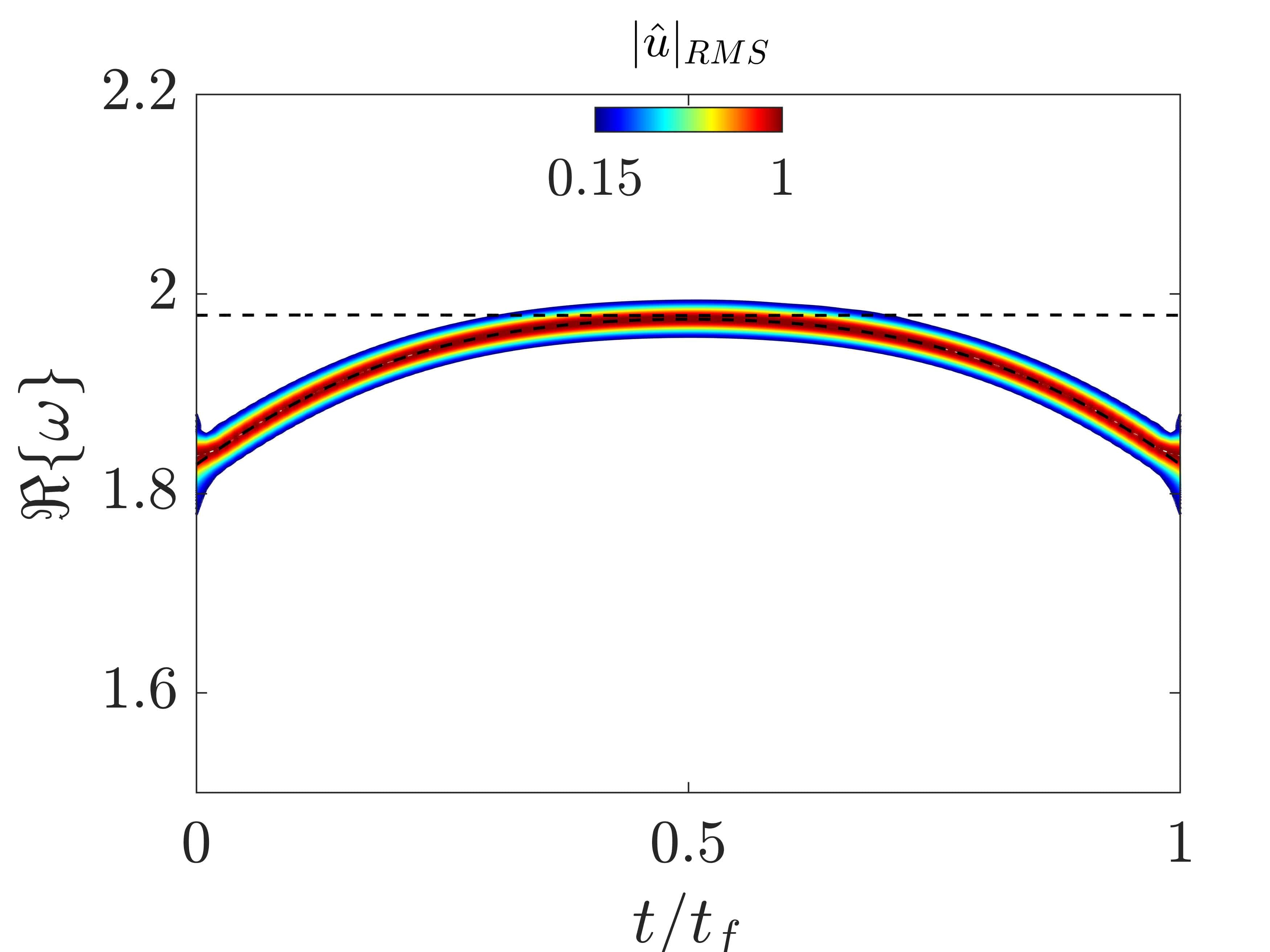}}
	\caption{(a-c) Displacement field of the lattice in response to initial conditions for (a) fast, (b) intermediate and (c) slow temporal modulations. The energy is injected in the topological mode localized on the right boundary. A complete edge-to-edge transfer is achieved only in case of slow modulation. (d-f) Corresponding spectrograms, showing that the edge-to-edge transition occurs along the topological branch in case of adiabatic transformation (f), while there is scattering of energy to bulk modes in case of (e) intermediate and (d) fast modulations.}
	\label{Fig5}    
\end{figure}
The propagating bands, highlighted with gray boxes in the figure, are populated by a number of bulk states. Interestingly, a right localized topological state is spanning the gap within the first half of parameter space $\phi/\pi\in\left[0,1\right]$ (represented with a red solid curve) and left localized for $\phi/\pi\in\left[1,2\right]$ (represented with a red dashed curve). In contrast, the central point $\phi/\pi=1$, corresponds to a motion that encompasses the entire chain, and constitutes a transition point between right and left localization when smooth modulations of $\phi$ are applied. 
Additional details on the dispersion relation, topological invariant, and existence of localized states are reported in Appendix B.\\
Consider now that the energy is initially injected in the right localized topological state, corresponding to a starting value $\phi_i/\pi=0.5$. For a sufficiently slow temporal modulation, a modal transformation takes place such that the evolution of the associated state follows the characteristics of the topological branch illustrated in Figure \ref{Fig4}(b). In other words, we pursue topological pumping in time, in the attempt to drive the transfer of the edge state from the right to the left boundary. 
The transient analysis is performed using three different modulation velocities $\Omega_1=10^{-3}\pi$, $\Omega_2=3.3\cdot10^{-4}\pi$, and $\Omega_3=10^{-4}\pi$, in the neighborhood of the limiting speed for adiabaticity, which is represented in Figure \ref{Fig4}(c) and evaluated through the procedure detailed in Appendix A. In case of fast modulation ($\Omega_1=10^{-3}\pi$), the right-to-left transfer of the edge state fails, according to the displacement field illustrated in Figure \ref{Fig5}(a). This is confirmed by the spectrogram in Figure \ref{Fig5}(d), which displays a spectral content initially located in the lower branch. In correspondence of the transition point (which is the most critical region) the energy is transferred to the bulk mode, instead of being frequency-transformed along the topological branch. When an intermediate modulation speed is applied ($\Omega_2=3.3\cdot10^{-4}\pi$), part of the energy is transferred to the opposite edge, and part is scattered to the bulk mode, which reflects on additional vibrational content in the central part of the chain, as shown in Figure \ref{Fig5}(b). As such, the spectral content in the lattice is split between the topological and the neighboring state. Figures \ref{Fig5}(c,f) instead display an adiabatic energy transfer that consists in a complete transformation from edge-to-edge of the topological state, corresponding to a modulation speed $\Omega_3=10^{-4}\pi$. Under this configuration, the scattering to bulk states is negligible and the process is said to be adiabatic. 

Consider now that each mass is endowed with a control force that is able to locally induce gain/loss and stiffness modulation in form of $\gamma_i,\beta_i$, according to the schematic in Figure \ref{Fig4}(a). We firstly assume to have a number of gain/loss elements that equals the number of masses, which is not strictly necessary to create a shortcut and represents only one of the available solutions. This assumption will be relaxed in the final part of the paper, in which a single gain/loss element is employed to pursue the same scope. 
\begin{figure}[t]
	\centering
	\subfigure[]{\includegraphics[width=0.3\textwidth]{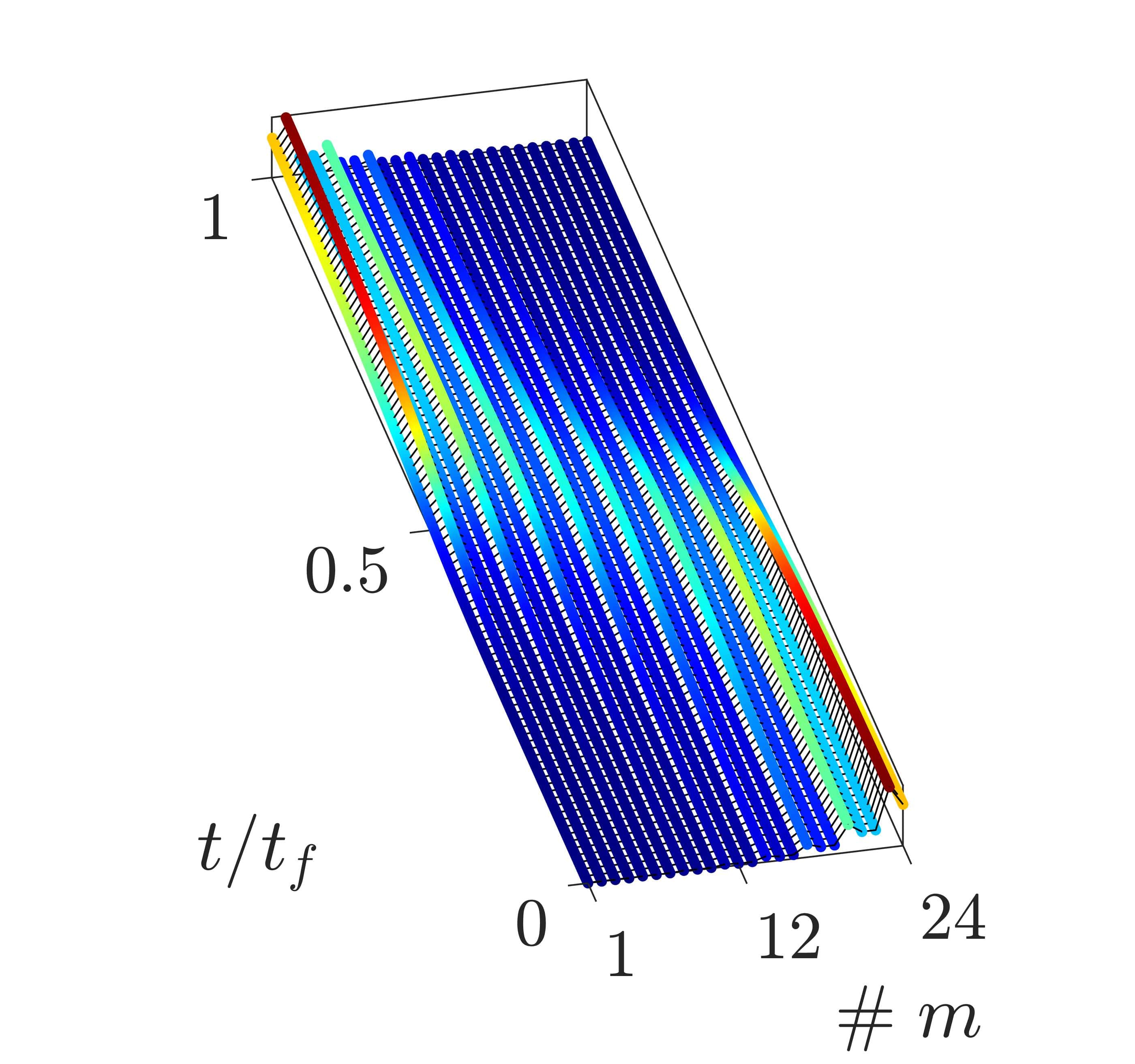}}\hspace{-0.25cm}
	\subfigure[]{\includegraphics[width=0.3\textwidth]{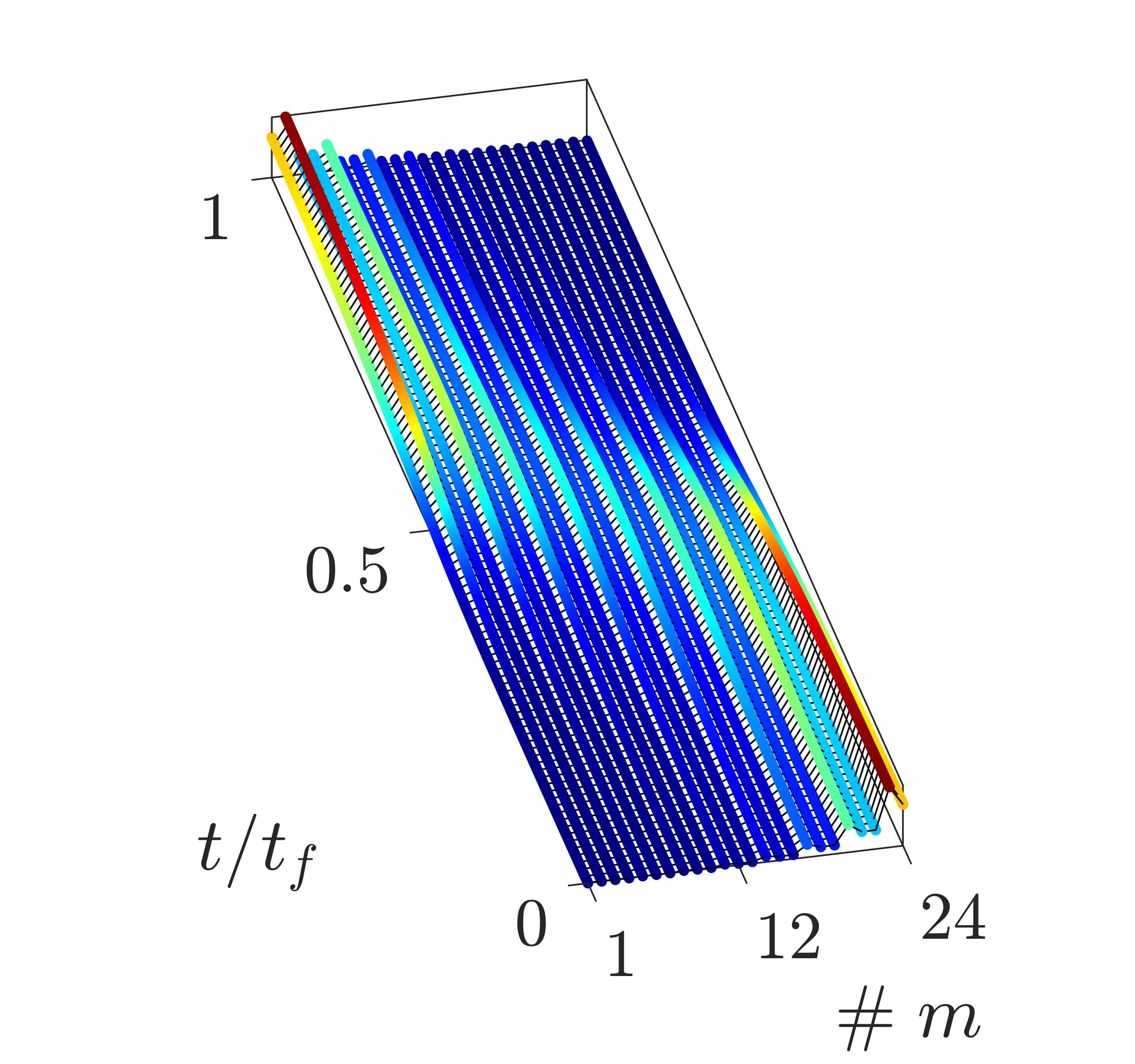}}\hspace{-0.25cm}
	\subfigure[]{\includegraphics[width=0.3\textwidth]{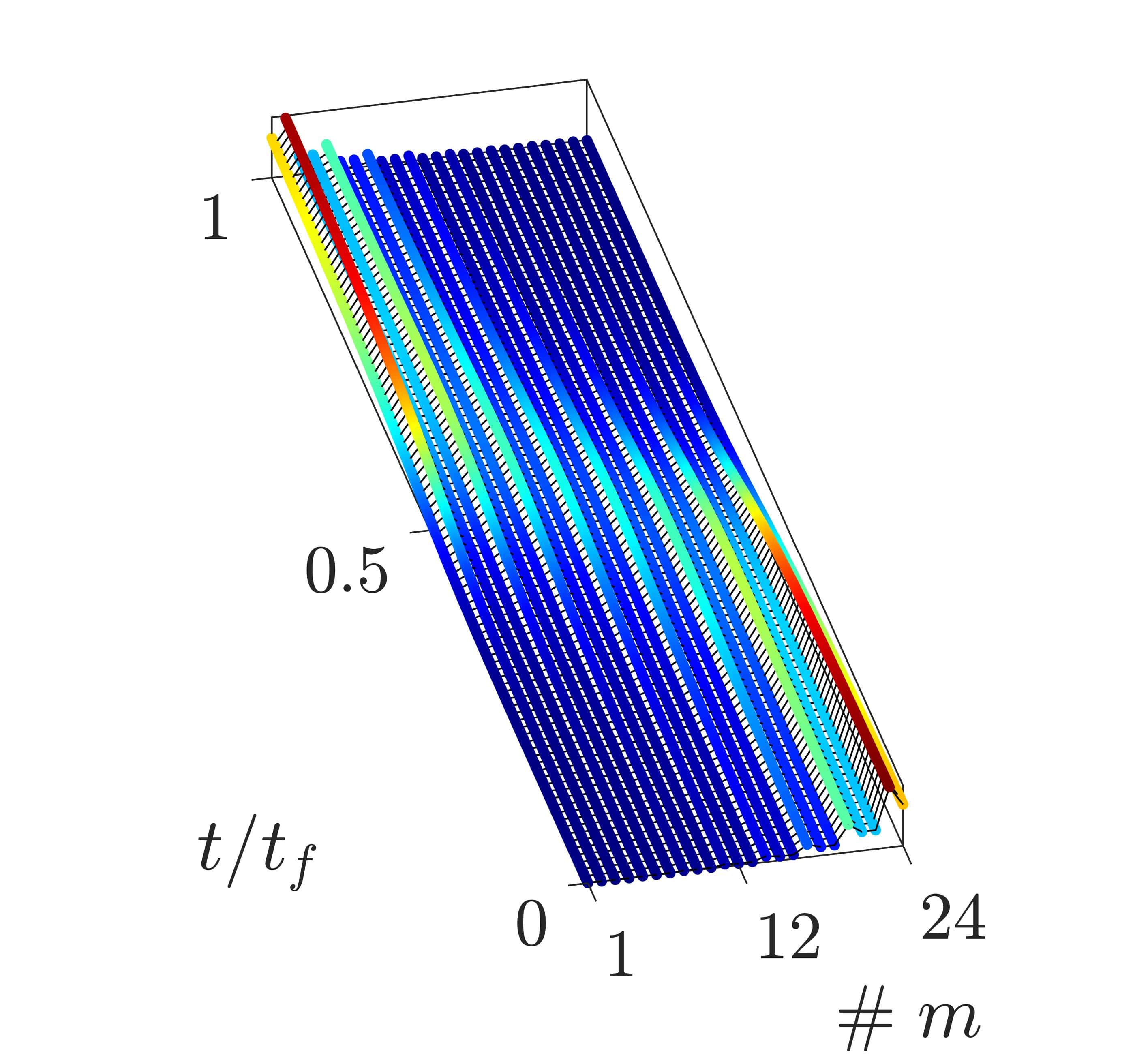}}\\
	\subfigure[]{\includegraphics[width=0.32\textwidth]{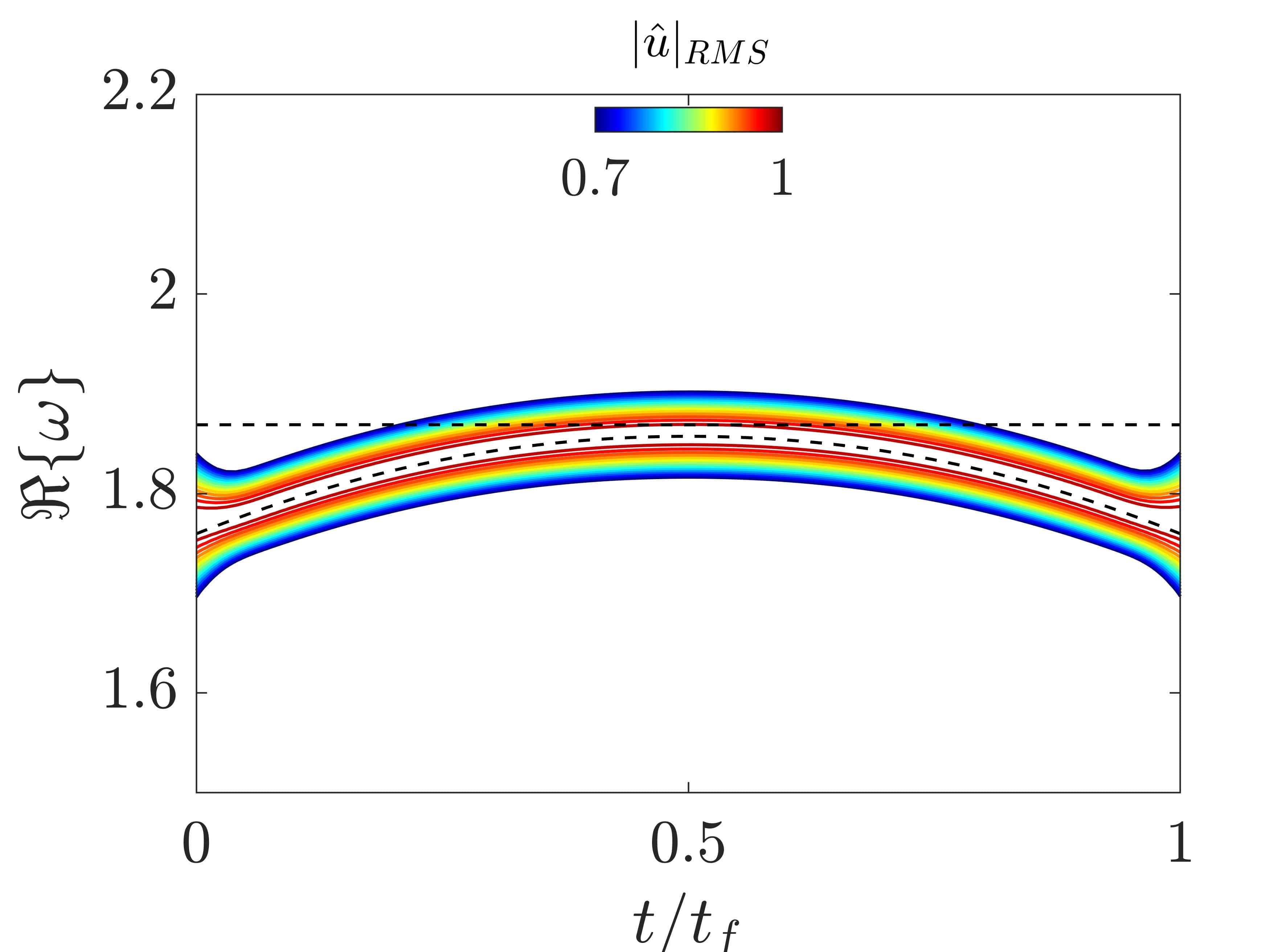}}\hspace{-0.25cm}	
	\subfigure[]{\includegraphics[width=0.32\textwidth]{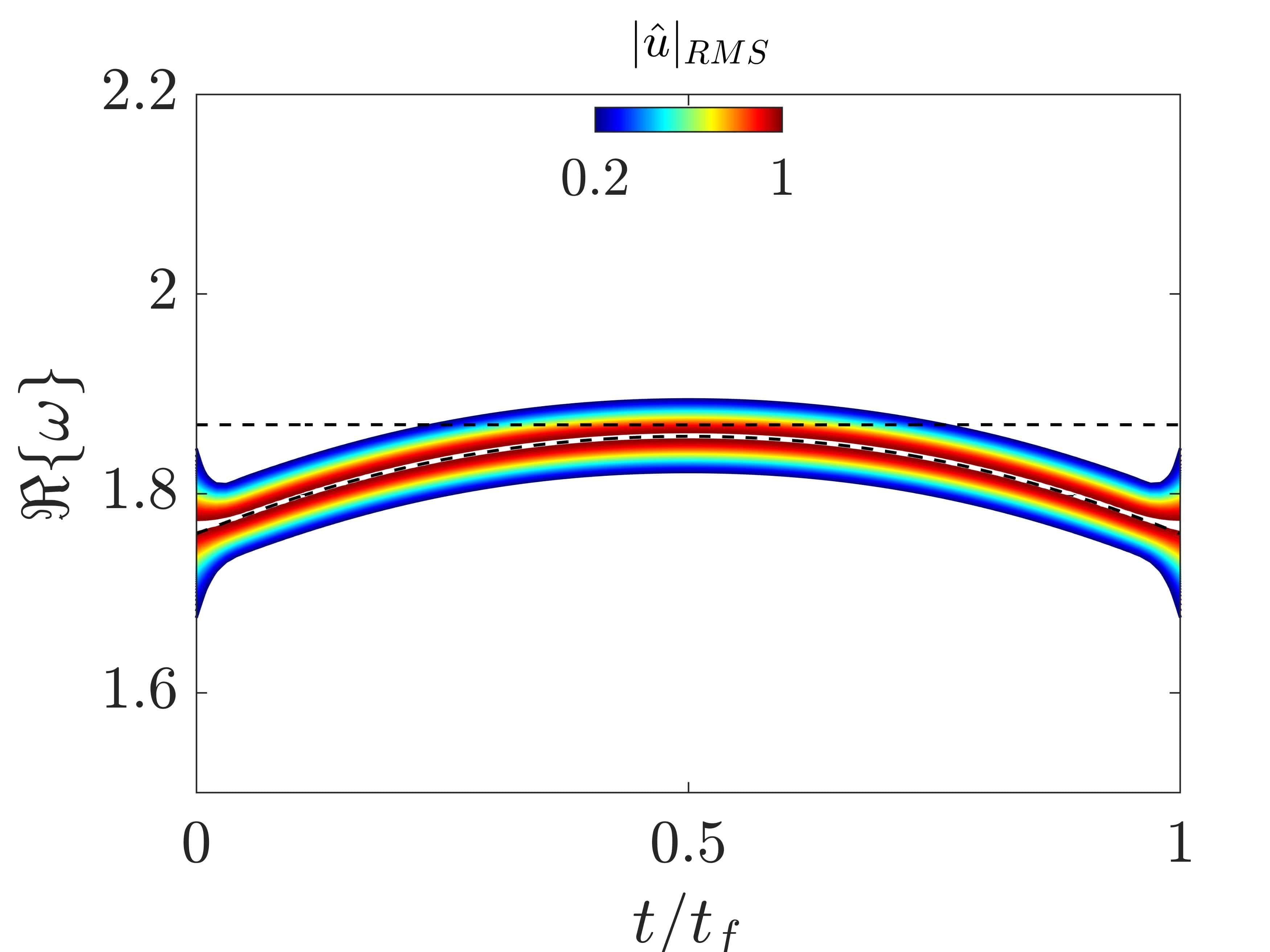}}\hspace{-0.25cm}	\subfigure[]{\includegraphics[width=0.32\textwidth]{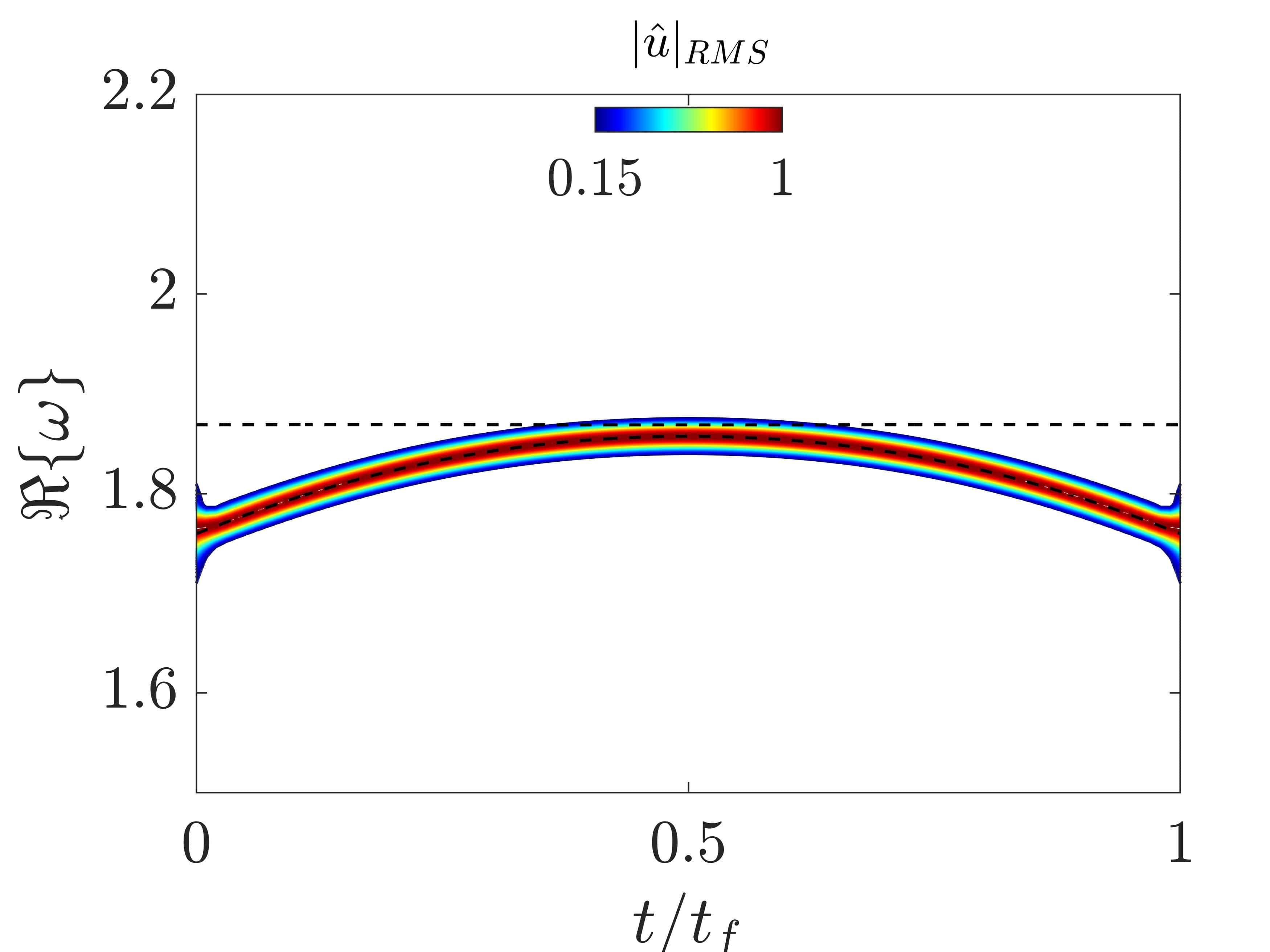}}
	\caption{(a-c) Displacement field for the lattice with a non-hermitian shortcut in the case of three relevant modulation velocities (fast-intermediate-slow). The three configurations exhibit the same edge-to-edge transformation, independently of the speed of modulation. (d-f) Corresponding spectrogram, that illustrate a frequency transformation along the topological branch.}
	\label{Fig6}    
\end{figure}
The dynamic equation projected in the modal coordinate set is equivalent to Equation \ref{eq:07}, except that now the $2\times2$ rotation matrix $R$ is replaced with the $N\times N$ orthonormal eigenvector matrix $\Psi$ associated with the quasi-static eigenvalue problem $\lambda\bm{u}=M^{-1}K\bm{u}$:
\begin{equation}
	\dot{\bm{z}}=H\bm{z}\hspace{1cm}H=\begin{bmatrix}
		-\Psi^{-1}\left(2\displaystyle\dot\Psi+D_1^\gamma \Psi\right)&-\Psi^{-1}\left(\displaystyle\ddot\Psi+D_1^\gamma\dot\Psi+D_2^\beta\Psi\right)\\[8pt]
		I&0
	\end{bmatrix}
	\label{eq:15}
\end{equation}
where now $D_1^\gamma=diag\left\{\gamma_1,\ldots,\gamma_i,\ldots,\gamma_N\right\}$ and $D_2^\beta=D_2+diag\left\{\beta_1,\ldots,\beta_i,\ldots,\beta_N\right\}$ are the dynamic stiffness matrices of the lattice. We consider the left top and right top blocks of the matrix $H$:
\begin{equation}
\begin{cases}
&-2\Psi^{-1}\dot\Psi-\Psi^{-1}D_1^\gamma \Psi\\[5pt]
&-\Psi^{-1}\ddot\Psi-\Psi^{-1}D_1^\gamma\dot\Psi-\Psi^{-1}D_2^\beta\Psi
\end{cases}
\label{eq:16}
\end{equation}
to nullify the coupling between two distinct modes $(r,s)$, we require that the coefficient in line $r$ and column $s$ of Equations \ref{eq:16} to be zero, yielding the following relations for $\gamma_i$ and $\beta_i$:
\begin{equation}
\begin{split}
&\gamma_i=-\frac{2\dot\psi_{i,s}}{\psi_{i,s}}\\[5pt]
&\beta_i=-\frac{\ddot\psi_{i,s}+\gamma_n\dot{\psi}_{i,s}}{\psi_{i,s}}
\end{split}
\label{eq:17}
\end{equation}
where $\psi_{i,s}$ is the $i^{th}$ amplitude coefficient associated to the $s^{th}$ eigenvector. Notice that the expressions are similar to the one reported in Equation \ref{eq:sol} for the pair of coupled oscillators and involve the time derivatives of the eigenvector coefficients. As a result, greater $\gamma_i$ and $\beta_i$ values are required for a faster modulation of the phason. 
The mathematical steps necessary to get to Equations \ref{eq:17} are reported in Appendix C.\\
To corroborate the above arguments, the system is simulated under the same initial conditions employed for the Hermitian configuration. Here, in contrast, each lattice site is augmented by a damping/stiffness parameter $\gamma_i,\beta_i$ that constitute the non-hermitian shortcut. 
\begin{figure}[t]
	\centering
	\subfigure[]{\includegraphics[width=0.3\textwidth]{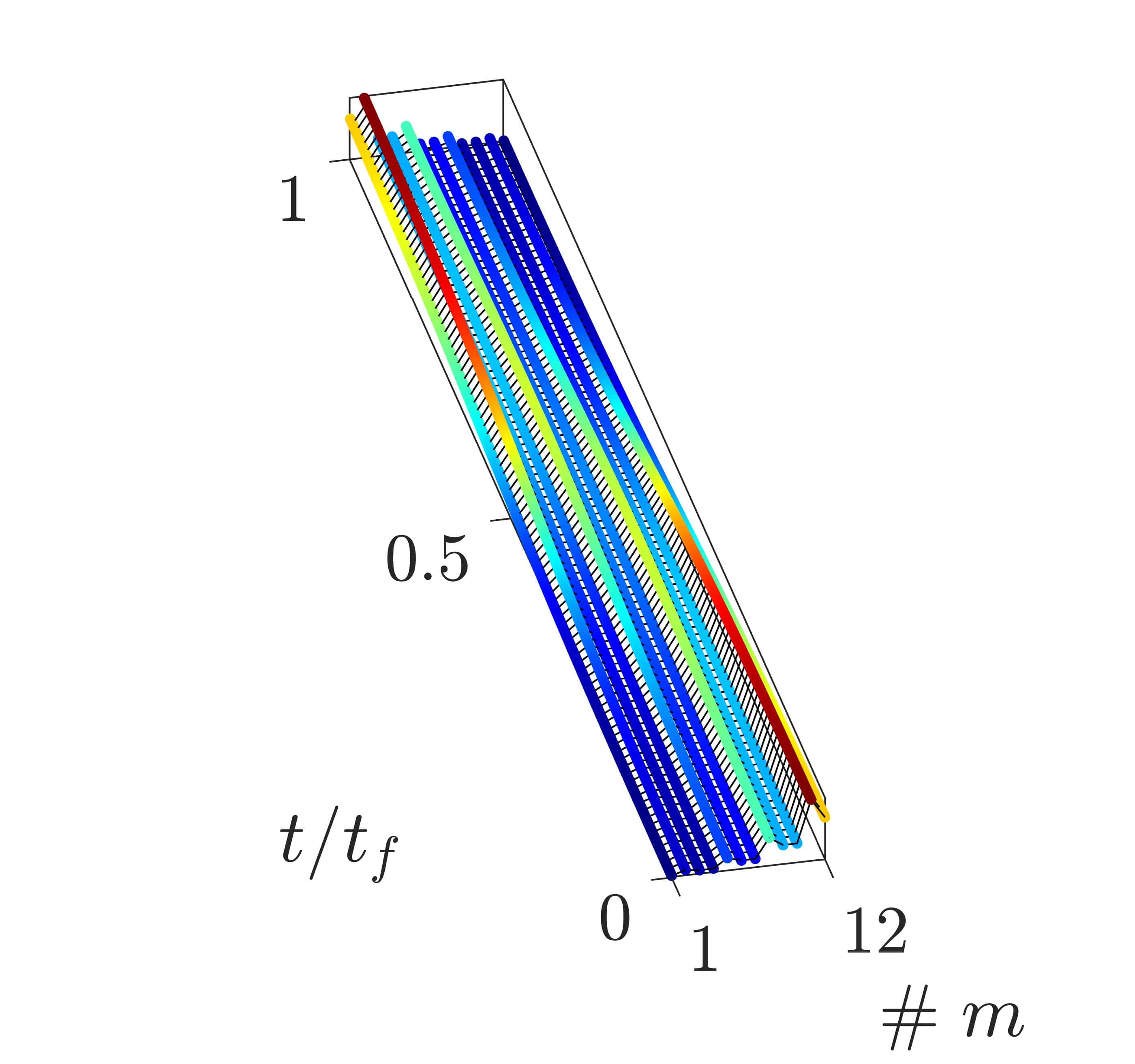}}\hspace{-0.25cm}
	\subfigure[]{\includegraphics[width=0.3\textwidth]{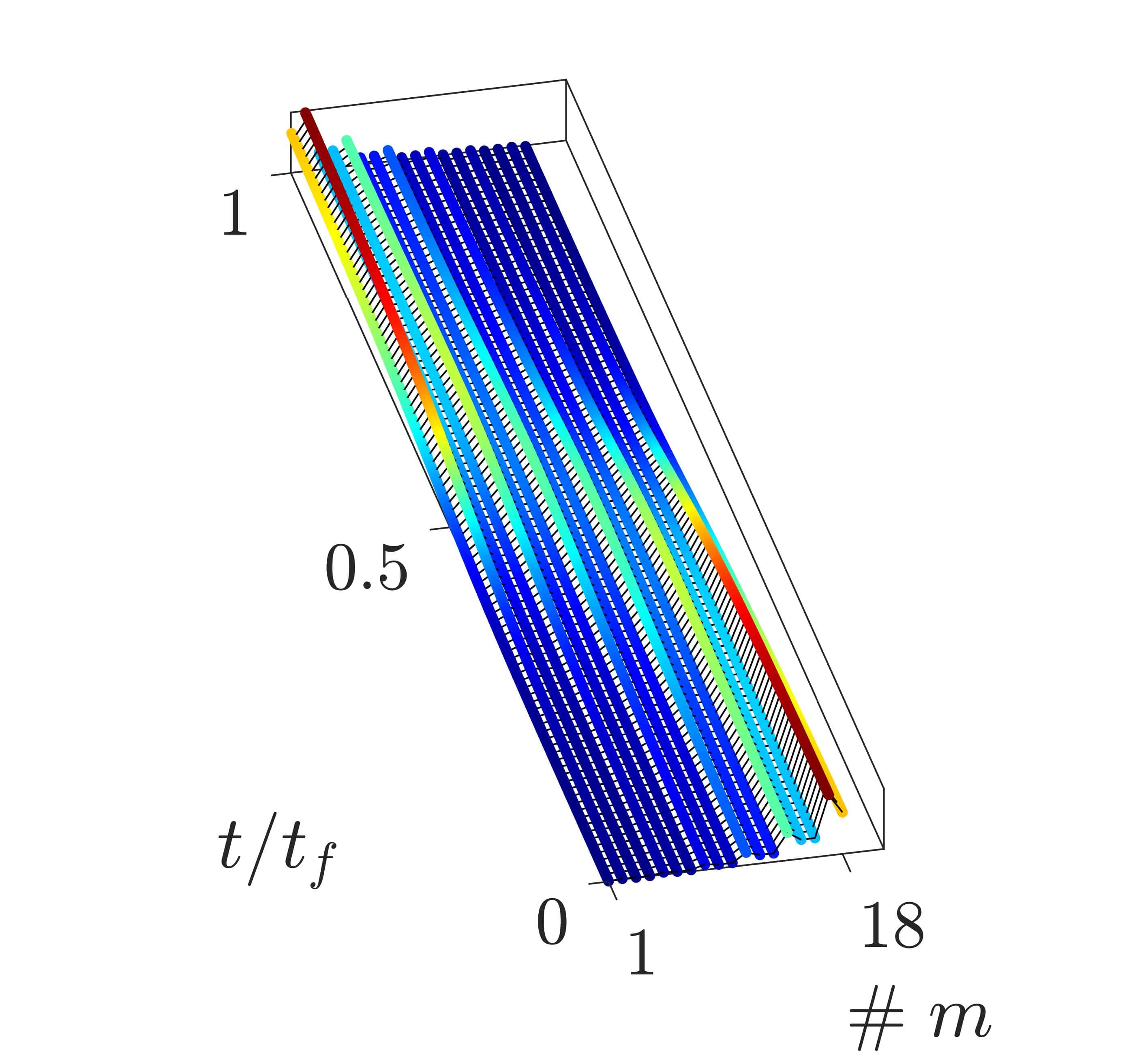}}\hspace{-0.25cm}
	\subfigure[]{\includegraphics[width=0.3\textwidth]{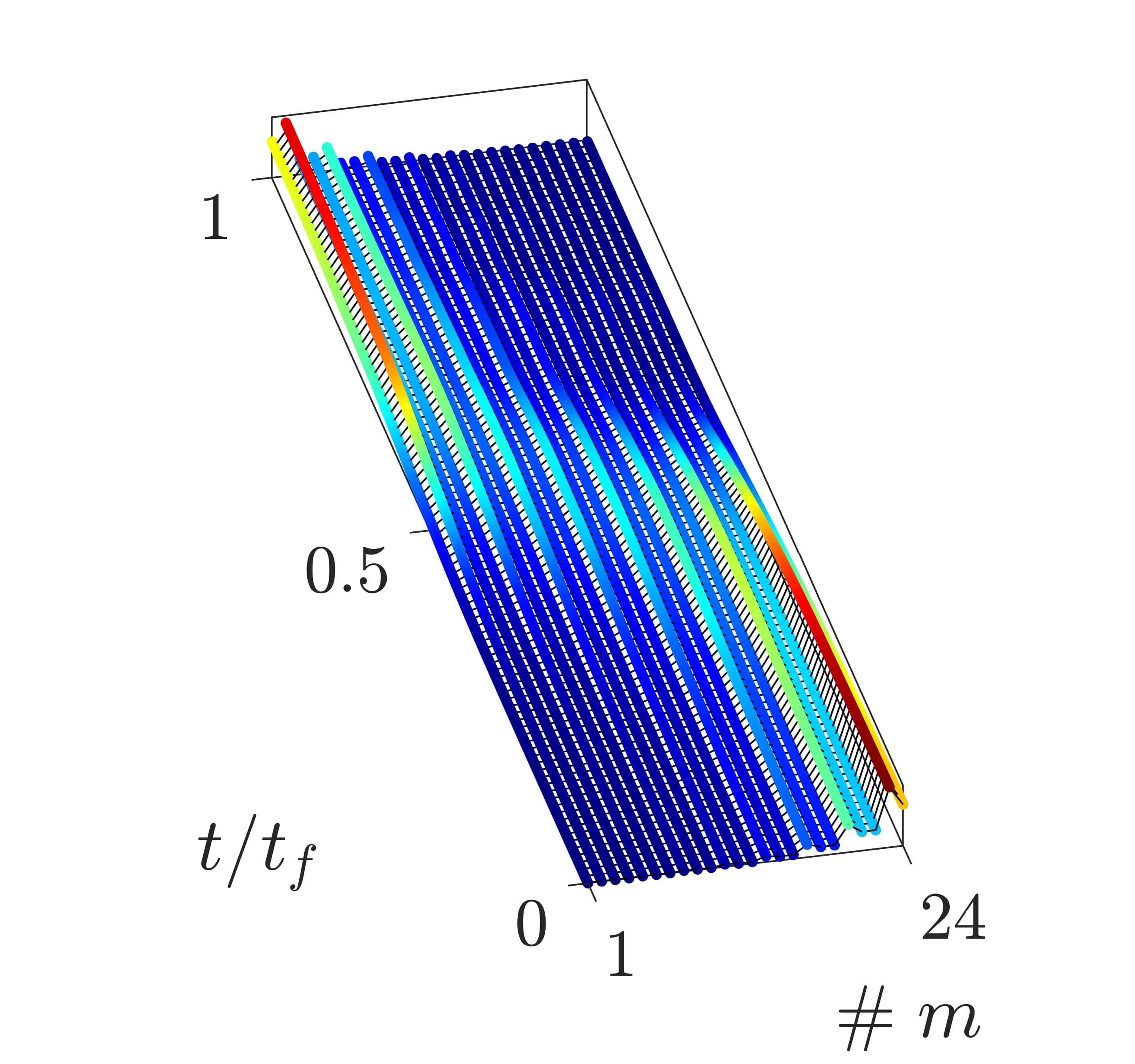}}\\
	\subfigure[]{\includegraphics[width=0.32\textwidth]{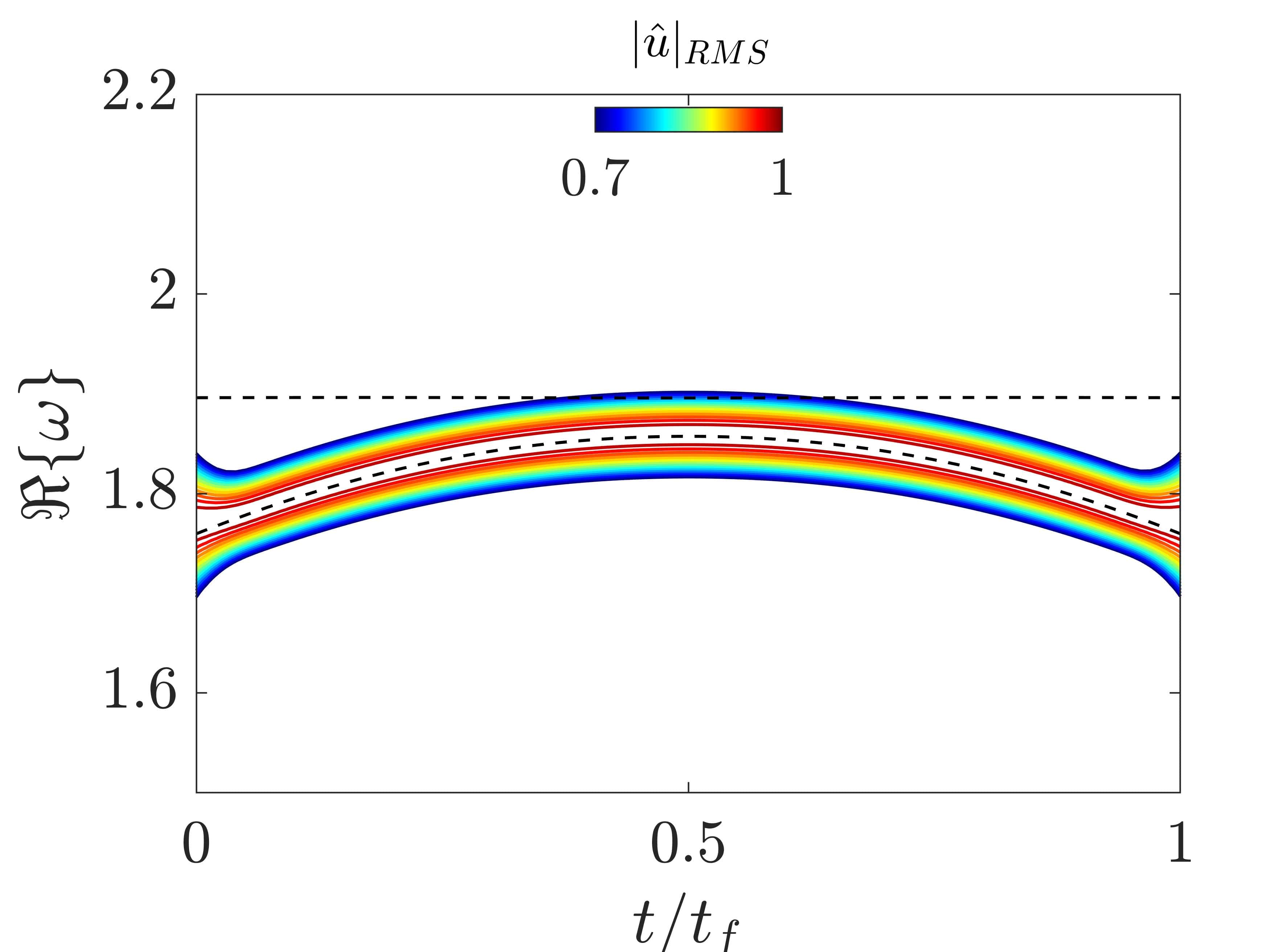}}\hspace{-0.25cm}	
	\subfigure[]{\includegraphics[width=0.32\textwidth]{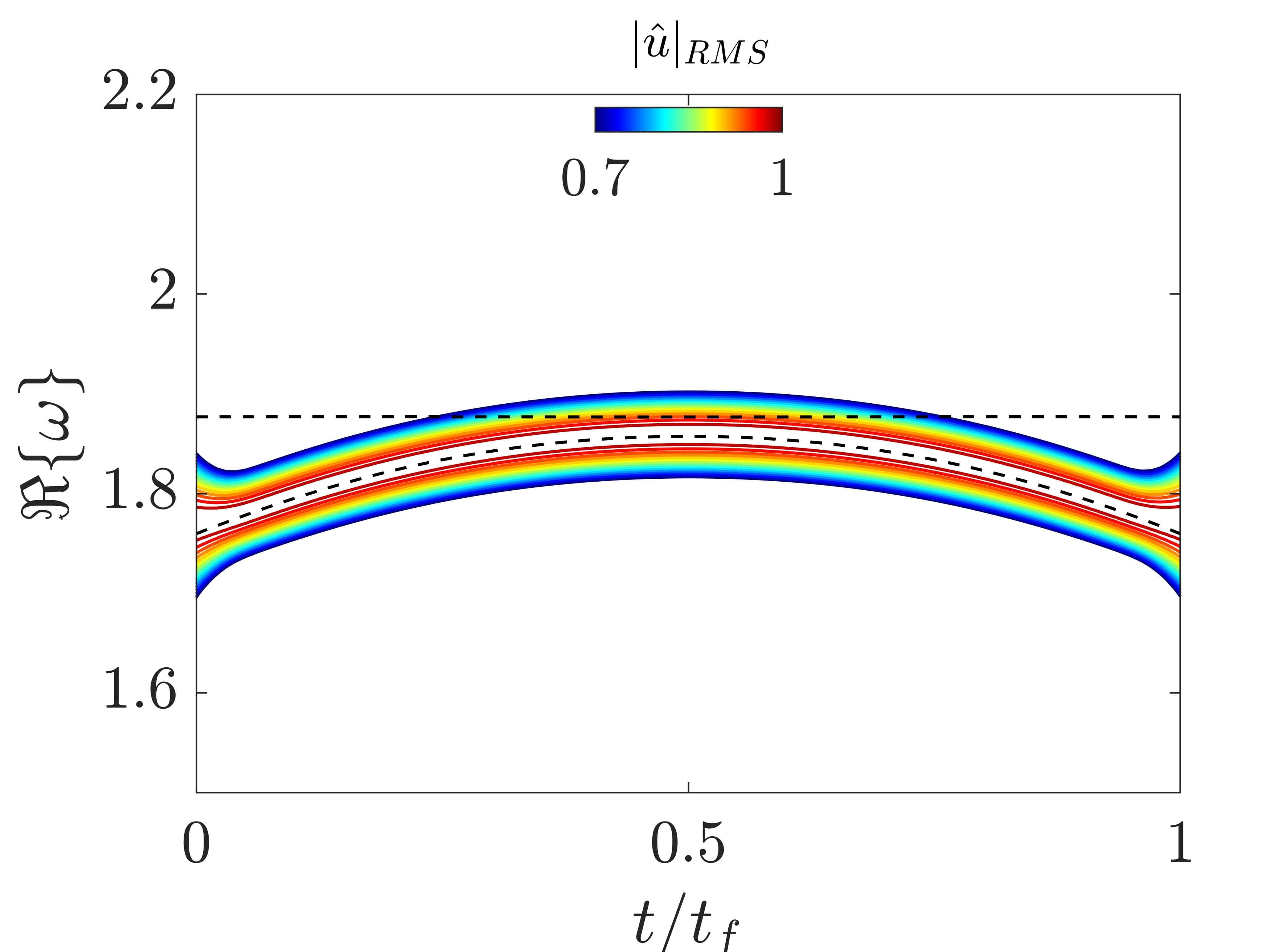}}\hspace{-0.25cm}	\subfigure[]{\includegraphics[width=0.32\textwidth]{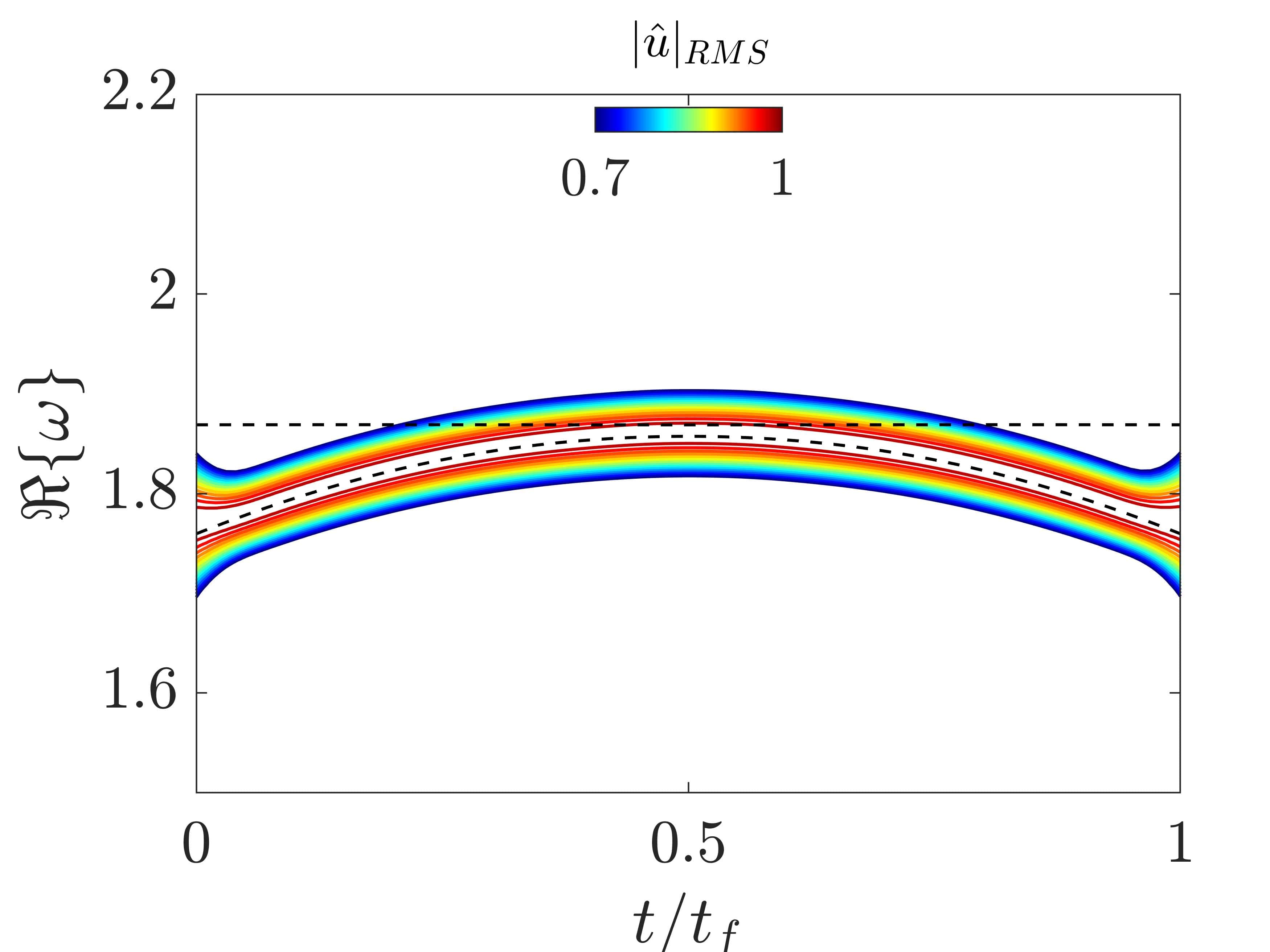}}
	\caption{(a-c) Displacement field of a lattice with (a) $N=12$, (b) $N=18$, (c) $N=24$ masses with applied non-hermitian shortcut, generated by way a single gain/loss pair. The simulations are performed under the same modulation velocity $\Omega_1=10^{-3}\pi$, which corresponds to nonadiabatic transitions in case the shortcut is not applied to the lattice. An edge-to-edge transformation is successfully achieved for the probed configurations. (d-f) corresponding spectrograms, illustrating the same frequency transition in time. }
	\label{Fig7}    
\end{figure}
The numerical results are displayed in Figure \ref{Fig6} for the three probed modulation velocities and, as expected, the achieved displacement field is invariant with $\Omega$, except for the time-scale, which is different (the final time of the simulations is $t_f=10^3,3\cdot10^3,10^4\;{\rm s}$ respectively). This difference reflects on a coarser resolution on the corresponding spectrograms that, however, describe the same transition along the topological branch and the same displacement field. That is, edge-to-edge temporal pumping is demonstrated to occur with a frequency transformation independently of the modulation velocity and, in particular, for values not allowed by the adiabatic limit.

Finally, we demonstrate that a single $\gamma_i,\beta_i$ pair applied to the $i^{th}$ and $(N-i+1)^{th}$ masses of the chain is sufficient to break the modal coupling between the neighboring states $\left(r,s\right)$, yielding the following values for $\gamma_i,\beta_i,\gamma_{N-i+1},\beta_{N-i+1}$:
\begin{equation}
\begin{split}
	&\gamma_i=-\gamma_{N-i+1}=-\frac{2\sum_{i=1}^{N}\Psi_{i,r}\dot\Psi_{i,s}}{\Psi_{i,r}\Psi_{i,s}-\Psi_{N-i+1,r}\Psi_{N-i+1,s}}\\[5pt]
	&\beta_i=-\beta_{N-i+1}=-\frac{\sum_{i=1}^{N}\Psi_{i,r}\ddot\Psi_{i,s}+\sum_{i=1}^{N}\Psi_{i,r}\gamma_i\dot\Psi_{i,s}}{\Psi_{i,r}\Psi_{i,s}-\Psi_{N-i+1,r}\Psi_{N-i+1,s}}
\end{split}
\label{eq:18}
\end{equation}
which is one of the available shortcut solutions as reported and discussed in Appendix C. The new expressions in Equation \ref{eq:18} are numerically tested through a fast modulation $\Omega_1=10^{-3}\pi$ and upon varying the size of the lattice, corresponding to a set of chains made of $N=12,18,24$ point masses. The position of the $i^{th}$ gain/loss pair can also be arbitrarily set and, without loss of generality, we chose $i=N/2$. Numerical results are reported in Figure \ref{Fig7}(a-f) in terms of displacement field and corresponding spectrogram. It is observed that, for all the probed configurations, the gain/loss pair activates a shortcut that is capable of violating the adiabatic limit and to perform a complete edge-to-edge transitions of the topological state.

\section{Conclusions}
We have presented a systematic procedure to generate a shortcut for a nonadiabatic transformation in a topological lattice. In the first part of the paper, the theoretical aspects are unfolded through the dynamic analysis of a pair of coupled oscillators, starting from the limit of adiabaticity to the evaluation of the necessary gain/loss action to generate the shortcut. 
The concept has been later generalized to mechanical lattices supporting topological edge states. We have shown a number of examples in which the barriers of adiabaticity have been violated to achieve superfast edge-to-edge transitions, expanding the available literature in the context of adiabatic pumping and modal transformations.
The concept can be also generalized to multi-dimensional systems, to the continuum and to make it applicable in different physical realms.

\section{Acknoweledgments}
The Italian Ministry of Education, University and Research is acknowledged for the support provided through the Project ``Department of Excellence LIS4.0 - Lightweight and Smart Structures for Industry 4.0".

\appendix
\newcommand{\beginappendix}{%
	\setcounter{table}{0}
	\renewcommand{\thetable}{A\arabic{table}}%
	\setcounter{figure}{0}
	\renewcommand{\thefigure}{A\arabic{figure}}%
}
\beginappendix
\section{Delineating the limiting speed for adiabaticity}
This section is intended to recall the general relationship describing the evolution of a state (in terms of amplitude) for a mechanical system with time-dependent parameters. The procedure hereafter presented yields a quantitative expression of the limiting speed for an adiabatic transformation, i.e. a transformation without scattering of energy to neighboring modes, which is preliminary for the numerical assessment reported in section II.\\
Consider a second order linear time varying dynamical system which is subjected to given initial conditions for the velocity $\dot{\bm{u}}_0$ and the position $\bm{u}_0$. Moreover, the general equation describing the free motion of the kinematic quantities writes:
\begin{equation}
	M\left(t\right)\ddot{\bm{u}}+C\left(t\right)\dot{\bm{u}}+K\left(t\right)\bm{u}=0
\end{equation}
where $M\left(t\right)$, $C\left(t\right)$ and $K\left(t\right)$ are the time-dependent mass, damping and stiffness matrices, respectively. It is convenient to express the dynamic equation in terms of a first order differential equation with augmented state vector $\bm{\psi}=\left[\dot{\bm{u}};\bm{u}\right]$. We get to:
\begin{equation}
	\dot{\bm{\psi}}=H\left(t\right)\bm{\psi}
	\label{eq:state_space}
\end{equation}
It is now assumed that the energy content is initially located in a single (reference) state $\bm{\psi}_0=\left[\dot{\bm{u}}_0;\bm{u}_0\right]$. When the temporal modulation is turned on, the neighboring states may be excited and, therefore, part of the energy eventually leak from the reference state. As such, the transient solution in an arbitrary temporal snapshot is described by a linear combination of the $\bm{\psi}_s^R$ available modes $\bm{u}\left(t\right)=\sum_{s}\bm{\psi}_s^R\left(t\right)q_s\left(t\right){\rm e}^{{\rm i}\theta_s\left(t\right)}$, where $\theta_s\left(t\right)=\int_{t_0}^{t}\omega_s\left(\tau\right){\rm d}\tau$ and $q_s$ are the corresponding unknown modal amplitudes. Here the eigenvalues $\omega_s\left(t\right)$ and right/left eigenvectors and $\bm{\psi}_s^{R,L}\left(t\right)$ are computed in the quasi-static sense, i.e. solving the instantaneous eigenvalue problems $H\bm{\psi}_s^R={\rm i}\omega_s\bm{\psi}_s^R$ and $\bm{\psi}_s^LH=\bm{\psi}_s^L{\rm i}\omega_s$. Moreover, $\bm{\psi}_s^R$ and $\bm{\psi}_r^L$ are normalized such that $\left\langle\bm{\psi}_r^L\right|\left.\bm{\psi}_s^R\right\rangle=\delta_{rs}$. For ease of notation, the time dependence is implicitly assumed. The substitution of the Ansatz $\bm{u}\left(t\right)$ into Eq. \ref{eq:state_space} gives:
\begin{equation}
	\sum_s\left(\dot{\bm{\psi}}_s^Rq_s+\bm{\psi}_s^R\dot{q}_s+{\rm i}\omega_s\bm{\psi}_s^Rq_s\right){\rm e}^{{\rm i}\theta_s}=H\sum_{s}\bm{\psi}_s^Rq_s{\rm e}^{{\rm i}\theta_s}
\end{equation}
which can be simplified as:
\begin{equation}
	\sum_s\left(\dot{\bm{\psi}}_s^Rq_n+\bm{\psi}_s^R\dot{q}_s\right){\rm e}^{{\rm i}\theta_s}=0
\end{equation}
Now, each side is multiplied by $\left\langle\psi_r^L\right|$ and the orthogonality $\left\langle\bm{\psi}_r^L\right|\left.\bm{\psi}_s^R\right\rangle=\delta_{rs}$ is exploited to get to:
\begin{equation}
	\dot{q}_r=-\left\langle\bm{\psi}_r^L\left.\right|\dot{\bm{\psi}}_r^R\right\rangle q_r-\sum_{s\neq r}\left\langle\bm{\psi}_r^L\left.\right|\dot{\bm{\psi}}_s^R\right\rangle q_s{\rm e}^{{\rm i}\left(\theta_s-\theta_r\right)}
	\label{eq:adiab1}
\end{equation}
Eq. \ref{eq:adiab1} can be further manipulated by differentiating $H\bm{\psi}_s^R={\rm i}\omega_s\bm{\psi}_s^R$, that is:
\begin{equation}
	\dot{H}\bm{\psi}_s^R+H\dot{\bm{\psi}_s^R}={\rm i}\dot{\omega}_s\bm{\psi}_s^R+{\rm i}\omega_s\dot{\bm{\psi}}_s^R
	\label{eq:evpdiff}
\end{equation}
Eq. \ref{eq:evpdiff} is now multiplied by $\left\langle\bm{\psi}_r^L\right|$ and combined with $\left\langle\bm{\psi}_r^L\left|H\right|\dot{\bm{\psi}}_s^R\right\rangle=\left\langle\bm{\psi}_r^L\left.\right|\dot{\bm{\psi}}_s^R\right\rangle{\rm i}\omega_r$:
\begin{equation}
	\begin{split}
		\left\langle\bm{\psi}_r^L\left|\dot{H}\right|\bm{\psi}_s^R\right\rangle+\left\langle\bm{\psi}_r^L\left|H\right|\dot{\bm{\psi}}_s^R\right\rangle&={\rm i}\dot{\omega}_s\left\langle\bm{\psi}_r^L\left.\right|\bm{\psi}_s^R\right\rangle+{\rm i}\omega_s\left\langle\bm{\psi}_r^L\left.\right|\dot{\bm{\psi}}_s^R\right\rangle\\[5pt]
		\left\langle\bm{\psi}_r^L\left|H\right|\dot{\bm{\psi}}_s^R\right\rangle&={\rm i}\omega_r\left\langle\bm{\psi}_r^L\left.\right|\dot{\bm{\psi}}_s^R\right\rangle
	\end{split}
\end{equation}
for $s\neq r$:
\begin{equation}
	\left\langle\bm{\psi}_r^L\left|\dot{H}\right|\bm{\psi}_s^R\right\rangle={\rm i}\left(\omega_s-\omega_r\right)\left\langle\bm{\psi}_r^L\left.\right|\dot{\bm{\psi}}_s^R\right\rangle
\end{equation}
which is finally substituted into Eq. \ref{eq:adiab1} to get to the general differential equation describing the temporal evolution of the $r^{th}$ modal amplitude $q_r$:
\begin{equation}
	\dot{q}_r=-\left\langle\bm{\psi}_r^L\left.\right|\bm{\dot{\psi}}_r^R\right\rangle q_r-\sum_{s\neq r}\displaystyle\frac{\left\langle\bm{\psi}_r^L\left|\dot{H}\right|\bm{\psi}_s^R\right\rangle}{{\rm i}\left(\omega_s-\omega_r\right)} q_s{\rm e}^{{\rm i}\left(\theta_s-\theta_r\right)}
	\label{eq:adiab2}
\end{equation}
assuming initial conditions for the reference state $q_r(0)=1$ and for $q_s(0)=0$ with $s\neq r$, i.e. all the energy is initially stored in $\bm{\psi}_r$, the temporal evolution of $q_r\left(t\right)$ is given by the integral $q_r\left(t\right)=\int_{0}^{t}\dot{q}_r\left(\xi\right)d\xi$. That is, the contribution of:
\begin{equation}
	\int_{0}^{t}\sum_{s\neq r}\displaystyle\frac{\left\langle\bm{\psi}_r^L\left|\dot{H}\right|\bm{\psi}_s^R\right\rangle}{{\rm i}\left(\omega_s-\omega_r\right)} q_s{\rm e}^{{\rm i}\left(\theta_s-\theta_r\right)}d\xi
	\label{eq:integral}
\end{equation}
is negligible for an adiabatic transition such that the energy present in $q_r$ do not couple in time with the other modes with amplitude $q_s$. 
Eq. \ref{eq:integral} can be integrated by parts:
\begin{equation}
	\int_{0}^{t}\sum_{s\neq m}\displaystyle\frac{\left\langle\bm{\psi}_r^L\left|\dot{H}\right|\bm{\psi}_s^R\right\rangle}{{\rm i}\left(\omega_s-\omega_r\right)} q_s{\rm e}^{{\rm i}\left(\theta_s-\theta_r\right)}d\xi=-\left.\sum_{s\neq r}\displaystyle\frac{\left\langle\bm{\psi}_r^L\left|\dot{H}\right|\bm{\psi}_s^R\right\rangle}{\left(\omega_s-\omega_r\right)^2} q_s{\rm e}^{{\rm i}\left(\theta_s-\theta_r\right)}\right|_0^t+\int_{0}^{t}\frac{d}{d\xi}\left(q_s\displaystyle\frac{\left\langle\bm{\psi}_r^L\left|\dot{H}\right|\bm{\psi}_s^R\right\rangle}{\left(\omega_s-\omega_r\right)^2}\right){\rm e}^{{\rm i}\left(\theta_s-\theta_r\right)}d\xi
	\label{eq:final1}
\end{equation}
It can be shown that the integral term on the right hand side is small and therefore can be neglected \cite{ibanez2014adiabaticity}. For adiabaticity we require that:
\begin{equation}
	\left|\displaystyle\frac{\left\langle\bm{\psi}_r^L\left|\dot{H}\right|\bm{\psi}_s^R\right\rangle}{\left(\omega_s-\omega_r\right)^2}\right|<<1
	\label{eq:final2}
\end{equation}
which is an expression for the limiting modulation velocity $\dot{H}$ for any $s\neq r$. Note that if the matrix $H$ is parametric with $\phi$, i.e. $H=H\left(\phi\left(t\right)\right)$, thus $\dot{H}=\frac{\partial H}{\partial\phi}\frac{\partial \phi}{\partial t}$. Eq. \ref{eq:final2} can be rewritten in a more convenient way:
\begin{equation}
	\left|\displaystyle\frac{\left\langle\bm{\psi}_r^L\left|\displaystyle\frac{\partial H}{\partial\phi}\right|\bm{\psi}_s^R\right\rangle}{\left(\omega_s-\omega_r\right)^2}\frac{\partial\phi}{\partial t}\right|<<1
	\label{eq:final3}
\end{equation}
This expression is employed in the paper to estimate the limiting speed $\Omega=\partial\phi/\partial t$ for a time-dependent parametric system of mechanical oscillators. 
\section{Numerical methods}
\begin{figure}[t]
	\centering
	\subfigure[]{\includegraphics[width=0.35\textwidth]{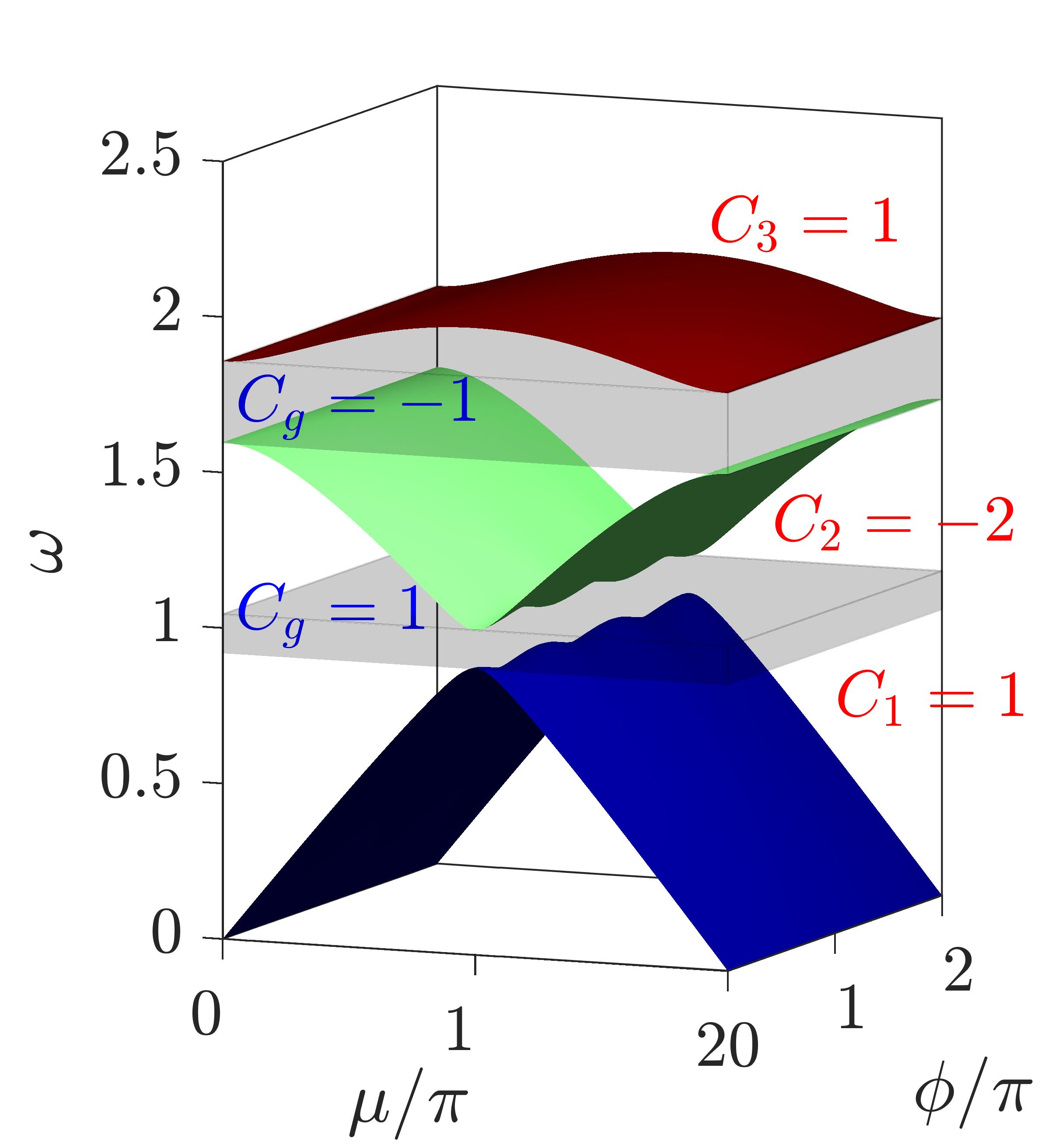}}
	\subfigure[]{\includegraphics[width=0.6\textwidth]{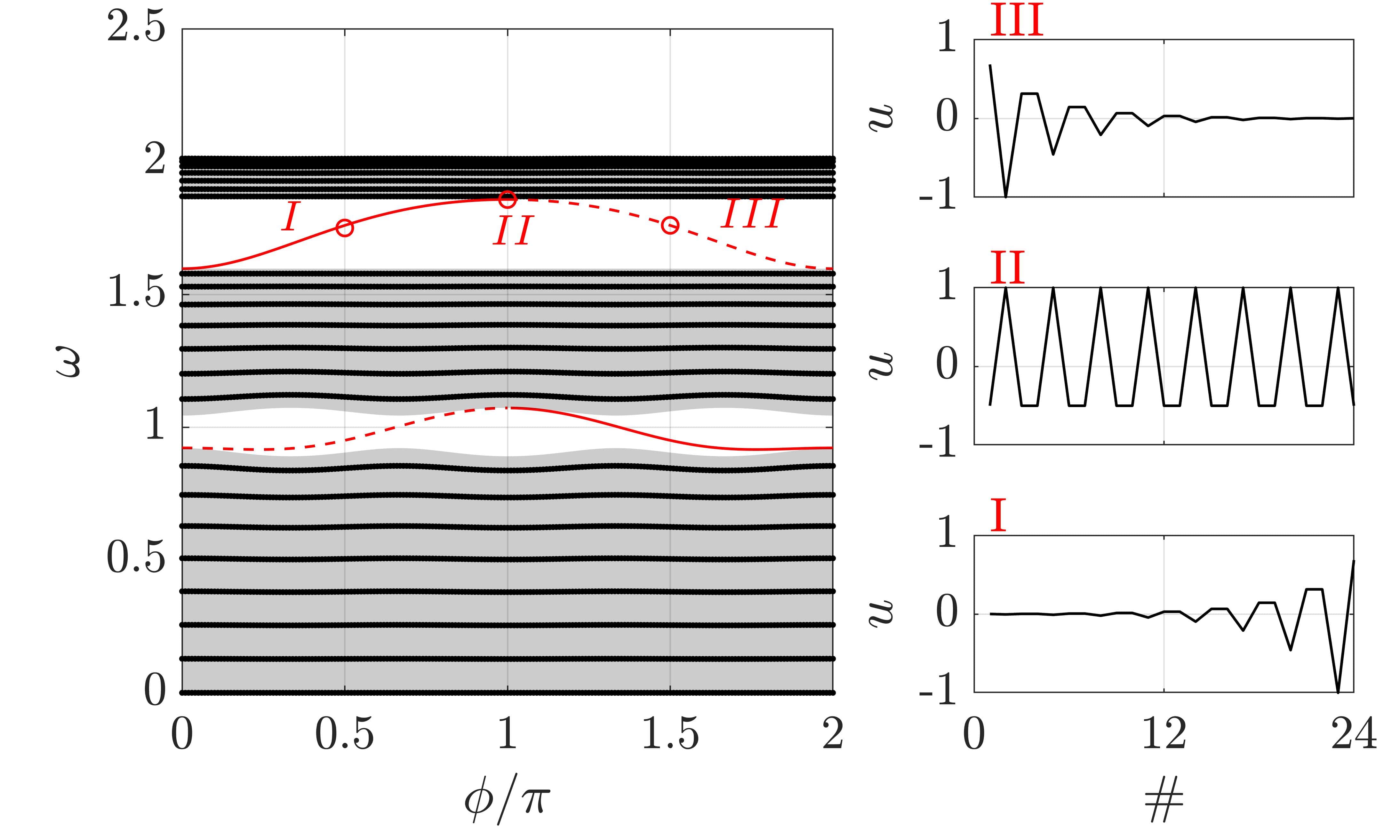}}\hspace{-0.25cm}	
	\caption{(a) Dispersion bands for a lattice with $\theta=1/3$ and corresponding topological invariant. (b) Spectrum of a finite lattice made of $N=24$ mass elements. The bulk and the topological modes are displayed with black and red lines, respectively. Dashed lines denote left localization, whereas solid lines denote right localization. The mode shapes for I-III are shown alongside the spectrum.}
	\label{FigS1}    
\end{figure}

The study of the band topology and existence of localized states is accomplished starting from the dispersion analysis of the unitary cell, which is evaluated through the following eigenvalue problem:
\begin{equation}
	K_R\left(\phi,\mu\right)\bm{u}=\omega^2M_R\left(\mu\right)\bm{u}
	\label{B01}
\end{equation}
where $K_R=W_A^{\dag}KW_A$, $M_R=W_A^{\dag}MW_A$ are the reduced stiffness and mass matrices, and $(\cdot)^\dag$ stands for transpose conjugate of the matrix. 
$W_A$ is the matrix that relates the displacements of the unit element via Bloch conditions $u\left(\omega,x+3a\right)=u\left(\omega,x\right){\rm e}^{{\rm i}3a}$, where $\mu=3a$. 
$M$ and $K$ are the mass and stiffness matrices of the unit cell. $W_A$, $M$, and $K$ are contextually evaluated as:
\begin{equation}
W_A=\begin{bmatrix}
	1&0&0\\
	0&1&0\\
	0&0&1\\
	{\rm e}^{{\rm i}\mu}&0&0
\end{bmatrix}\hspace{0.5cm}
M=\begin{bmatrix}
	m/2&0&0&0\\
	0&m&0&0\\
	0&0&m&0\\
	0&0&0&m/2
\end{bmatrix}\hspace{0.5cm}
K=\begin{bmatrix}
	 k_1&-k_1&0&0\\
	-k_1&k_1+k_2&-k_2&0\\
	0&-k_2&k_2+k_3&-k_3\\
	0&0&-k_3&k_3
\end{bmatrix}
\end{equation}
Since the stiffness modulation is characterized by $\theta=1/3$, the dispersion relation $\omega\left(\phi,\mu\right)$ features three dispersion surfaces that are separated by a gap, as reported in Figure \ref{FigS1}(a) for the entire wavenumber-parameter space. The emergence of localized modes spanning nontrivial gaps is inherently linked with the topological characteristics of the band structure, where the relevant topological invariant is the Chern number $C$:
\begin{equation}
	C_n=\frac{1}{2\pi {\rm i}}\int_D\nabla\times\left(\bm{u}^*\cdot\nabla\bm{u}\right)dD
\end{equation} 
here, the integration domain $D$ spans the two-dimensional torus  $\left[\mu,\phi\right]\in D=\left[0,2\pi\right]\times\left[0,2\pi\right]$ and $\bm{u}$ are the eigenvectors relative to the eigenvalue problem into Equation \ref{B01}. The integration is accomplished for the three bands, which results into the Chern numbers $C_1=1$, $C_2=-2$, and $C_3=1$, respectively. A label for the $r^{th}$ gap is instead assigned to qualify the existence of topological states, which corresponds to the sum of the Chern numbers below the gap $C_g^{(r)}=\sum_{n=1}^rC_n$. In the case at hand, $C_{g}^{(1)}=1$ for the first gap and $C_g^{(2)}=-1$ for the second gap, which describe the mode localization of a topological mode that is spanning the gap upon  varying the phase $\phi$. This is confirmed by the eigenfrequency analysis of a finite lattice made of $N=24$ mass elements under free-free boundary conditions. The corresponding spectrum upon varying the modulation phase $\phi$ is displayed in Figure \ref{FigS1}(b). In the figure, the bulk states are represented by black curves. Such states populate the dispersion bands, whose limits are highlighted with the gray regions. Interestingly, a pair of modes, highlighted in red, are spanning the first and the second gaps. Due to the nontrivial nature of the band structure, such states are topological and can be either localized at one of the boundaries, depending on the modulation parameter $\phi$. 
For instance, the mode spanning the first gap, corresponding to a gap label $C_g^{(1)}=1$ is a left localized mode (dashed line) that transform into a right localized state (solid line) for a positive variation of $\phi$. In contrast, the second gap has $C_g^{(2)}=-1$, which denotes a transition from right to left localization, as shown from the mode shapes displayed in Figure \ref{FigS1} I-III alongside the spectrum.

\section{Generalized shortcut equations}
The necessary condition to break the coupling between two distinct modes $r$ and $s$ is to nullify the $r^{th}$ line and $s^{th}$ column of the matrices reported into Equation \ref{eq:16}. We firstly observe that $\Psi$ is an orthonormal basis, which implies that $\Psi^{-1}=\Psi^T$. Equation \ref{eq:16} can be conveniently rewritten as:
\begin{equation}
	\begin{cases}
		&-2\Psi^T\dot\Psi-\Psi^TD_1^\gamma \Psi\\[5pt]
		&-\Psi^T\ddot\Psi-\Psi^TD_1^\gamma\dot\Psi-\Psi^TD_2^\beta\Psi
	\end{cases}\hspace{0.5cm}with\hspace{0.5cm}
	\Psi=\begin{bmatrix}
	\Psi_{1,1}&\ldots&\ldots&\ldots&\Psi_{N,1}\\
	\vdots&\Psi_{r,r}&\ldots&\Psi_{r,s}&\vdots\\
	\vdots&\vdots&\ddots&\vdots&\vdots\\
	\vdots&\Psi_{s,r}&\ldots&\Psi_{s,s}&\vdots\\
	\Psi_{N,1}&\ldots&\ldots&\ldots&\Psi_{N,N}\\
	\end{bmatrix}
\label{eq:C1}
\end{equation}
In analogy to section II, the first matrix equation in \ref{eq:C1} allows for the evaluation of the damping/anti-damping contributions $\gamma_i$, which are then used in the second matrix equation to compute the stiffness modulation terms $\beta_i$. In particular, the coefficient in position $(r,s)$ is $-2\sum_{i=1}^{N}\Psi_{i,r}\dot\Psi_{i,s}$ for the matrix term $-2\Psi^T\dot\Psi$, and $-\sum_{i=1}^N\Psi_{i,r}\gamma_i\Psi_{i,s}$ is obtained for the matrix term $-\Psi^TD_1^\gamma\Psi$. That is, the first equality to be satisfied is:
\begin{equation}
-2\sum_{i=1}^{N}\Psi_{i,r}\dot\Psi_{i,s}-\sum_{i=1}^N\Psi_{i,r}\gamma_i\Psi_{i,s}=0
\label{eq:C2}
\end{equation}
For the second matrix equation in \ref{eq:C1} we have the coefficient $-\sum_{i=1}^{N}\Psi_{i,r}\ddot\Psi_{i,s}$ from the matrix term $-\Psi^T\ddot\Psi$ in position $(r,s)$, $-\sum_{i=1}^{N}\Psi_{i,r}\gamma_i\dot\Psi_{i,s}$ from the term $-\Psi^TD_1^\gamma\dot\Psi$, and $-\sum_{i=1}^{N}\Psi_{i,r}\beta_i\Psi_{i,s}$. The second equality to be satisfied is therefore:
\begin{equation}
-\sum_{i=1}^{N}\Psi_{i,r}\ddot\Psi_{i,s}-\sum_{i=1}^{N}\Psi_{i,r}\gamma_i\dot\Psi_{i,s}-\sum_{i=1}^{N}\Psi_{i,r}\beta_i\Psi_{i,s}=0
\label{eq:C3}
\end{equation}
Equations \ref{eq:C2} and \ref{eq:C3} must be simultaneously satisfied for the shortcut to occur. One of the $\infty^{2N-2}$ available solutions for $\gamma_i$ and $\beta_i$ is:
\begin{equation}
\begin{split}
&\gamma_i=-\frac{2\dot\Psi_{i,s}}{\Psi_{i,s}}\\[5pt]
&\beta_i=\frac{-\ddot\Psi_{i,s}-\gamma_i\dot\Psi_{i,s}}{\Psi_{i,s}}
\end{split}
\end{equation}
which corresponds to a number $N$ of gain/loss elements distributed along the chain. A easier implementation instead is assuming that only one gain/loss pair is non-null, in correspondence of the $i^{th}$ and ${(N-i+1)}^{th}$ masses:
\begin{equation}
\begin{split}
&\gamma_i=-\gamma_{N-i+1}=-\frac{2\sum_{i=1}^{N}\Psi_{i,r}\dot\Psi_{i,s}}{\Psi_{i,r}\Psi_{i,s}-\Psi_{N-i+1,r}\Psi_{N-i+1,s}}\\[5pt]
&\beta_i=-\beta_{N-i+1}=-\frac{\sum_{i=1}^{N}\Psi_{i,r}\ddot\Psi_{i,s}+\sum_{i=1}^{N}\Psi_{i,r}\gamma_i\dot\Psi_{i,s}}{\Psi_{i,r}\Psi_{i,s}-\Psi_{N-i+1,r}\Psi_{N-i+1,s}}
\end{split}
\end{equation}
yielding the second shortcut solution employed in the final part of the paper.

\bibliography{References}
\end{document}